\pdfoutput=1
\documentclass[12pt,a4paper]{book}
\usepackage[vmarginratio=1:1,hmarginratio=1:1]{geometry} % to have a centered title page
\usepackage{emptypage} % to remove headers and footers from empty pages
\usepackage[utf8]{inputenc}
\usepackage{xcolor}
\usepackage{graphicx}
\graphicspath{{./figures/}
              {./figures/chapter1/}
              {./figures/chapter2/}
              {./figures/chapter3/}
              {./figures/chapter2/DoublonDecay/}}
\usepackage[fleqn,tbtags]{mathtools}
\usepackage{amssymb}
\usepackage{libertine}
\usepackage{libertinust1math}
\usepackage[T1]{fontenc}
\usepackage{amsbsy}
\usepackage[spanish,english]{babel}
\usepackage{csquotes}
\usepackage{microtype}
\usepackage{relsize}

\usepackage{titlesec}

\newcommand{\chapnumfont}{%     % define font for chapter number
  \fontsize{100}{100}%          % font size 100pt, baselineskip 100pt
  \selectfont%                  % activate font
}
\colorlet{chapnumcol}{gray!75}  % color for chapter number

\titleformat{\chapter}[display]
{\filleft}
{\filleft\chapnumfont\scshape\textcolor{chapnumcol}{\thechapter}}
{-24pt}
{\Huge\scshape}
\titlespacing*{\chapter}{0pt}{-50pt}{40pt}

\titleformat{name=\chapter, numberless}[block]
{\Huge\bfseries\filright}
{} {0pt}
{\thispagestyle{empty}}

\titleformat{\section}{\Large\scshape}{\thesection.}{0.5em}{}
\titleformat{\subsection}{\large\scshape}{\thesubsection.}{0.5em}{}

\usepackage[
  backend=biber,
  style=numeric-comp,
  sorting=none,
  giveninits=true,
  maxbibnames=99,
  doi=false,
  url=false,
  isbn=false,
  eprint=false
]{biblatex}
 
\DeclareSourcemap{
  \maps[datatype=bibtex, overwrite]{
    \map{
      \perdatasource{mypapers.bib}
      \step[fieldset=KEYWORDS, fieldvalue=own, append]
    }
  }
}

\usepackage{appendix}
\usepackage{chngcntr}
\usepackage{etoolbox}
\AtBeginEnvironment{subappendices}{%
\chapter*{Appendices}
\addtocontents{toc}{\vspace{\normalbaselineskip}}
\addcontentsline{toc}{section}{Appendices}
\counterwithin{figure}{section}
\counterwithin{table}{section}
}
\AtEndEnvironment{subappendices}{%
\counterwithout{figure}{section}
\counterwithout{table}{section}
}

\usepackage{placeins}
\let\Oldsection\section
\renewcommand{\section}{\FloatBarrier\Oldsection}
\let\Oldsubsection\subsection
\renewcommand{\subsection}{\FloatBarrier\Oldsubsection}
\let\Oldsubsubsection\subsubsection
\renewcommand{\subsubsection}{\FloatBarrier\Oldsubsubsection}

\usepackage{hyperref}
\hypersetup{colorlinks,
  linkcolor={blue!80!black},
  citecolor={blue!80!black},
  urlcolor={blue!80!black}
}

\DeclarePairedDelimiter\abs{\lvert}{\rvert}
\DeclarePairedDelimiter\floor{\lfloor}{\rfloor}
\DeclarePairedDelimiter\norm{\lVert}{\rVert}
\DeclarePairedDelimiter\ket{\lvert}{\rangle}
\DeclarePairedDelimiter\bra{\langle}{\rvert}
\DeclarePairedDelimiter\kket{\lvert}{\rangle\!\rangle}
\DeclarePairedDelimiter\bbra{\langle\!\langle}{\rvert}
\DeclarePairedDelimiter\mean{\langle}{\rangle}
\DeclareMathOperator{\tr}{tr}
\DeclareMathOperator{\sign}{sign}
\DeclareMathOperator{\re}{Re}
\DeclareMathOperator{\im}{Im}

\newcommand{\mbf}[1]{\mathbf{#1}}
\newcommand{\myvec}[1]{\begin{pmatrix}#1\end{pmatrix}}
\newcommand{\HC}{\mathrm{H.c.}}
\newcommand{\inn}{\mathrm{in}}
\newcommand{\out}{\mathrm{out}}
\newcommand{\bk}{\mbf k}
\newcommand{\vac}{\mathrm{vac}}
\newcommand{\ES}{\mathrm{ES}}
\newcommand{\EBS}{E_\mathrm{BS}}
\newcommand{\EBSsub}[1]{E_{\mathrm{BS},#1}}

\newcommand{\up}{\mathsmaller{\uparrow}}
\newcommand{\dn}{\mathsmaller{\downarrow}}
\newcommand{\Jeff}{J^\mathrm{eff}}
\newcommand{\jeff}{J_\mathrm{eff}}
\newcommand{\Heff}{H_\mathrm{eff}}
\newcommand{\U}{\mathcal{U}}
\newcommand{\A}{\tfrac{E\abs{\mbf d_{ij}}}{\omega}}
\newcommand{\bes}[1]{\mathcal{J}_{#1}}
\newcommand{\supi}[1]{^{(#1)}}
\newcommand{\ol}{\bar}
\newcommand{\overbar}[1]{\mkern 1.5mu\overline{\mkern-1.5mu#1\mkern-1.5mu}\mkern 1.5mu}
\newcommand{\reseq}{\stackrel{*}{=}} % restricted equality
\newcommand{\diff}[1]{d\mkern-1.5mu{#1}\,}
\newcommand{\diffE}{d\mkern-2mu E\,}
\newcommand{\summ}{\mathop{\mathlarger{\sum}}}

\begin{document}

% \title{Quantum dynamics in low dimensional topological systems}
% \author{Miguel Bello}
% \date{\today}
% \maketitle
\begin{titlepage}

  % \vspace*{\fill} % to center text vertically in page
  \begin{center}

    {\Huge\scshape Quantum dynamics in\\[0.2em] low-dimensional 
    topological\\[0.2em] systems
    
    } % for some reason that empty line is necessary

    \vspace{2.5em}
    Memoria de la tesis presentada por\\[0.5em]
    \large Miguel Bello Gamboa\\[0.5em]
    \normalsize para optar al grado de Doctor en Ciencias Físicas\\

    \vspace{2.5em}
    \includegraphics[width=0.5\linewidth]{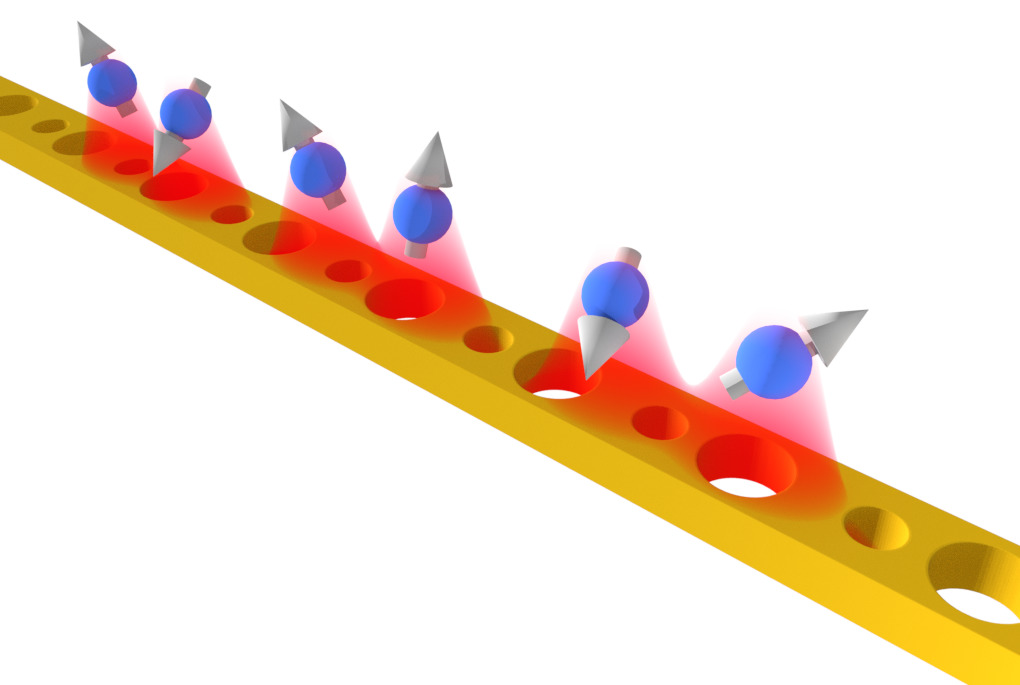}

    \large
    \vspace{2.5em}
    Universidad Autónoma de Madrid\\
    Instituto de Ciencia de Materiales de Madrid (CSIC)\\

    \vspace{2.5em}
    \emph{Directora:} Gloria Platero Coello\\
    \emph{Tutor:} Carlos Tejedor de Paz\\

    \vspace{2.5em}
    Madrid, Septiembre 2019

  \end{center}
  \vspace*{\fill} 

\end{titlepage}

% \newgeometry{hmarginratio=2:3} % to have unequal margins as in a normal book

\setcounter{secnumdepth}{-1} % these are unnumbered chapters
\chapter*{Abstract/\\\textcolor{gray}{Resumen}}
The discovery of topological matter has revolutionized the field of 
condensed matter physics giving rise to many interesting phenomena, and 
fostering the development of new quantum technologies. In this thesis we 
study the quantum dynamics that take place in low dimensional topological 
systems, specifically 1D and 2D lattices that are instances of topological 
insulators. First, we study the dynamics of doublons, bound states of two 
fermions that appear in systems with strong Hubbard-like interactions. We 
also include the effect of periodic drivings and investigate how the 
interplay between interaction and driving produces novel phenomena. 
Prominent among these are the disappearance of topological edge states in 
the SSH-Hubbard model, the sublattice confinement of doublons in certain 2D 
lattices, and the long-range transfer of doublons between the edges of any 
finite lattice. Then, we apply our insights about topological insulators to 
a rather different setup: quantum emitters coupled to the photonic analogue 
of the SSH model. In this setup we compute the dynamics of the emitters, 
regarding the photonic SSH model as a collective structured bath. We find 
that the topological nature of the bath reflects itself in the photon bound 
states and the effective dipolar interactions between the emitters. Also, 
the topology of the bath affects the single-photon scattering properties. 
Finally, we peek into the possibility of using these kind of setups for 
the simulation of spin Hamiltonians and discuss the different ground states 
that the system supports./\\

\begin{otherlanguage}{spanish}
{\color{gray}
\noindent El descubrimiento de la materia topológica ha revolucionado el 
campo de la física de la materia condensada, dando lugar a muchos fenómenos 
interesantes y fomentando el desarrollo de nuevas tecnologías cuánticas. En 
esta tesis estudiamos la dinámica cuántica que tiene lugar en sistemas 
topológicos de baja dimensión, concretamente en redes 1D y 2D que son 
aislantes topológicos. Primero estudiamos la dinámica de dublones, estados 
ligados de dos fermiones que aparecen en sistemas con interacciones fuertes 
de tipo Hubbard. También incluimos el efecto de modulaciones periódicas en 
el sistema e investigamos los fenómenos que produce la acción conjunta de 
estas modulaciones y la interacción entre partículas. Entre ellos, cabe 
destacar la desaparición de estados de borde topológicos en el modelo 
SSH-Hubbard, el confinamiento de dublones en una única subred en 
determinadas redes 2D, y la transferencia de largo alcance de dublones 
entre los bordes de cualquier red finita. Después, aplicamos nuestros 
conocimientos sobre aislantes topológicos a un sistema bastante distinto: 
emisores cuánticos acoplados a un análogo fotónico del modelo SSH. En este 
sistema calculamos la dinámica de los emisores, considerando el modelo SSH 
fotónico como un baño estructurado colectivo. Encontramos que la naturaleza 
topológica del baño se refleja en los estados ligados fotónicos y en las 
interacciones dipolares efectivas entre los emisores. Además, la topología 
del baño afecta a las propiedades de scattering de un fotón. Finalmente 
echamos un breve vistazo a la posibilidad de usar este tipo de sistemas para
la simulación de Hamiltonianos de spin y discutimos los distintos estados 
fundamentales que el sistema soporta.
}
\end{otherlanguage}

\chapter*{Acknowledgements/\\\textcolor{gray}{Agradecimientos}}
\selectlanguage{english}
I would like to thank Gloria Platero, my thesis advisor, for taking me into 
her research group and providing me with all the necessary means for doing
this thesis. I also thank her for her trust, proximity and constant 
encouragement. Thanks to Charles E. Creffield for helping me take the first
steps into research, and to Sigmund Kohler for being always willing to talk 
about physics. I thank J. Ignacio Cirac for giving me the opportunity to do a 
stay for 3 months at the Max Planck Institute of Quantum Optics in Garching. 
My time there was very productive and rewarding. I thank all the people there 
for their warm welcome, specially to Javier and Johannes, with whom I have 
spent great times and done very fun trips. There I also met Geza Giedke, who 
has always been keen to answer my questions, and Alejandro González Tudela, 
whom I thank for proposing very interesting problems, and also for his 
closeness and dedication. 
I thank Klaus Richter for inviting me for a short stay at the University of 
Regensburg, and for the enlightening discussions we held there together with 
other members of his team.
I thank all my workmates from the ICMM: Mónica, Fernando, Yue, 
Jordi, Chema, Álvaro, Beatriz, Jesús, Jose Carlos, Guillem, Sigmund and Tobias
for creating such a relaxed working environment, the outings and good moments 
together. I thank specially Mónica for encouraging me to keep going; Álvaro, for 
convincing me to go to music concerts I will never forget; and Beatriz for 
making me laugh so much. I thank all of them for teaching me directly or 
indirectly many of the things I have learned during these four years. Thanks 
to my uncle Jose Manuel for being a constant inspiration. Thanks to Darío for 
being there in the good and bad moments. Last, I want to thank my parents, 
this thesis is specially dedicated to them.

The works here presented were supported by the Spanish Ministry of Economy and 
Competitiveness through grant no. BES-2015-071573. I thank the Institute of 
Materials Science of Madrdid (ICMM-CSIC) for letting me use their facilities 
and the Autonomous University of Madrid for accepting me in their doctorate program.
/\\

\begin{otherlanguage}{spanish}
{\color{gray}
\noindent Quiero agradecerle a Gloria Platero, mi directora de tesis, el haberme
acogido en su grupo de investigación y el haberme dado todos los medios 
necesarios para hacer esta tesis. También le agradezco su confianza, cercanía 
y estímulo constantes. Gracias a Charles E. Creffield por ayudarme a dar los 
primeros pasos en la investigación, y a Sigmund Kohler por estar siempre 
dispuesto a hablar de física. Quiero agradecerle a J. Ignacio Cirac el haberme 
brindado la oportunidad de realizar una estancia de 3 meses en el instituto 
Max Planck de Óptica Cuántica en Garching. Mi tiempo allí fue muy productivo 
y enriquecedor. A toda la gente de allí le agradezco su calurosa acogida, en 
especial a Javier y Johannes con quienes pasé muy buenos ratos e hice viajes 
muy divertidos. Allí también conocí a Geza Giedke, que siempre ha estado 
dispuesto a responder mis dudas, y a Alejandro González Tudela, a quien 
agradezco el haberme propuesto problemas muy interesantes, y su cercanía y 
dedicación. A Klaus Richter le agradezco el haberme invitado a la Universidad 
de Regensburg y las interesantes discusiones sobre física que allí mantuvimos 
junto con otros miembros de su grupo. 
Le agradezco a todos mis compañeros del ICMM: Mónica, Fernando, 
Yue, Jordi, Chema, Álvaro, Beatriz, Jesús, Jose Carlos, Guillem, Sigmund y 
Tobias el crear un ambiente de trabajo tan distendido y las salidas y buenos 
ratos que hemos pasado juntos. Le agradezco especialmente a Mónica el motivarme
a seguir adelante; a Álvaro, por convencerme para ir a conciertos de música que 
nunca olvidaré; y a Beatriz, por hacerme reír tanto. A todos ellos les 
agradezco el haberme enseñado directa o indirectamente muchas de las cosas que 
he aprendido durante estos cuatro años. 
Gracias a mi tío Jose Manuel por ser una fuente constante de inspiración. 
Gracias a Darío, por estar ahí siempre en los buenos y malos momentos. 
Por último, quiro darle las gracias a mis padres, esta tesis está dedicada
especialmente a ellos.

Los trabajos aquí presentados fueron posibles gracias al apoyo económico del 
Ministerio de Economía y Competitividad a través de la beca n.º BES-2015-071573.
Le agradezco al Instituto de Ciencia de Materiales de Madrid (ICMM-CSIC)
el haberme permitido utilizar sus instalaciones y a la Universidad Autónoma de 
Madrid el haberme aceptado en su programa de doctorado. 
}
\end{otherlanguage}

\microtypesetup{protrusion=false} % solves extra dots in toc problem due to microtype
\tableofcontents                  
\microtypesetup{protrusion=true}

\setcounter{secnumdepth}{2} % to start regular chapters
\chapter{Introduction \label{chap:intro}}
Topology is the field of mathematics that studies the properties of 
spaces that are preserved under continuous transformations. It is not 
concerned about the particular details of those spaces, but on their 
most general aspects, like their number of connected components, the 
number of holes they have, etc. With such a broad point of view, it is 
not surprising that it has many applications in other sciences beyond 
mathematics. Its application to the field of condensed matter physics 
is relatively new. It began around the 1980s, when scientists such as 
Michael Kosterlitz, Duncan Haldane and David Thouless started using 
topological concepts to explain exotic features of newly discovered 
phases of matter, such as the quantized Hall conductance of certain 2D 
materials at very low temperatures, the so-called integer Quantum Hall 
effect. In 2016 they were awarded the Nobel Prize in physics for this 
and other works~\cite{kosterlitz1973,thouless1982,thouless1983,haldane1983} 
which opened the field of \emph{topological matter}. Since then, the 
interest on this topic has grown exponentially, and so has the number 
of applications harnessing the exotic properties of topological phases.

Broadly speaking, topological matter is a new type of matter characterized
by global topological properties. These properties stem, e.g., from 
the pattern of long-range entanglement in the ground 
state~\cite{wen2017,kitaev2006b} or, in the case of topological insulators 
and superconductors, from the electronic wavefunction in the whole 
Brillouin zone~\cite{bernevig2013topo,hasan2010,qi2011}. Importantly, these 
new phases of matter display edge modes, which are conducting states 
localized at the edges of the material. They are said to be topologically 
protected, that is, they are robust against perturbations which do not break 
certain symmetries of the system. For example, edge states in the integer 
quantum Hall effect are protected against backscattering, unpaired Majorana 
fermions are protected against any perturbation that preserves fermion 
parity~\cite{kitaev2001}, etc. This makes topological phases of matter 
very interesting for developing applications. Among them, perhaps the most
exciting is the realization of fault-tolerant quantum 
computers~\cite{nayak2008}.

This new point of view in condensed matter physics puts forward many 
interesting challenges. On a fundamental level, our understanding of 
topological phases of matter is not complete yet. How many different 
phases are there? How to characterize them? These questions have only 
been answered partially, mostly for systems of non-interacting 
particles~\cite{ryu2010}. On a practical level, we would like to predict 
which materials display topological properties and be able to probe 
them in experiment. To date, several topological materials have been 
demonstrated in experiments~\cite{konig2007,hsieh2008}, and systematic
searches have been carried out, unveiling that a large percentage of all 
known materials are expected to have non-trivial topological 
properties~\cite{bradlyn2017,vergniory2019,tang2019,zhang2019}.

While looking for real materials displaying topological properties is 
one possibility, an alternative is to simulate them in 
the lab. A quantum simulator is a device that can be tuned in a way to 
mimic the behavior of another quantum system or theoretical model that we 
want to investigate~\cite{feynman1982,georgescu2014}. Of course, a fully 
fledged quantum computer would allow for the simulation any other quantum 
system~\cite{lloyd1996}. But, we are yet far from building a proper, 
fault-tolerant, quantum computer. Thus, analogue quantum simulation is a 
more reachable goal for the time being. One of the best platforms for doing 
quantum simulation are ultracold atoms trapped in optical 
lattices~\cite{bloch2008,goldman2016}. In these experiments an atomic gas 
of neutral atoms cooled to temperatures near absolute zero is loaded into a 
high vacuum chamber and trapped by dipolar forces at the maxima or minima 
of the electromagnetic field generated by interfering lasers. 
The optical nature of the trapping potential allows for the generation of 
virtually any desired lattice geometry. Adding a periodic driving 
considerably enriches the physics of these systems and provides a means for 
controlling and manipulating them. Such driving can produce effects, such as
coherent destruction of tunneling~\cite{dunlap1986,grifoni1998}, and can 
even be used to design artificial gauge 
fields~\cite{dalibard2011,creffield2011,creffield2014,goldman2014}. Using 
these techniques some topological models have already been demonstrated in 
experiment~\cite{struck2011,atala2013,jotzu2014}. 

% However, despite the many advances in the field, driven interacting 
% systems have been less studied than non-interacting ones. Understanding
% the role interactions play in these set-ups is a hard task of 
% fundamental importance, since the behavior of the system may change 
% drastically compared with the noninteracting case. 
% In chapter \ref{chap:doublons} we address this issue investigating the
% dynamics of two strongly interacting particles in topological lattices 
% under the influence of periodic drivings. 

Quantum dots (QD) are another candidate technology for doing quantum 
information processing~\cite{loss1998,petta2005,hanson2007}, and quantum 
simulation~\cite{barthelemy2013,hensgens2017}. Laterally-defined QDs are 
made patterning electrodes on top of a sandwich of two semiconductors 
hosting a two-dimensional electron gas (2DEG) at their interface. Applying 
particular voltages to the electrodes, a specific potential landscape can be
generated, which depletes the 2DEG, trapping just a few electrons. In this 
manner, artificial atoms and molecules can be created in solid-state 
devices~\cite{oosterkamp1998}, whose energy levels can be tuned by 
electrostatic means. Nowadays, increasingly larger arrays of QDs are being 
fabricated~\cite{zajac2016,volk2019}, which would allow for the simulation 
of simple topological lattice models. Also hybrid systems with improved 
capabilities are being created combining QDs with superconducting 
cavities~\cite{mi2018}.

% The high controllability and isolation from the environment achieved 
% in cold atom experiments makes it a great platform for observing 
% doublons, but it is not the only option. Nowadays, solid-state devices 
% such as quantum dot arrays are being investigated as platforms for 
% doing quantum computation and quantum simulation~\cite{hensgens2017}. 
% One of the benefits of this technology as compared with others is the 
% easy scalability of applications and fast operation times. Nonetheless,
% one of its weakness is the short coherence times due to the coupling 
% with environmental degrees of freedom. What happens, then, with 
% doublons in these more noisy experiments? Can doublons be observed, for
% example, in quantum dot arrays? These are the questions we try to 
% address in this section.

Another research direction exports ideas from topological matter to other 
fields of physics. For example, topological insulators can be used as a 
guide for the design of novel metamaterials with interesting mechanical, 
acoustic and optical properties~\cite{lu2016,huber2016}. In particular, the 
application of topological ideas to photonics has given rise to a new field
known as topological quantum optics~\cite{ozawa2019}. This is a rather young
field where many open questions are waiting to be answered.

% The field of topological matter is still a rather young field of physics, 
% and many important developments are yet to come from the crossover of this
% and other fields. For example, topological insulators are being used as a 
% guide for the design of novel metamaterials with interesting mechanical 
% and optical properties~\cite{lu2016,huber2016}. 
% Interplay between topology and other stuff

The outline of this thesis is as follows: In chapter~\ref{chap:theory} we 
review different mathematical techniques that we use in subsequent chapters;
minor details are left as appendices at the end of each chapter. In 
chapter~\ref{chap:doublons} we investigate the dynamics of interacting 
fermions in driven topological lattices. In 
chapter~\ref{chap:topologicalQED} we explore the dynamics of quantum 
emitters coupled to a 1D topological photonic lattice, namely a photonic 
analogue of the SSH model. Last, we summarize our findings and give 
an overview of the possible experimental implementations in 
chapter~\ref{chap:conclu}.

\chapter{Theoretical preliminaries \label{chap:theory}}
In an attempt to keep the text as much self-contained as possible, here we 
briefly introduce the most important theoretical tools used to derive the 
results presented in this thesis. First, we comment on topological
matter, placing special emphasis on the SSH model, a key player in 
subsequent chapters. Second, we give a primer on Floquet theory, a basic 
tool for studying driven systems with some periodic time dependence; 
we will use it in section~\ref{sec:doublondynamics}. Last, we discuss two
different approaches to open quantum systems: master equations and 
resolvent operator techniques. They are relevant for 
section~\ref{sec:doublondecay} and the whole of
chapter~\ref{chap:topologicalQED}. 

\section{Topological phases of matter \label{sec:SSH}}

Towards the middle of the last century, Landau proposed a theory for matter
phases and phase transitions where different phases are characterized by 
symmetry breaking and local order parameters. Then, the discovery of the 
integer~\cite{klitzing1980} and fractional~\cite{tsui1982} Quantum Hall 
effects in the 1980s revolutionized the field of condensed matter physics. 
These new phases could not be understood in terms of local order 
parameters, and posed a problem for the established theory. After that, 
many other phases that fall beyond Landau's paradigm have been 
discovered and a new picture has started to emerge, one where topology 
plays a major role~\cite{wen2017}.

Topological phases can be divided in two large families:
\emph{topologically ordered phases} and \emph{symmetry-protected 
topological phases} (SPT). Both refer to gapped phases, i.e., with a 
nonzero energy gap above the ground state in the thermodynamic limit at 
zero temperature. The fundamental classifying principle that operates in
all this discussion is as follows: Two systems 
belong to the same phase if one can be deformed adiabatically into the 
other without closing the gap, and they belong to different phases 
otherwise.
Phases with true topological order are long-range entangled with a 
degenerate 
ground state whose degeneracy depends on the topology of the underlying 
space, and quasiparticles with fractional statistics called anyons. Examples
include the fractional Quantum Hall effect~\cite{laughlin1983}, chiral spin 
liquids~\cite{savary2016}, or the toric code~\cite{kitaev2006a}, to name a 
few. SPT phases, on the other hand, are short-range entangled and they 
require the presence of certain symmetries to be well-defined. A 
characteristic of all topological phases is the presence of edge modes or 
excitations that are said to be ``topologically protected'', meaning that 
they are robust against certain types of disorder.

SPT phases can be further subdivided into interacting and non-interacting 
phases, depending on whether particles interact with each other or not. 
Examples of the first type are Haldane's spin-1 chain and the AKLT 
model~\cite{affleck1987}. Examples of the second type include 
Kitaev's chain~\cite{kitaev2001} and Chern insulators. Although a complete 
classification of all SPT phases is not known yet, some classification 
schemes have been elucidated for particular cases. For example, 1D 
interacting phases can be characterized by the projective representations of
their symmetry groups~\cite{chen2011,schuch2011,verresen2017}, and a 
complete classification of non-interacting fermionic phases, also known as 
topological insulators and superconductors, has been obtained in terms of a 
topological band theory, which assigns topological invariants to the energy 
bands of their spectrum~\cite{hasan2010,qi2011,bernevig2013topo}. For their 
classification the relevant 
symmetries are~\cite{ryu2010}: \emph{Time-reversal} symmetry, which 
corresponds to an antiunitary operator that commutes with the Hamiltonian, 
\emph{charge-conjugation} (also known as \emph{particle-hole}), which 
corresponds to an antiunitary operator that anticommutes with the 
Hamiltonian, and \emph{chiral} (also known as \emph{sublattice}) symmetry, 
which corresponds to a unitary operator that anticommutes with the 
Hamiltonian. The presence of these symmetries imposes restrictions on the 
single-particle Hamiltonian matrix, $H$, as shown in the table below
\begin{center}
  {\renewcommand{\arraystretch}{1.5}%
  \begin{tabular}{cc}
    Symmetry & Definition \\
    \hline\hline
    time-reversal & $U_\mathcal{T}H^*U^\dagger_\mathcal{T}=H$\\
    charge-conjugation & $U_\mathcal{C}H^*U^\dagger_\mathcal{C}=-H$\\
    sublattice & $U_\mathcal{S}HU^\dagger_\mathcal{S}=-H$ \\
    \hline
  \end{tabular}}
  \vspace*{1em}
\end{center}
where $U_\alpha$, $\alpha\in\{\mathcal{T},\mathcal{C},\mathcal{S}\}$, 
are unitary matrices and ``$^*$'' denotes complex conjugation.
Note that having two of them, automatically implies that all three 
symmetries are present. As it turns out, there are ten different classes 
depending on whether the aforementioned symmetries are present or absent 
and whether they square to $\pm 1$ if 
present. For each class, the classification scheme determines how many 
distinct phases are possible depending on the dimensionality of the system.
There are essentially three possibilities: there may be just the trivial 
phase; two phases, one trivial and another topological, distinguished 
by a $\mathbb{Z}_2$ topological invariant; or there may be infinitely many 
distinct phases distinguished by a $\mathbb{Z}$ topological invariant.

A word of caution is in order. Throughout the literature, the term 
``topological'' is used somewhat vaguely. In many cases it is just a label 
to refer to any phase other than the trivial. But, what is a trivial phase?
Well, it is just a phase that can be connected adiabatically with that 
of the vacuum, or the phase of a disconnected lattice.

\subsection{The SSH model}

\begin{figure}[!htb]
  \centering
  \includegraphics{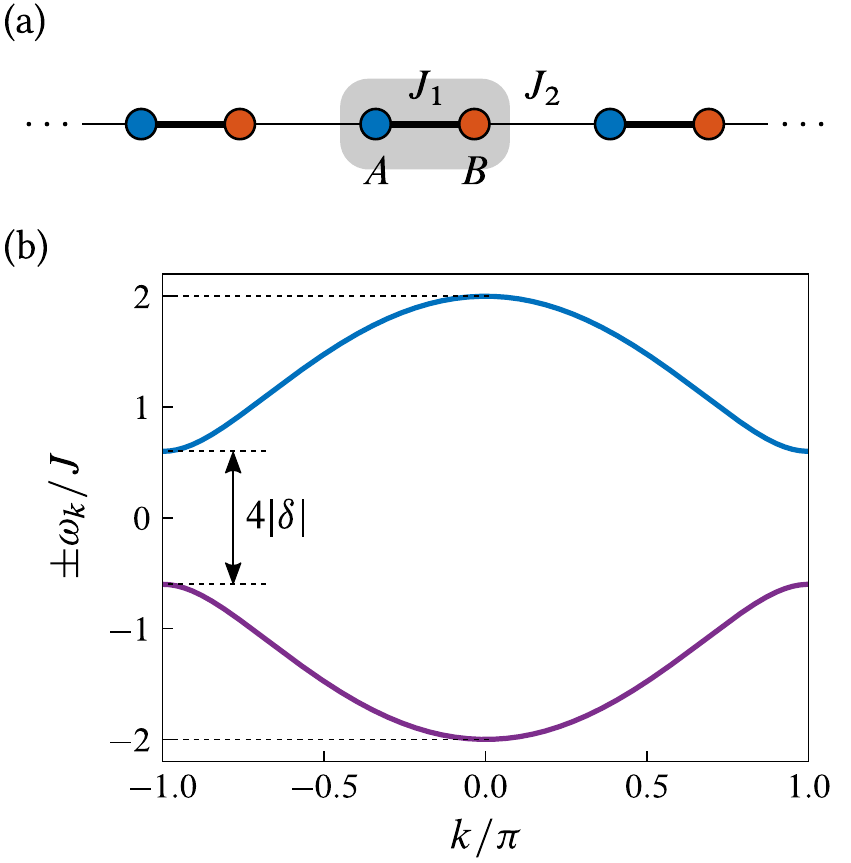}
  \caption{(a) Schematic drawing of the SSH model: a 1D lattice with
  two sites per unit cell (in gray), labelled $A$ and $B$, with alternating 
  hopping amplitudes $J_1=J(1+\delta)$ and $J_2=J(1-\delta)$.
  (b) Upper and lower energy bands of the SSH model.}
  \label{fig:ssh_scheme_bands}
\end{figure}

One of the simplest models featuring a non-interacting SPT phase is the 
SSH model, named after Su, Schrieffer and Heeger, who first studied it in 
the 1970s~\cite{su1979,asboth2016short}. It describes non-interacting 
particles hopping on a 1D lattice with staggered nearest-neighbour hopping 
amplitudes $J_1$ and $J_2$. The lattice consists of $N$ unit cells, each 
one hosting two sites that we label $A$ and $B$, see 
Fig.~\ref{fig:ssh_scheme_bands}(a). Its Hamiltonian can be written as 
\begin{equation}
  H_\mathrm{SSH}=-\summ_j \left(J_1c^\dagger_{jA}c_{jB}
  +J_2c^\dagger_{j+1A}c_{jB}+\HC\right)\,,
  \label{eq:SSH}
\end{equation}
where $c_{j\alpha}$ annihilates a particle (boson or fermion) in the 
$\alpha\in\{A,B\}$ sublattice at the $j$th unit cell. 
The hopping amplitudes are usually parametrized as $J_1={J(1+\delta)}$ and 
$J_2=J{(1-\delta)}$, where $\delta\in[-1,1]$ is the so-called dimerization 
constant. Assuming periodic boundary conditions, the Hamiltonian can be 
written in momentum space as $H_\mathrm{SSH}=\sum_k V^\dagger_k H_k V_k$, 
with $V_k=(c_{kA},c_{kB})^T$, and
\begin{equation}
  H_k=\myvec{0 & f(k) \\ f^*(k) & 0}\,, \quad
  f(k)=-J(1+\delta)-J(1-\delta)e^{-ik}\,.
  % V_k=\myvec{c_{kA} \\ c_{kB}}\,,
\end{equation}
Here, $f(k)$ denotes the coupling between the 
modes $c_{k\alpha}=\sum_j e^{-ikj}c_{j\alpha}/\sqrt{N}$. This 
Hamiltonian can be easily diagonalized as 
\begin{equation}
  H_\mathrm{SSH}=\summ_k\omega_k\left(u^\dagger_ku_k-l^\dagger_kl_k\right)\,,
\end{equation}
with
\begin{gather}
  \omega_k=\abs{f(k)}=J\sqrt{2(1+\delta^2)+2(1-\delta^2)\cos(k)}\,,\\
  u_k=\frac{1}{\sqrt{2}}\left(e^{-i\phi_k}c_{kA}+c_{kB}\right)\,,
  \quad
  l_k=\frac{1}{\sqrt{2}}\left(e^{-i\phi_k}c_{kA}-c_{kB}\right)\,,
  \label{eq:uandl}
\end{gather}
and $\phi_k=\arg(f(k))$. Its spectrum consists of two bands with dispersion 
relation $-\omega_k$ (lower band), and $\omega_k$ (upper band), spanning 
the ranges $[-2J,-2|\delta|J]$ and $[2|\delta|J,2J]$ respectively, see 
Fig.~\ref{fig:ssh_scheme_bands}(b).

The SSH model has all three symmetries mentioned in previous paragraphs: 
time-reversal, charge-conjugation and chiral symmetry. Thus, according to 
the classification, it belongs to the BDI class, which in 1D supports 
distinct topological phases characterized by a $\mathbb{Z}$ topological 
invariant. To find this invariant, let us point out that chiral symmetry 
is represented in momentum space by the $z$-Pauli matrix $\sigma_z$, that 
is, $\sigma_z H_k\sigma_z=-H_k$. Expressing $H_k$ 
in the basis of Pauli matrices, $H_k=h_0(k)I+\mbf h(k)\cdot \pmb\sigma$, 
this symmetry constraint forces $h_0(k)=0$ and $h_z(k)=0$, so the vector 
$\mbf h(k)$ lies on a plane. Also, the existence of a gap requires 
$\mbf h(k)\neq 0$. Therefore, $\mbf h(k)$ is a map from $S^1$ (the first 
Brillouin zone) to $\mathbb{R}^2\setminus\{0\}$. Such maps can be 
characterized by a topological invariant that only takes integer values: 
the winding number, $\mathcal{W}$, of the curve $\mbf h(k)$ around the 
origin. Furthermore, as long as symmetries are preserved it is impossible 
to change the winding number of the curve without making it pass through 
the origin, in accordance with the fact that distinct topological phases 
are separated by phase transitions in which the gap closes.

\begin{figure}[!htb]
  \centering
  \includegraphics{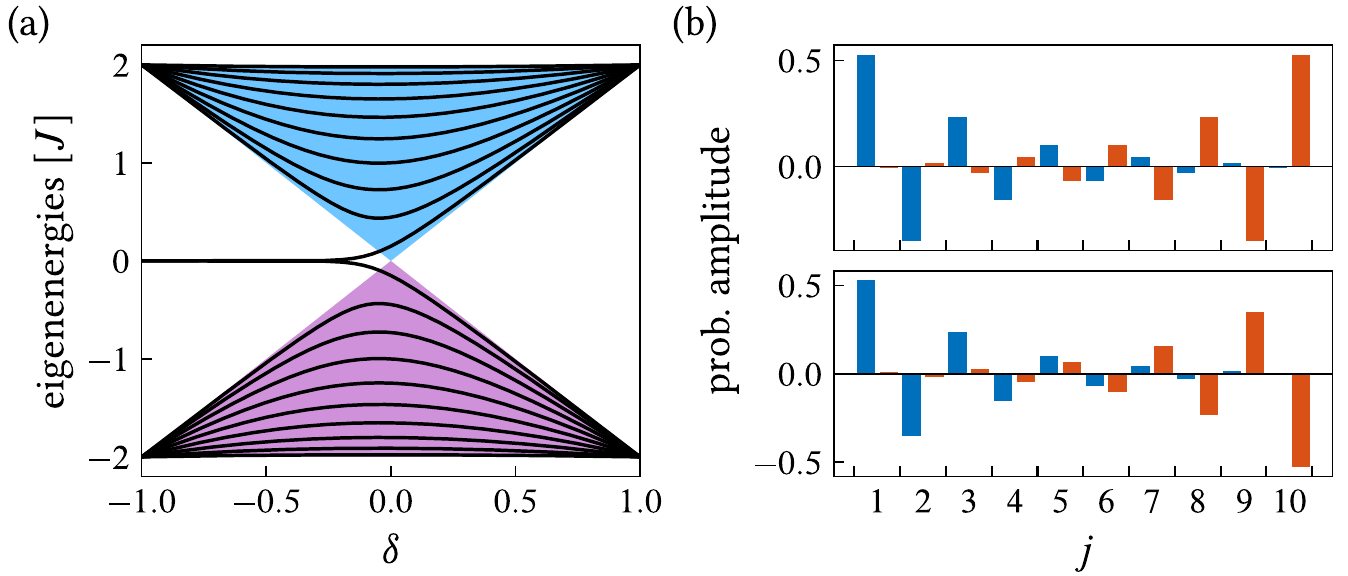}
  \caption{(a) Single-particle spectrum as a function of the dimerization 
  constant for a finite chain with $N=10$ unit cells. The shaded areas 
  correspond to the ranges of the energy bands in the thermodynamic limit
  (b) Edge states for the same chain at $\delta=-0.2$; blue (red) bars 
  correspond to the amplitudes in sublattice $A$ ($B$)}
  \label{fig:finiteSSH}
\end{figure}

The SSH model can be in two distinct phases: a topological phase 
with $\mathcal{W}=1$, for $J_1<J_2$ ($\delta<0$), or a trivial phase with 
$\mathcal{W}=0$, for $J_1>J_2$ ($\delta>0$). We remark that higher winding 
numbers can be achieved if longer-range hoppings are 
included~\cite{perezgonzalez2019a,perezgonzalez2019b}. By the 
\emph{bulk-boundary correspondence}, the value of $|\mathcal{W}|$ 
corresponds to the number of pairs of edge states supported by the 
system~\cite{bohungchen2018}. These states are exponentially localized at the 
edges of the chain and its energy is pinned on the middle of the band gap. 
In Fig.~\ref{fig:finiteSSH}(a) we show the energy spectrum of a finite 
chain consisting of $N=10$ dimers. There, it can be seen how for $\delta<0$ 
the energies of two states detach from the bulk energy bands (shaded areas) 
and quickly converge to zero. By the energy spectrum, in 
Fig.~\ref{fig:finiteSSH}(b), we show the amplitudes of the two midgap 
states for a particular value of $\delta<0$, proving that they are 
exponentially localized to the edges of the chain.
To better understand the features of the energy spectrum and the edge states
let us turn again to chiral symmetry. For systems defined on a lattice, we 
say that the system is bipartite if the lattice can be divided in two 
sublattices ($A$ and $B$) such that hopping processes only connect sites 
belonging to different sublattices. As it turns out, any bipartite lattice has 
chiral symmetry embodied in the transformation $c_{jA}\to c_{jA}$, 
$c_{jB}\to - c_{jB}$. Its action over a single-particle wavefunction
is to reverse the sign of the amplitudes on one sublattice---hence, the name
sublattice symmetry---which changes the sign of the Hamiltonian. This 
implies that the eigenvalues either come in pairs with opposite energies or 
have energy equal to zero and are also eigenvalues of the chiral symmetry 
operator. Note that the eigenstates in a pair have the same wavefunction, 
except for a change of the sign of the amplitudes in one sublattice. On the 
other hand, eigenstates with zero energy have support on a single sublattice. 
Edge states in the thermodynamic limit have zero energy, but in a finite system 
they hybridize forming symmetric and antisymmetric combinations which 
constitute a chiral symmetric pair with an energy splitting $\Delta\epsilon$
that decreases exponentially with increasing chain size, 
$\Delta\epsilon\propto e^{-N/\lambda}$. 

For chains with an odd number of sites there must be an odd number of 
single-particle states precisely at zero energy due to chiral symmetry. 
Indeed, they support only one state at zero energy, exponentially localized
on a single edge, with weight on just one sublattice. Note that in this 
case, changing the sign of $\delta$ amounts to inverting the chain spatially. 
Thus, depending on the sign of $\delta$ this edge state is localized on the 
left or right edge.

\section{Floquet theory}

Floquet theory is concerned with the solution of periodic linear 
differential equations. In quantum mechanics, it is used to solve 
systems modelled by Hamiltonians with some explicit periodic time 
dependence, that is, solutions of the Schr\"odinger equation
\begin{equation}
  i\partial_t\ket{\psi(t)}=H(t)\ket{\psi(t)}\,
  \label{eq:schrodinger}
\end{equation}
with $H(t+T)=H(t)$ (here and throughout the text we work in units such 
that $\hbar=1$). For simplicity, we will just consider systems with 
a Hilbert space of finite dimension. Floquet's theorem states that there 
is a fundamental set of solutions $\{\ket{\psi_\alpha(t)}\}$ to this 
equation of the form 
$\ket{\psi_\alpha(t)}=e^{-i\epsilon_\alpha t}\ket{\phi_\alpha(t)}$,
with $\epsilon_\alpha$ real and $\ket{\phi_\alpha(t)}$ periodic with the 
same periodicity as the Hamiltonian,
$\ket{\phi_\alpha(t+T)}=\ket{\phi_\alpha(t)}$~\cite{holthaus2015}. 
Therefore, any solution of Eq.~\eqref{eq:schrodinger} can be expanded as 
\begin{equation}
  \ket{\psi(t)}=\summ_\alpha c_\alpha 
  e^{-i\epsilon_\alpha t}\ket{\phi_\alpha(t)}\,,
\end{equation}
with time independent coefficients $c_\alpha$. In analogy with 
Bloch's Theorem, $\epsilon_\alpha$ is called \emph{quasienergy} and 
$\ket{\phi_\alpha(t)}$ is called \emph{Floquet mode}. Introducing 
$\ket{\psi_\alpha(t)}$ in~\eqref{eq:schrodinger}, we can see that Floquet 
modes satisfy the eigenvalue equation
\begin{equation}
  \mathcal{H}\ket{\phi_\alpha}\equiv
  \left(H-i\partial_t\right)\ket{\phi_\alpha}
  =\epsilon_\alpha\ket{\phi_\alpha}\,,
  \label{eq:eigenFloquet}
\end{equation}
where $\mathcal{H}$ is the so-called Floquet operator, and it is Hermitian.

Equivalently, Floquet's theorem states that for time-periodic systems the 
unitary time evolution operator can be factorized as
\begin{equation}
  U(t_2,t_1)=\summ_\alpha e^{-i\epsilon_\alpha (t_2-t_1)}
  \ket{\phi_\alpha(t_2)}\bra{\phi_\alpha(t_1)}
  =e^{-i\Heff (t_2-t_1)}P(t_2,t_1)\,,
\end{equation}
with $\Heff=\sum_\alpha\epsilon_\alpha\ket{\phi_\alpha(t_2)}\bra{\phi_\alpha(t_2)}$
being an effective time-independent Hamiltonian, and 
$P(t_2,t_1)=\sum_\alpha\ket{\phi_\alpha(t_2)}\bra{\phi_\alpha(t_1)}$ a
unitary operator, $T$-periodic in both of its arguments. The long-term 
dynamics is governed by $\Heff$, wile the dynamics within a period, 
also known as the \emph{micromotion}, is given by $P$. Note that 
there is a gauge freedom and we could also have written 
\begin{equation}
  U(t_2,t_1)=P^\dagger(t_0,t_2)e^{-i\Heff(t_2-t_1)}P(t_0,t_1)\,,
\end{equation}
setting $\Heff=\sum_\alpha\epsilon_\alpha
\ket{\phi_\alpha(t_0)}\bra{\phi_\alpha(t_0)}$.
In some texts the micromotion operator is written as 
$P(t_0,t)=e^{iK(t)}$, with $K(t)$ Hermitian and $T$-periodic, depending 
implicitly on the choice of $t_0$.

Floquet modes and quasienergies play a central role in the study of periodic
systems. There are several ways to compute them. One can integrate 
numerically the dynamics of some states forming an orthonormal basis at 
$t_1=0$, obtaining $U(t,0)$ for $t\in[0,T]$. The eigenvalues of $U(T,0)$ are
the complex phases $e^{-i\epsilon_\alpha T}$, with associated eigenvectors 
$\ket{\phi_\alpha(T)}=\ket{\phi_\alpha(0)}$. Their full time dependence 
can be reconstructed as 
$\ket{\phi_\alpha(t)}=e^{i\epsilon_\alpha t}U(t,0)\ket{\phi_\alpha(0)}$.
Note that, since the quasienergies appear in a complex phase, they are 
defined modulo $2\pi/T=\omega$. Given a quasienergy $\epsilon_\alpha$, we 
can add to it a multiple of the frequency 
$\epsilon_{\alpha n}=\epsilon_\alpha + n\omega$ 
and the corresponding Floquet mode 
$\ket{\phi_{\alpha n}(t)}=e^{in\omega t}\ket{\phi_\alpha(t)}$
will also satisfy Eq.~\eqref{eq:eigenFloquet} with 
eigenvalue $\epsilon_{\alpha n}$. In fact any of these Floquet modes,
$\ket{\phi_{\alpha n}(t)}$, $n\in\mathbb{Z}$, yields the same solution to 
the Schr\"odinger equation. So, the quasienergy spectrum is periodic, 
much like quasimomentum in crystalline solids, and it suffices to consider 
the quasienergies within a range of width $\omega$.

Another procedure exploits the fact that that Floquet modes are periodic in 
time, so they can be expanded in Fourier series,
\begin{equation}
  \ket{\phi_\alpha}=\summ_n e^{-in\omega t}\ket{c_{\alpha n}}\,, \quad
  \ket{c_{\alpha n}}=\frac{1}{T}\int_0^T \diff{t}
  e^{in\omega t}\ket{\phi_\alpha(t)}\,.
\end{equation}
The time-independent vectors $\ket{c_{\alpha n}}$ can themselves be 
expanded into some basis, $\{\ket{\beta}\}$, of the system's Hilbert space. 
Now, we can identify $e^{-in\omega t}\ket{\beta}\equiv\ket{n\beta(t)}$ as 
basis states of an enlarged Hilbert space built as the product of the 
system's Hilbert space and the space of $T$-periodic 
functions~\cite{sambe1973}. We will denote the elements of this space as 
$\kket{\phi}$, so that in ``time representation'' they are 
$\langle t\kket{\phi}=\ket{\phi(t)}$. An inner product can be defined in 
this enlarged space as the composition of the usual inner products in each 
of the constituent spaces
\begin{equation}
  \langle\!\langle\phi\vert\phi'\rangle\!\rangle=\frac{1}{T}\int_0^T
  \diff{t} \mean{\phi(t)\vert\phi'(t)} \,.
\end{equation}
With respect to this basis, the matrix elements of the Floquet operator 
are
\begin{align}
  \bbra{n'\beta'}\mathcal{H}\kket{n\beta} &=
  \frac{1}{T}\int_0^T \diff{t} e^{in'\omega t}
  \bra{\beta} H(t)-i\partial_t\ket{\beta}e^{-in\omega t} \nonumber\\
  &= \bra{\beta'}H_{n'-n}\ket{\beta} 
  - n\omega\delta_{n'n}\delta_{\beta'\beta}\,,
  \label{eq:floquetElements}
\end{align}
where $H_{n'-n}$ is the $(n'-n)$th Fourier component of the Hamiltonian. 
Thereby, we have transformed a time-dependent problem into a 
time-independent one with an extra (infinite-dimensional) degree of 
freedom. This extra degree of freedom is directly related to the apparent 
redundancy of the Floquet modes mentioned earlier. 
Frequently, a time dependence in the Hamiltonian stems from the coupling to 
the modes of, e.g., a laser field in the semiclassical 
limit~\cite{kohler2006}. In this limit the basis states $\kket{\beta n}$ 
can be interpreted as having a definite number of photons, and Floquet 
modes can be viewed as dressed states. Based on this analogy, the indices
$(n',n)$ in Eq.~\eqref{eq:floquetElements} are referred to as the 
photon indices. Furthermore, the matrix elements of $H_m$ are said to 
describe $m$-photon processes. 

The benefit of this technique is that it allows the direct application 
of methods used for the diagonalization of time-independent Hamiltonians 
to compute quasienergies and Floquet modes. In the high frequency regime, 
blocks with different photon number are far apart in energy, and it makes 
sense to use perturbation theory to block diagonalize the Floquet 
operator~\cite{eckardt2015,mikami2016,bukov2015}. This provides a 
series expansion of $\Heff$ in powers of $\omega^{-1}$, known as the 
high-frequency expansion (HFE). In some situations, computing a few terms 
in this expansion is easier than computing the Floquet modes and 
quasienergies, and provides more physical insight. In section 
\ref{sec:doublondynamics}, we use this technique to investigate the dynamics
of a pair of strongly-interacting fermions under the action of an ac field.

\section{Open quantum systems}

Quantum mechanics has allowed us to explain many interesting phenomena
at the cost of a more complicated description of fundamental particles 
and interactions. For example, the number of variables needed to 
describe the state of an ensemble of particles grows exponentially with 
the number of particles in the ensemble. This makes it very hard to analyze
large systems involving many particles, or systems that interact with 
external, uncontrolled degrees of freedom. However, finding ways to 
tackle these problems is a necessity, since more often than not this 
is the situation we face in real experiments. 

As opposed to \emph{closed} quantum systems, \emph{open} quantum systems 
are systems that interact with an environment, also called bath or 
reservoir, which is a collection of infinitely many degrees of freedom. 
Through this interaction the system exchanges information, energy and 
particles with the environment. As a result, dissipation and decoherence 
are introduced into the system~\cite{marquardt2008}. The former is the 
phenomenon by which the system exchanges energy with the environment, 
eventually reaching thermal equilibrium with it, while the latter is the 
phenomenon by which coherent superpositions of states are lost over time. 
Below we summarize two distinct approaches to the study of open quantum 
systems.

\subsection{Master equations}

A proper description of the system taking into account the effects of 
decoherence and dissipation is given by a density matrix, which besides 
pure states also includes statistical mixtures of them. The usual way to 
study this type of problems involves tracing out the bath degrees of 
freedom, obtaining a first order linear differential equation for the 
system's reduced density matrix, the so-called master equation. There are 
different ways to do this depending on the regime and approximations that 
apply to the system under consideration \cite{petruccione2002open}.

We now proceed to show how to obtain a master equation, valid in 
the regime of weak system-bath coupling.
The combination of system and bath is described as a whole by the 
Hamiltonian $H=H_S+H_B+H_I$, where $H_S$ and $H_B$ are the free 
Hamiltonians of system and bath respectively, and $H_I$ is the Hamiltonian 
that describes the interaction between them. We restrict the interaction
to the case where $H_I$ is linear in both system and bath operators, 
$\{X_j\}$ and $\{B_j\}$ respectively, $H_I=\sum_j X_j\otimes B_j$.
The entire system evolves according to the von-Neumann equation:
\begin{equation} 
  \dot{\tilde{\rho}}(t)=-i[\tilde{H_I}(t),\tilde{\rho}(t)] \,,
  \label{eq:von-Neumann}
\end{equation}
Here, $\rho$ is the full density matrix of the system plus the bath; 
the tilde over an operator denotes the interaction picture,
\begin{equation}
  \tilde O(t)\equiv e^{i(H_S+H_B)t}Oe^{-i(H_S+H_B)t} \,,
\end{equation}
where $O$ is the corresponding operator in the Schr\"odinger picture.
Notice that at time $t=0$, both the Schr\"odinger and the interaction 
picture representations coincide. The integral form of 
Eq.~\eqref{eq:von-Neumann} is
\begin{equation}
  \tilde{\rho}(t)=\rho(0)
  -i\int_0^t \diff{s} [\tilde{H_I}(s),\tilde\rho(s)] \,.
\end{equation}
Inserting this expression for $\rho(r)$ back in the right hand side of 
Eq.~\eqref{eq:von-Neumann}, tracing over the bath degrees of freedom, 
we get
\begin{equation}
	\dot{\tilde{\rho_S}}(t)=-i\tr_B [\tilde{H_I}(t),\rho(0)]
  -\int_0^t \diff{s}\tr_B 
  [\tilde{H_I}(t),[\tilde{H_I}(s),\tilde{\rho}(s)]] \,. 
  \label{eq:no-approx}
\end{equation}
To obtain a closed equation for the reduced density matrix of the 
system, we need to make some approximations: 
\begin{itemize}
  \item \emph{Born approximation}: We assume that at all times, the total 
    density matrix can be factorized as 
    $\rho(t)\approx \rho_S(t)\otimes\rho_B$. This amounts to neglect any 
    entanglement between the system and the bath, which is justified if 
    the coupling between them is small enough. Furthermore we will 
    always consider a thermal state for the bath
    $\rho_B\propto e^{-\beta H_B}$.
  \item \emph{Markov approximation}: We replace $\rho_S(s)$ by $\rho_S(t)$ 
    in the integrand, such that the time evolution of the system only 
    depends on its current state but not on its previous states. 
    Furthermore, we let the lower integration bound go to $-\infty$ and 
    make a change of variable $s=t-\tau$. This approximation is justified if
    the timescale in which the reservoir correlations decay is small 
    compared to the timescale in which the system varies noticeably. 
\end{itemize}
Last, if $\mean{\tilde{B}_j(t)}=0$, where we denote
$\mean{x}\equiv\tr_B(x\rho_B)$, as is the case in all the models 
presented in this thesis, we can neglect the first term, obtaining
\begin{equation}
  \dot{\tilde{\rho_S}}(t)=-\int_0^\infty \diff{\tau} \tr_B
  [\tilde{H_I}(t),[\tilde{H_I}(t-\tau),\tilde{\rho_S}(t)\otimes\rho_B]] 
  \,. \label{eq:BornMarkov}
\end{equation}
Transforming it back to the Schr\"odinger picture, we get
\begin{equation}
  \begin{split}
    \dot{\rho_S}&=-i[H_S,\rho_S]- \int_0^\infty \diff{\tau} \tr_B 
    [H_I,[\tilde{H_I}(-\tau),\rho_S\otimes \rho_B]]\\
    &\equiv -i[H_S,\rho_S] + \mathcal{L}[\rho_S]\,,
  \end{split}
  \label{eq:BornMarkovSpicture}
\end{equation}
where the first term accounts for the coherent unitary dynamics, whereas 
the second includes dissipation and decoherence. 

This equation, known in the literature as the Bloch-Redfield master 
equation~\cite{redfield1957}, 
is the one we will use for studying the decay of doublons in 
noisy environments, see section~\ref{sec:doublondecay} and 
appendix~\ref{app:BlochRedfield}. 
Note that this equation does not necessarily preserve the positivity of 
the density matrix. One has to perform a further 
\emph{rotating-wave approximation}, which involves averaging over the 
rapidly oscillating terms in Eq.~\eqref{eq:BornMarkov} to obtain an 
equation that does preserve it, i.e., one that can be put in Lindblad 
form~\cite{petruccione2002open}. This other equation, known as the quantum 
optical master equation in some contexts, is the one  we will use for 
studying the dynamics of quantum emitters coupled to a topological bath
, see chapter \ref{chap:topologicalQED} and 
appendix~\ref{app:QuantumOpticalME}.

\subsection{Resolvent formalism \label{sec:resolvent}}

Quite often in quantum physics we want to know what happens to a particular
system of interest (e.g. a qubit or atom) when coupling it to an external 
driving, or a bath. The coupling may induce transitions between the bare 
eigenstates of the system, whose probability can be computed exactly in 
some cases with resolvent operator 
techniques~\cite{cohen1992atom,steck2019atomoptics}.

Let us consider a system with free Hamiltonian $H_0$, perturbed such that 
the actual Hamiltonian of the system is $H=H_0+V$. The time evolution 
operator satisfies
\begin{equation}
  i\partial_t U(t,t_0)=HU(t,t_0) \,.
  \label{eq:U}
\end{equation}
We can split the time evolution operator into two operators $G^\pm(t,t_0)$ 
that evolve the state of the system at time $t=t_0$ forwards and backwards 
in time respectively, 
\begin{equation}
  G^\pm(t,t_0)=\pm U(t,t_0)\Theta\left(\pm(t-t_0)\right)\,.
  \label{eq:propagators}
\end{equation}
Here, $\Theta(t)$ denotes Heaviside's step function. Differentiating 
Eq.~\eqref{eq:propagators} we can see that they satisfy the same 
equation,
\begin{equation}
  \left(i\partial_t - H\right)G^\pm(t,t_0)=i\delta(t-t_0)\,.
\end{equation}
They are usually referred to as the \emph{retarded} (``$+$'') and 
\emph{advanced} (``$-$'') \emph{Green's functions} or \emph{propagators}. 
For a time-independent system $G^\pm(t,t_0)$ depends only on 
$\tau=t-t_0$. Let us now introduce their Fourier transform
\begin{gather}
  G^\pm(\tau, 0)=-\frac{1}{2\pi i}\int_{-\infty}^\infty \diffE
  e^{-iE\tau}G^\pm(E)\,,\\
  G^\pm(E)=\frac{1}{i}\int_{-\infty}^\infty \diff{\tau}
  e^{iE\tau}G^\pm(\tau, 0)\,,
  \label{eq:propagator_energy}
\end{gather}
% where the prefactor $-1/i$ is included for convenience.  
Substituting 
$G^+(\tau,0)=e^{-iH\tau}\Theta(\tau)$ in Eq.~\eqref{eq:propagator_energy} 
we find
\begin{align}
  G^+(E)&=\frac{1}{i}\int_{0}^\infty \diff{\tau} e^{i(E-H)\tau}
  =\lim_{\eta\to 0^+}
  \frac{1}{i}\int_{0}^\infty \diff{\tau} e^{i(E-H+i\eta)\tau}\nonumber\\
  &=\lim_{\eta\to 0^+} \frac{1}{E-H+i\eta}\,,
\end{align}
and similarly,
\begin{equation}
  G^-(E)=\lim_{\eta\to 0^+}\frac{1}{E-H-i\eta}\,.
\end{equation}
These expressions suggest the definition of the operator
\begin{equation}
  G(z)=\frac{1}{z-H}\,,
\end{equation}
which is a function of a complex variable $z$, such that 
$G^\pm(E)=G(E\pm i0^+)$. $G(z)$ is called the \emph{resolvent} of the 
Hamiltonian $H$.

From the definition~\eqref{eq:propagators}, it is clear that the transition 
amplitudes from an initial state $\ket{\alpha}$ to a final state 
$\ket{\beta}$ after a period $\tau$ can be computed as
\begin{equation}
  \bra{\beta}U(\tau)\ket{\alpha}=-\frac{1}{2\pi i}\int_{-\infty}^\infty
  \diffE e^{-iE\tau}\bra{\beta}G(E+i0^+)\ket{\alpha}\,.
\end{equation}
Thus, the analytical properties of the matrix elements of the resolvent play
a crucial role in determining the dynamics of the system. It can be shown
that the matrix elements of $G(z)$ are analytic in the whole complex plane
except for the real axis, where they have poles and branch cuts at the 
discrete and continuous spectrum of $H$ respectively. Furthermore, it is 
possible to continue analytically $G(z)$ bridging the cuts in the real axis,
exploring other Riemann sheets of the function, where it may no longer be 
analytic and may contain poles with a nonzero imaginary part, the so-called
unstable poles.

% It turns out that the resolvent of the perturbed system is related to that 
% of the unperturbed system by a simple algebraic equation
% \begin{equation}
%   G(z) = G_0(z) + G_0(z)VG(z)\,,
% \end{equation}
% which can be derived directly from the identity
% \begin{equation}
%   \frac{1}{A}=\frac{1}{B}+\frac{1}{B}(B-A)\frac{1}{A}\,,
% \end{equation}
% substituting $A=z-H$, and $B=z-H_0$. Iterating this equation, one can obtain
% a perturbative expansion for $G$ in powers of $V$ and $G_0$. 

% However, if the system under study is simple enough, it is possible to 
% in 
% some cases, this is not necessary. 

We will now see how to obtain explicit formulas for the relevant matrix 
elements of the resolvent. Suppose we want to know what happens to the 
states of a particular subspace spanned by $\{\ket{\alpha}\}$, which are 
eigenstates of $H_0$. Let us denote the projector onto that subspace as
$P=\sum_\alpha\ket{\alpha}\bra{\alpha}$, and the projector on the 
complementary subspace as $Q=1-P$. Then, from the defining equation of the 
resolvent $\left(z-H\right)G=1$, multiplying it from the right by $P$, and 
from the left by $P$ and $Q$, we get the equations
\begin{gather}
  P(z-H)P\left[PGP\right]-PVQ\left[QGP\right]=P\,,\label{eq:proyector1}\\
  -QVP\left[PGP\right]+Q(z-H)Q\left[QGP\right]=0\,.\label{eq:proyector2}
\end{gather}
Solving for $QGP$ in Eq.~\eqref{eq:proyector2} 
\begin{equation}
  QGP=\frac{Q}{z-QH_0Q-QVQ}VPGP \label{eq:QGP}\,,
\end{equation}
and substituting back in Eq.~\eqref{eq:proyector1}, we obtain
\begin{equation}
  P\left[z-H_0-V-V\frac{Q}{z-QH_0Q-QVQ}V\right]PGP=P\,.
\end{equation}
Introducing the operator 
\begin{equation}
  R(z)=V+V\frac{Q}{z-QH_0Q-QVQ}V\,,\label{eq:level-shift}
\end{equation}
which is known as the \emph{level-shift operator}, we can express
\begin{equation}
  PG(z)P=\frac{P}{z-PH_0P-PR(z)P}\,.\label{eq:pgp}
\end{equation}
From Eq.~\eqref{eq:QGP} we can now obtain
\begin{equation}
  QG(z)P=\frac{Q}{z-QH_0Q-QVQ}V\frac{P}{z-PH_0P-PR(z)P}.
\end{equation}

In section \ref{sec:QEdynamics} and appendix \ref{app:SelfEnergies} 
we use these formulas for computing the survival probability 
amplitude of the excited state of one and two quantum emitters.

\chapter{Doublon dynamics \label{chap:doublons}}
Recent experimental advances have provided reliable and tunable setups to 
test and explore the quantum mechanical world. Paradigmatic examples are 
ultracold atoms trapped in optical lattices~\cite{bloch2008,goldman2016}, 
quantum dots~\cite{petta2005,hanson2007,gaudreau2012,forster2014}, and 
photonic 
crystals~\cite{sansoni2012,guzmansilva2014,mukherjee2015a,mukherjee2015b}.
In these setups, quantum coherence is responsible for many exotic phenomena,
in particular, the transfer of quantum information between different 
locations, a process known as quantum state transfer. Even particles
themselves, which may carry quantum information encoded in their internal 
degrees of freedom, can be transferred in a controlled manner in these 
setups. Given its importance in quantum information processing applications,
many theoretical and experimental works have studied these processes in 
recent years~\cite{bose2003,yung2005,busl2013,sanchez2014,benseny2016}. 

Floquet engineering, that is, the use of periodic drivings in order
to modify the properties of a system, has become an essential technique in 
the cold-atom toolbox, which has enabled the simulation of 
some topological models~\cite{struck2011,atala2013,jotzu2014}. However, most
of the studies carried out so far utilize this technique aimed at the 
single-particle level. In this chapter we investigate the dynamics of two 
strongly interacting fermions bound together, forming what is termed a 
``doublon'', under the action of periodic drivings. We show how to harness 
the topological properties of different lattices to transfer 
doublons~\cite{bello2016}, and demonstrate a phenomenon by which the driving
confines doublon dynamics to a particular sublattice~\cite{bello2017a}. 
Afterwards, we address the question whether doublons can be observed in 
noisy systems such as quantum dot arrays~\cite{bello2017b}.

\section{What are doublons? \label{sec:whataredoublons}}

An ubiquitous model in the field of Condensed Matter is the Hubbard model. 
Despite its seeming simplicity, it captures a great variety of phenomena 
ranging from metallic behavior to insulators, magnetism and 
superconductivity. The Hamiltonian of this model consists of two 
contributions: a hopping term $H_J$ that corresponds to the kinetic energy 
of particles moving in a lattice, and an on-site interaction term $H_U$ 
that corresponds to the interaction between particles occupying the same 
lattice site. For spin-1/2 fermions, this Hamiltonian can be written as
\begin{equation}
  H=-\summ_{i,\,j,\,\sigma}J_{ij}c^\dagger_{i\sigma}c_{j\sigma}
  +U\summ_in_{i\up}n_{i\dn}\equiv H_J+H_U\,.
  \label{eq:Hubbard}
\end{equation}
Here, $c_{i\sigma}$ annihilates a fermion with spin $\sigma\in\{\up,\dn\}$ 
at site $i$, and $n_{i\sigma}=c^\dagger_{i\sigma}c_{i\sigma}$ is the usual 
number operator. $J_{ij}$ is the, possibly complex, hopping amplitude 
between sites $i$ and $j$ (Hermiticity of $H$ requires $J_{ij}=J_{ji}^*$). 
The interaction strength $U$ corresponds to the energy cost of the double 
occupancy.

As we shall see, in the strongly interacting limit of the Hubbard model 
($U\gg\abs{J_{ij}}$) particles occupying the same lattice site can bind 
together, even for repulsive interactions. This happens due to energy 
conservation, and the fact that, in a lattice, the maximum kinetic energy a 
particle can have is limited to the width of the energy bands. Therefore, an
initial state where the particles occupy the same site cannot decay to a 
state where the particles are separated, since they would not have enough 
kinetic energy on their own to compensate for the large interaction energy. 
In principle, both bosons~\cite{valiente2008,compagno2017,gorlach2017} and 
fermions~\cite{creffield2010,hofmann2012} can form such $N$-particle bound 
states. While the former allow for any occupation number, for fermions with 
spin $s$ the occupation of one site is restricted to at most $2s+1$ 
particles. In particular, two spin-1/2 fermions may be in a singlet spin 
configuration on the same lattice site and form a \emph{doublon}. 
Over the last years they have been investigated experimentally, mostly 
with cold atoms in optical 
lattices~\cite{winkler2006,folling2007,strohmaier2010,preiss2015,tai2017}.

\begin{figure}
  \centering
  \includegraphics{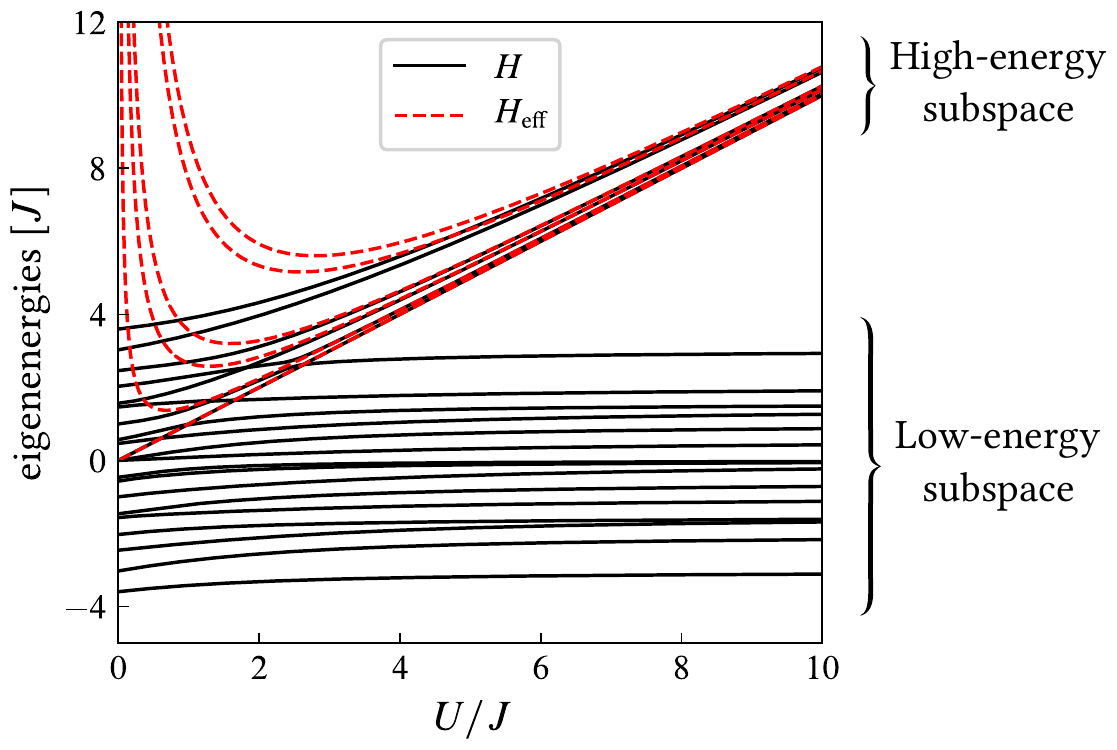}
  \caption{Energy spectrum of $H$ and $\Heff$ for two spin-1/2 fermions
  forming a singlet in the SSH-Hubbard model (chain with $N=3$ dimers and 
  $\delta=-0.2$, see next section) as a function of the interaction 
  strength. For large interactions, the effective Hamiltonian correctly 
  reproduces the eigenenergies of the high-energy subspace.}
  \label{fig:heffspectrum}
\end{figure}

The Hilbert space of two particles in a singlet configuration is spanned by 
two types of states: \emph{single-occupancy states}
\begin{equation}
  \frac{1}{\sqrt{2}}
  \left(c^\dagger_{i\up}c^\dagger_{j\dn}-c^\dagger_{i\dn}c^\dagger_{j\up}\right)
  \ket{0}\,,\quad 1\leq i < j \leq N\,, 
  \label{eq:single_occup}
\end{equation}
and \emph{double-occupancy states}, also known as doublons,
\begin{equation}
  c^\dagger_{j\up}c^\dagger_{j\dn}\ket{0}\,, \quad j=1,\dots,N\,.
  \label{eq:double_occup}
\end{equation}
Both are eigenstates of $H_U$ with eigenvalues $0$ and $U$ respectively. 
The hopping term couples both types of states, so that they no longer are 
eigenstates of the full Hamiltonian. However, for sufficiently large values 
of $U$ the eigenstates also discern in two groups, namely, $N(N-1)/2$ states
with energies $|\epsilon_n|\lesssim 4J$, which have a large overlap with the
single-occupancy states, and $N$ states with energies 
$|\epsilon_n|\simeq U$, which have a large overlap with the doublon states. 
We will refer to the span of the former as the \emph{low-energy subspace} 
(they are also referred to as \emph{scattering eigenstates}), and the span 
of the latter as the \emph{high-energy subspace} (also known as two-particle
\emph{bound states}). This distinction can be clearly appreciated in 
Fig.~\ref{fig:heffspectrum}. In this regime, a state initially having a high
double occupancy will remain like that as it evolves in time. In this sense,
we can say that the total double occupancy is an approximate conserved 
quantity in the strongly-interacting regime.  

Treating the tunneling as a perturbation it is possible to obtain an 
effective Hamiltonian for the high-energy subspace~\cite{macdonald1988}. 
The method, which goes by the name of Schrieffer-Wolff 
transformation~\cite{bravyi2011}, provides a unitary transformation that 
block-diagonalizes the Hamiltonian perturbatively in blocks of states with 
different number of doubly occupied sites. To achieve this, we first split 
the kinetic term into hoppings that increase, decrease, and leave unaltered 
the double occupancy, $H_J=H^+_J+H^-_J+H^0_J$, where
\begin{gather}
  H^+_J = -\summ_{i,\,j,\,\sigma}J_{ij} n_{i\ol\sigma}c^\dagger_{i\sigma}
  c_{j\sigma}h_{j\ol\sigma}\,,\\
  H^-_J = -\summ_{i,\,j,\,\sigma}J_{ij} h_{i\ol\sigma}c^\dagger_{i\sigma}
  c_{j\sigma}n_{j\ol\sigma}\,,\\
  H^0_J = -\summ_{i,\,j,\,\sigma}J_{ij} 
  \left(n_{i\ol\sigma}c^\dagger_{i\sigma}c_{j\sigma}n_{j\ol\sigma}
  +h_{i\ol\sigma}c^\dagger_{i\sigma}c_{j\sigma}h_{j\ol\sigma}\right)\,.
\end{gather}
In these equations $\ol\sigma$ denotes the opposite value of $\sigma$ and 
$h_{i\sigma}\equiv 1-n_{i\sigma}$. The transformation of $H$ by any unitary 
$e^{iS}$ ($S^\dagger=S$) can be computed as
\begin{equation}
  \tilde H=e^{iS} H e^{-iS}=H+\frac{[iS,H]}{1!}
  +\frac{[iS,[iS,H]]}{2!}+\cdots\,.
\end{equation}
Noting that 
$\left[H_U,H^\alpha_J\right]=\alpha U H^\alpha_J$, $\alpha\in\{\pm,0\}$, 
one can readily see that in order to remove the terms that couple the two 
sectors at zeroth order, $H^\pm_J$, one has to choose 
$iS={\left(H^+_J-H^-_J\right)/U}$. Then,
\begin{equation}
  \tilde H = H^0_J+H_U+\frac{\left[H^+_J,H^-_J\right]
  +\left[H^0_J,H^-_J\right]+\left[H^+_J,H^-_J\right]}{U}
  +O\left(U^{-2}\right)\,.
\end{equation}
Keeping terms up to order $U^{-1}$ that act non-trivially on doublon 
states, we obtain the following effective Hamiltonian for doublons
\begin{equation}
  \Heff=H_U + \frac{1}{U}H^+_JH^-_J\,.
\end{equation}
Since we are concerned with states with no singly occupied sites, in 
$H^+_JH^-_J$ we have to consider just those hopping processes where a 
doublon splits and then recombines again to the same site or to an adjacent 
site,
\begin{equation}
  H^+_JH^-_J \reseq
  \summ_{i,\,j,\,\sigma} \abs{J_{ij}}^2 
  n_{i\ol\sigma}c^\dagger_{i\sigma}c_{j\sigma}h_{j\ol\sigma}
  h_{j\ol\sigma}c^\dagger_{j\sigma}c_{i\sigma}n_{i\ol\sigma}
  + \summ_{i,\,j,\,\sigma}J_{ij}^2
  n_{i\ol\sigma}c^\dagger_{i\sigma}c_{j\sigma}h_{j\ol\sigma}
  h_{i\sigma}c^\dagger_{i\ol\sigma}c_{j\ol\sigma}n_{j\sigma}\,.
\end{equation}
The asterisk on top of the equal sign denotes equality when restricted 
to the doublon subspace. The terms in the first sum can be rewritten as
\begin{equation}
  n_{i\ol\sigma}c^\dagger_{i\sigma}c_{j\sigma}h_{j\ol\sigma}
  h_{j\ol\sigma}c^\dagger_{j\sigma}c_{i\sigma}n_{i\ol\sigma}=
  n_{i\ol\sigma}n_{i\sigma}-n_{i\ol\sigma}n_{i\sigma}
  n_{j\ol\sigma}n_{j\sigma}\,,
  \quad i\neq j\,.
\end{equation}
Here, we have used the fact that for a doublon state, if a site is occupied
it has to be double occupied, i.e., 
$n_{i\sigma}\reseq n_{i\sigma}n_{i\ol\sigma}\reseq n_{i\ol\sigma}$. As for 
the terms in the second sum,
\begin{align}
  n_{i\ol\sigma}c^\dagger_{i\sigma}c_{j\sigma}h_{j\ol\sigma}
  h_{i\sigma}c^\dagger_{i\ol\sigma}c_{j\ol\sigma}n_{j\sigma}=
  c^\dagger_{i\ol\sigma}c^\dagger_{i\sigma}c_{j\sigma}c_{j\ol\sigma} \,,
  \quad i\neq j\,.
\end{align}
Finally, using doublon creation and annihilation operators,
$d^\dagger_i=c^\dagger_{i\up}c^\dagger_{i\dn}$ and $d_i=c_{i\dn}c_{i\up}$,
the effective Hamiltonian for doublons can be written as
\begin{equation}
  \Heff=\summ_{i,\,j}\frac{2J_{ij}^2}{U}d^\dagger_i d_j
  +\summ_i\mu_i d^\dagger_i d_i 
  -\summ_{i,\,j}\frac{2\abs{J_{ij}}^2}{U}d^\dagger_i d_i
  d^\dagger_j d_j\,,
  \label{eq:Heffdoublons}
\end{equation}
with $\mu_i=U+\sum_j 2\abs{J_{ij}}^2/U$. The first term 
corresponds to an effective hopping for doublons, the second to an 
effective on-site chemical potential, and the third to an attractive 
interaction between doublons.

We remark that for a 1D lattice with homogeneous nearest-neighbor hoppings, 
the two-body spectrum can be computed exactly with Bethe Ansatz 
techniques~\cite{essler2005hubbard}. 

In appendix~\ref{app:EffectiveHamiltonian} we derive an effective 
Hamiltonian for doublons including also the effect of an external 
periodic driving.

\section{Doublon dynamics in 1D and 2D lattices\label{sec:doublondynamics}}

After understanding what doublons are, and what makes them stable 
quasiparticles, we are now in a good position to discuss their dynamics.
In this section we show how the interplay between topology, interactions, 
and driving, leads to surprising effects that constrain the motion of 
doublons.

\subsection{Dynamics in the SSH chain}

Interestingly, the topological properties of the SSH model can be harnessed 
to produce the transfer of non-interacting particles between the ends of a 
finite chain, without them occupying the intervening sites. Key to this 
process is the presence of edge states. Let us recall their properties. A 
finite dimer chain supports two edge states $\ket{\ES_\pm}$, with energies 
$\pm\epsilon/2$, when it is in the topological phase ($\delta<0$). They are 
well separated energetically from the rest of (bulk) states, and there is a 
small energy splitting between them that decreases exponentially with 
increasing chain size, $\epsilon\propto e^{-N/\lambda}$. 
Each of them is exponentially localized on both edges of the chain, and 
they are even and odd under space inversion; thus, they can be regarded as 
a non-local two level system. As a consequence, a particle in a 
superposition of both edge states will oscillate between the ends of the 
chain as it evolves in time. For example, if the particle is initially on 
the first site of the chain 
$\ket{\psi(0)}=\ket{1}\simeq\left(\ket{\ES_+}+\ket{\ES_-}\right)/\sqrt{2}$, 
the probability to find it on the last site of the chain 
$\ket{2N}\simeq\left(\ket{\ES_+}-\ket{\ES_-}\right)/\sqrt{2}$ is
\begin{equation}
  \abs{\mean{2N\vert\psi(t)}}^2=
  \frac{1}{4}\abs*{e^{-i\epsilon t/2} - e^{i\epsilon t/2}}^2
  =\frac{1-\cos(\epsilon t)}{2}\,,
\end{equation}
thus, in a time period $T_0=\pi/\epsilon$ the particle will be transferred
with certainty to the other end of the chain. 

We want to know how interaction between particles affects this process, 
and see whether the controlled transfer of doublons is possible. The 
Hamiltonian of the system corresponds to Eq.~\eqref{eq:Hubbard} with 
\begin{equation}
  J_{ij}=\begin{cases}
    J\left[1+\delta(-1)^{\max(i,j)}\right]\,,&\abs{i-j}=1 \\
    0\,,&\text{otherwise}
  \end{cases}\,.
\end{equation}
This is just a different way to express $H_\mathrm{SSH}$ 
[Eq.~\eqref{eq:SSH} in section~\ref{sec:SSH}] including the spin degree
of freedom and the on-site interaction between particles. We will refer to 
this model as the SSH-Hubbard model.

\begin{figure}[!htb]
  \centering
  \includegraphics{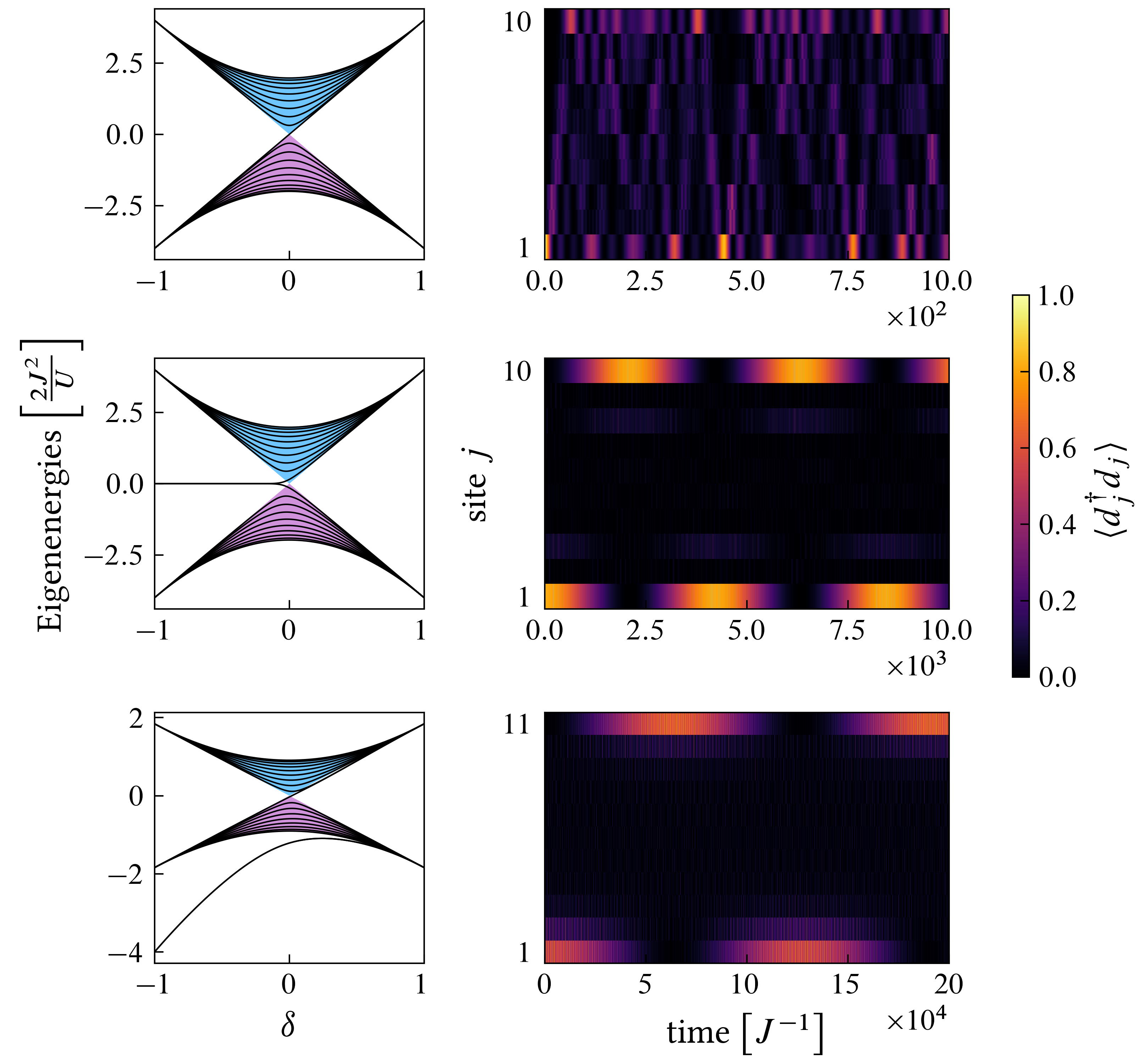}
  \caption{Spectrum of the effective Hamiltonian (left), and associated 
  doublon dynamics (right), computed numerically solving the Schr\"odinger
  equation in the two-particle singlet subspace. In all the cases shown, 
  a doublon is initially on the 1st site of the lattice. 
  Case without any external influence (top row). The spectrum shows 
  the absence of edge states for any value of $\delta$. The parameters
  for the dynamics are: $\delta=-0.3$, $U=10J$ and $N=5$. 
  Case with a compensating local potential $V_\mathrm{gate}$ on the ending 
  sites of the chain (middle row). A pair of topological edge states 
  appears for negative values of $\delta$. The parameters for the dynamics 
  are the same as in the previous case. 
  Case with an external ac-field with parameters: $E=3.2 J$, $\omega=2J$ 
  and $b=1/2$ (bottom row). The spectrum shows a pair of Shockley edge 
  states that separate from the bottom of the lower band. The dynamics is 
  for a chain with $11$ sites, $\delta=0$, and $U=16J$.}
  \label{fig:dynamicsSSH}
\end{figure}

In Fig.~\ref{fig:dynamicsSSH} (top row) we plot the dynamics of a doublon 
starting on the first site of a small chain. As can be observed, the 
edge-to-edge oscillations are lost in the strongly-interacting regime. We 
can understand why looking at the effective Hamiltonian for doublons
\begin{equation}
  H_\mathrm{eff} =
  \summ_j\left(\Jeff_1 d^\dagger_{2j}d_{2j-1} 
  + \Jeff_2 d^\dagger_{2j+1}d_{2j} + \HC\right)
  +\summ_j\mu_j d^\dagger_jd_j\,,
  \label{eq:HeffSSH}
\end{equation}
which can be obtained directly from Eq.~\eqref{eq:Heffdoublons}, neglecting 
the interaction between doublons, since we are considering just one of them.
The effective hopping amplitudes are $\Jeff_1=2J^2(1+\delta)^2/U$ and 
$\Jeff_2=2J^2(1-\delta)^2/U$. Importantly, the effective local chemical 
potential $\mu_j$ is different for the ending sites, as they have one fewer 
neighbor,
\begin{equation}
  \mu_j=\begin{cases}
    \mu_\mathrm{bulk}=\Jeff_1+\Jeff_2+U\,, & 2\leq j\leq 2N-1 \\
    \mu_\mathrm{edge}=\Jeff_1+U\,, & j=1,2N
  \end{cases}\,.
\end{equation}
This chemical potential difference at the edges spoils the chiral symmetry 
of the model, and is able to shift the energy of the edge states that 
appear in the topological phase of the unperturbed SSH model. In fact,
for the particular value 
$\Delta\mu\equiv\mu_\mathrm{bulk}-\mu_\mathrm{edge}=\Jeff_2$, they 
completely merge into the bulk bands, see Fig.~\ref{fig:dynamicsSSH} 
(top row). This explains why the mechanism that produces the edge-to-edge 
oscillations in the single particle case does not apply directly to the 
doublon case. 

We can now think of different ways to restore the edge states in order 
to produce the transfer of doublons. One possibility is is to add a local 
potential at the edges of the chain so as to compensate for the 
difference in chemical potential,
\begin{equation}
  H = H_J + H_U + 
  V_\mathrm{gate}\summ_\sigma\left(n_{1\sigma}+n_{2N\sigma}\right)
  \,,
\end{equation}
with $V_\mathrm{gate}=J^2(1-\delta)^2/U$. The resulting effective 
Hamiltonian has a homogeneous chemical potential $\mu_j=\mu_\mathrm{gate}$ 
for all $j$, and is formally identical to that of the SSH model. Indeed, 
this produces the desired dynamics, see Fig.~\ref{fig:dynamicsSSH}
(middle row). Another possibility is to take advantage of this chemical 
potential difference, which can localize states on the edges of the chain 
of the Shockley type.  This requires the hopping 
amplitudes to be smaller than $\Delta\mu$. Usually this is not the case, 
however there is an efficient way to induce such states by driving the 
system with a high-frequency ac field. The ac field renormalizes the 
hoppings, which become smaller than in the undriven 
case~\cite{dunlap1986,grossmann1991,grifoni1998}. This cannot be achieved, 
for example, by simply reducing the hoppings $J_1$ and $J_2$ by hand, 
since this will also affect the effective chemical potential which still 
will be of the same order of $\Jeff_1$ and $\Jeff_2$. To model the ac 
field we add a periodically oscillating potential that rises linearly 
along the lattice,
\begin{equation}
  H(t)=H_J+H_U+E\cos(\omega t)\summ_j x_j\left(n_{j\up}+n_{j\dn}\right)\,.
  \label{eq:HtSSH}
\end{equation}
Here, $E$ and $\omega$ are the amplitude and frequency of the driving, and 
$x_j$ is the spatial coordinate along the chain. The geometry of the chain
is determined by the lattice constant, which we set as the unit of distance,
and the intracell distance $b\in[0,1]$, such that 
$x_j=\floor{(j-1)/2} + b{(j-1\mod 2)}$. Since the Hamiltonian 
is periodic in time we can apply Floquet theory and obtain a 
time-independent effective Hamiltonian in the high-frequency regime, see 
appendix~\ref{app:EffectiveHamiltonian}. As it turns out, when the 
leading energy scale is that of the of the interaction between particles, 
the effective Hamiltonian for doublons is the same as the one in 
Eq.~\eqref{eq:HeffSSH} with renormalized hopping amplitudes
\begin{gather}
  \Jeff_1=\bes{0}\left(\frac{2Eb}{\omega}\right)\frac{2J^2(1+\delta)^2}{U}
  \,, \\
  \Jeff_2=\bes{0}\left(\frac{2E(1-b)}{\omega}\right)\frac{2J^2(1-\delta)^2}{U}
  \,,
\end{gather}
where $\bes{0}$ denotes the zeroth order Bessel function of the first kind. 
The factor $2$ in the argument of $\bes{0}$ comes from the doublon's 
twofold electric charge. We show the effect of this renormalization in 
Fig.~\ref{fig:dynamicsSSH} (bottom row); as the effective tunneling 
reduces in magnitude the bulk bands become narrower, and two Shockley edge 
states are pulled from the bottom of the lower band. The geometry, which 
so far did not played any role, now becomes important in this 
renormalization of the hoppings. The simplest case is for $b=1/2$, in which 
both hoppings are renormalized by the same factor. We remark that the 
on-site effective chemical potential, being a local operator, commutes with 
the periodic driving potential, and so it is not renormalized. 

This approach has the peculiarity that it only relies on the $\Delta\mu$
produced by the interaction, and not on the topology of the chain. Thus, 
it can be used to produce the transfer of doublons in normal chains with 
an odd number of sites, see Fig.~\ref{fig:dynamicsSSH} (bottom row), 
contrary to the static approach. Notice that the high-frequency effective 
Hamiltonian preserves space inversion symmetry, which guarantees that 
the Shockley edge states hybridize forming even and odd combinations under 
space inversion, just like topological edge states do. However, Shockley
edge states lack the protection against certain types of disorder, such 
as off-diagonal disorder (see section~\ref{sec:SingleQEdynamics}), that 
topological edge states have. 

\begin{figure}[!htb]
  \centering
  \includegraphics{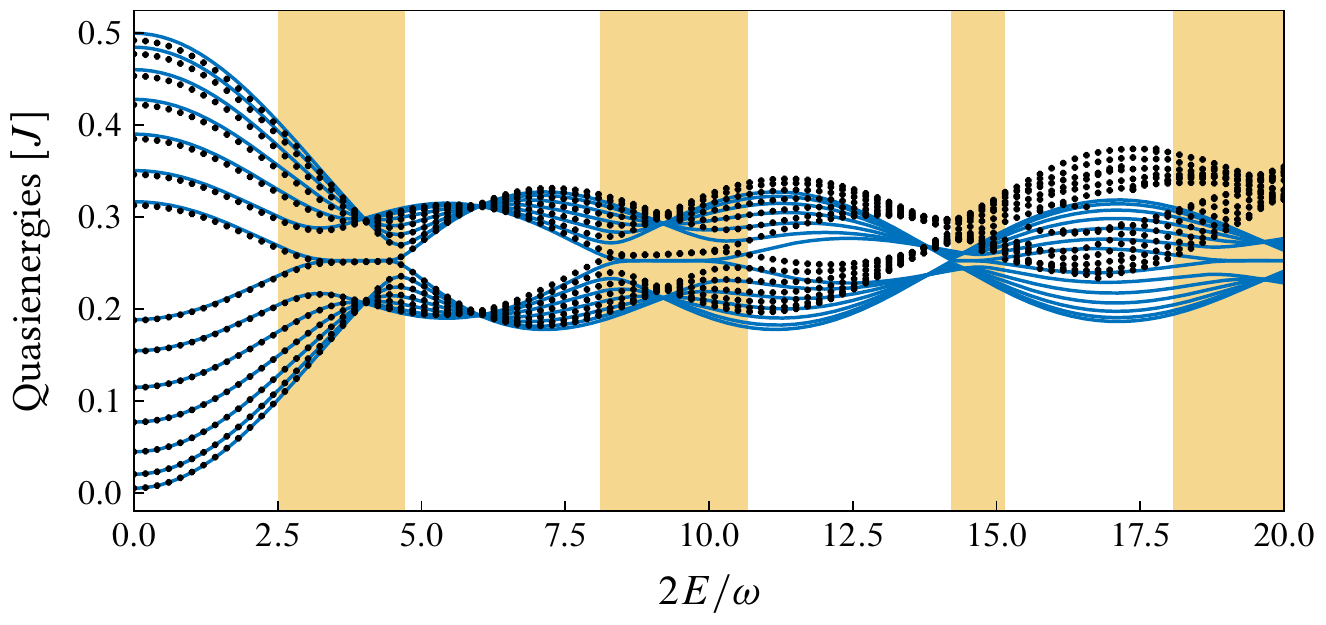}
  \caption{Quasienergy spectrum for a chain with a compensating local 
  potential $V_\mathrm{gate}$ and an external ac field, as a function of 
  the ac field amplitude. The parameters are $\delta=0.1$, $N=7$, $U=16J$, 
  $b=0.6$, and $\omega=2J$. Exact quasienergies (dots) have been 
  obtained diagonalizing the time evolution operator for one period. We 
  only show the quasienergies corresponding to the $2N$ Floquet modes with
  largest total double occupancy. The lines correspond to the approximate 
  quasienergies given by the effective Hamiltonian. Shaded areas mark 
  the regions where the system is in the non-trivial phase, according
  to the values of $\abs{\Jeff_1}$ and $\abs{\Jeff_2}$.}
  \label{fig:quasienergies}
\end{figure}

Last, we can combine both approaches, restoring the symmetries of the SSH
model, and being able to modify the system's topology with the ac field, 
provided $b\neq 1/2$~\cite{gomezleon2013}. In 
Fig.~\ref{fig:quasienergies} we compare the exact quasienergies of the 
doublon states with the quasienergies given by the effective Hamiltonian. 
Both agree as long as photon-resonance effects are negligible, i.e., 
$E\lesssim U$ (see appendix~\ref{app:EffectiveHamiltonian}). For 
field parameters such that $\abs{\Jeff_1}<\abs{\Jeff_2}$, the system 
supports a pair of topological edge states, which allow for the transfer 
of doublons.

\subsection{Dynamics in the \texorpdfstring{$\mathcal{T}_3$}{T3} and Lieb lattices}
% \subsection{Dynamics in the $\mathcal{T}_3$ and Lieb lattices}

We will now analyze the consequences of the effective local chemical 
potential on the doublon dynamics in 2D lattices.
We will consider lattices with homogeneous hopping amplitudes threaded by a 
static magnetic flux in the presence of a circularly polarized ac field. 
They are modelled by the Hamiltonian
\begin{equation}
  H(t)=-J\summ_{\mean{i,j},\,\sigma}e^{i\phi_{ij}}
  c^\dagger_{i\sigma}c_{j\sigma} + U\summ_j n_{j\up}n_{j\dn}
  + \summ_{j,\,\sigma} V_j(t)n_{j\sigma}\,,
\end{equation}
with $V_j(t)=x_jE\cos(\omega t)+y_jE\sin(\omega t)$, where 
$(x_j,y_j)\equiv \mbf r_j$ are 
the coordinates of site $j$. The sum in the hopping term runs over each 
oriented pair of nearest-neighbor sites. The magnetic flux induces complex
phases in the hoppings such that the sum of the phases around a closed loop 
equals $2\pi\Phi/\Phi_0$, where $\Phi$ is the total flux threading the loop 
and $\Phi_0$ is the magnetic flux quantum.

The effective Hamiltonian for doublons generalizes in a straightforward 
manner for 2D lattices (see appendix~\ref{app:EffectiveHamiltonian}),
\begin{equation}
  \Heff=\jeff\summ_{\mean{i,j}}e^{i2\phi_{ij}}d^\dagger_id_j
  +\summ_j\mu_jd^\dagger_jd_j\,,
  \label{eq:Heff2D}
\end{equation}
with effective hopping and chemical potentials
\begin{equation}
  \jeff=\frac{2J^2}{U}\bes{0}\left(\frac{2Ea}{\omega}\right)\,,\quad
  \mu_j=\frac{2J^2}{U}z_j\,,
\end{equation}
where $z_j$ is the coordination number (the number of nearest neighbors) of
site $j$. In this effective Hamiltonian we have neglected again
interaction terms, since we are considering just one doublon.
Importantly, the hopping renormalization is isotropic because
the ac field polarization is circular and all neighboring sites are the 
same distance apart; $\abs{\mbf r_i - \mbf r_j}=a$ for all neighboring $i$ 
and $j$. 

\begin{figure}[!htb]
  \centering
  \includegraphics{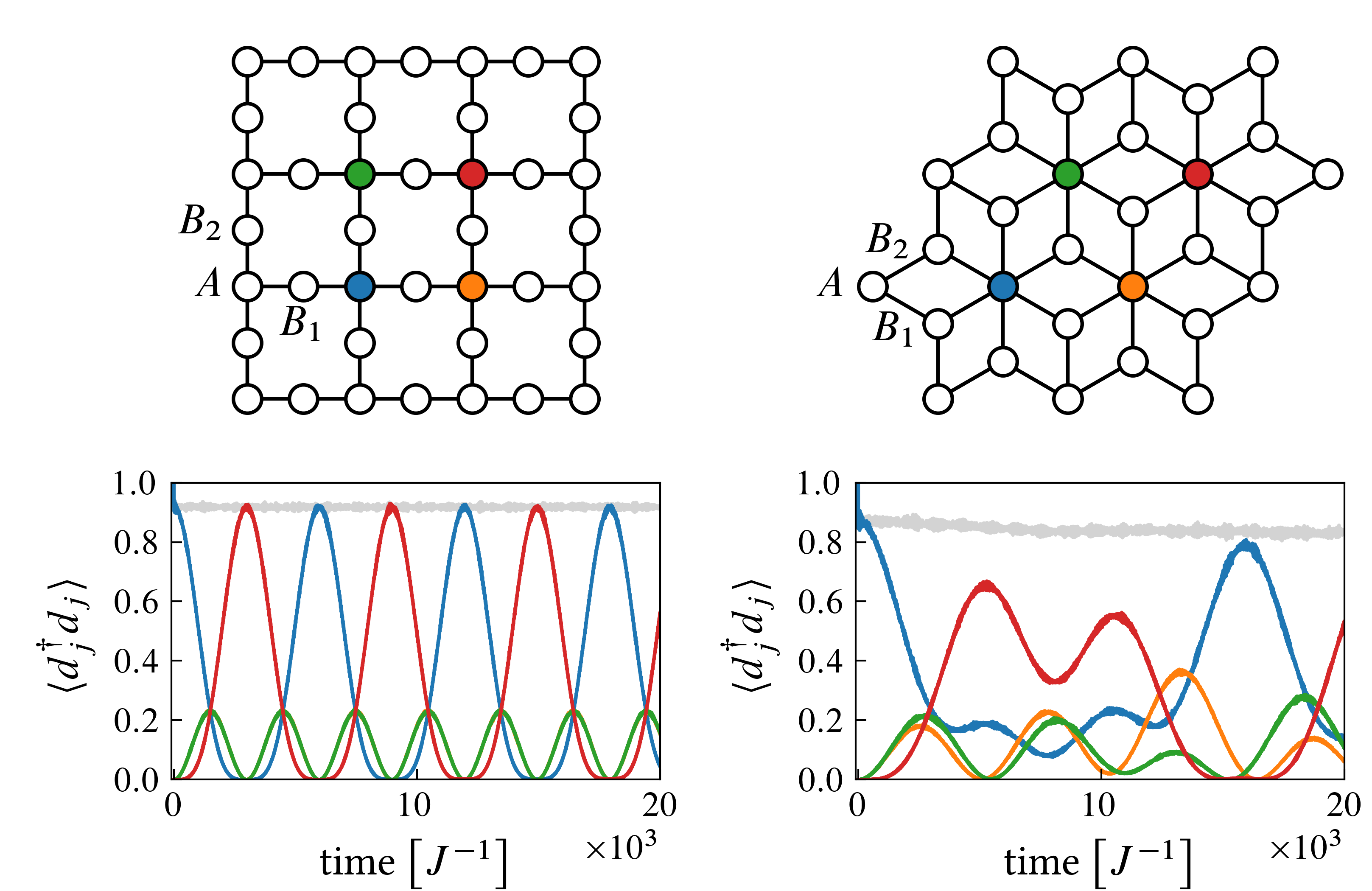}
  \caption{Doublon dynamics on a finite piece of a Lieb lattice 
  (a) and a $\mathcal{T}_3$ lattice (b). In both cases the 
  doublon is initially on the blue dot. Parameters: 
  $\Phi=0$, $U=16J$, $\omega=2J$, and $E=4.8J$ (a), 
  $E=4.2J$ (b). The grey line is the sum of the double 
  occupancy on the four colored sites. In the Lieb lattice, the 
  double occupancy on the green and yellow sites is the same.
  The dynamics have been obtained solving numerically the 
  time-dependent Schr\"odinger equation in the two-particle 
  singlet subspace.}
  \label{fig:dynamics2D}
\end{figure}

As we can see, the ac driving allows us to independently tune the effective
hopping amplitude with respect to the effective local chemical potential. 
This has a big impact on the dynamics of doublons in lattices that can be 
divided into sublattices with different coordination numbers, such as the 
Lieb lattice, and the $\mathcal{T}_3$ lattice, shown in 
Fig.~\ref{fig:dynamics2D}. As can be seen in the dynamics, the doublon 
moves mostly through sites with the same coordination number, an effect we 
have termed \emph{sublattice dynamics}. 

\begin{figure}[!htb]
  \centering
  \includegraphics{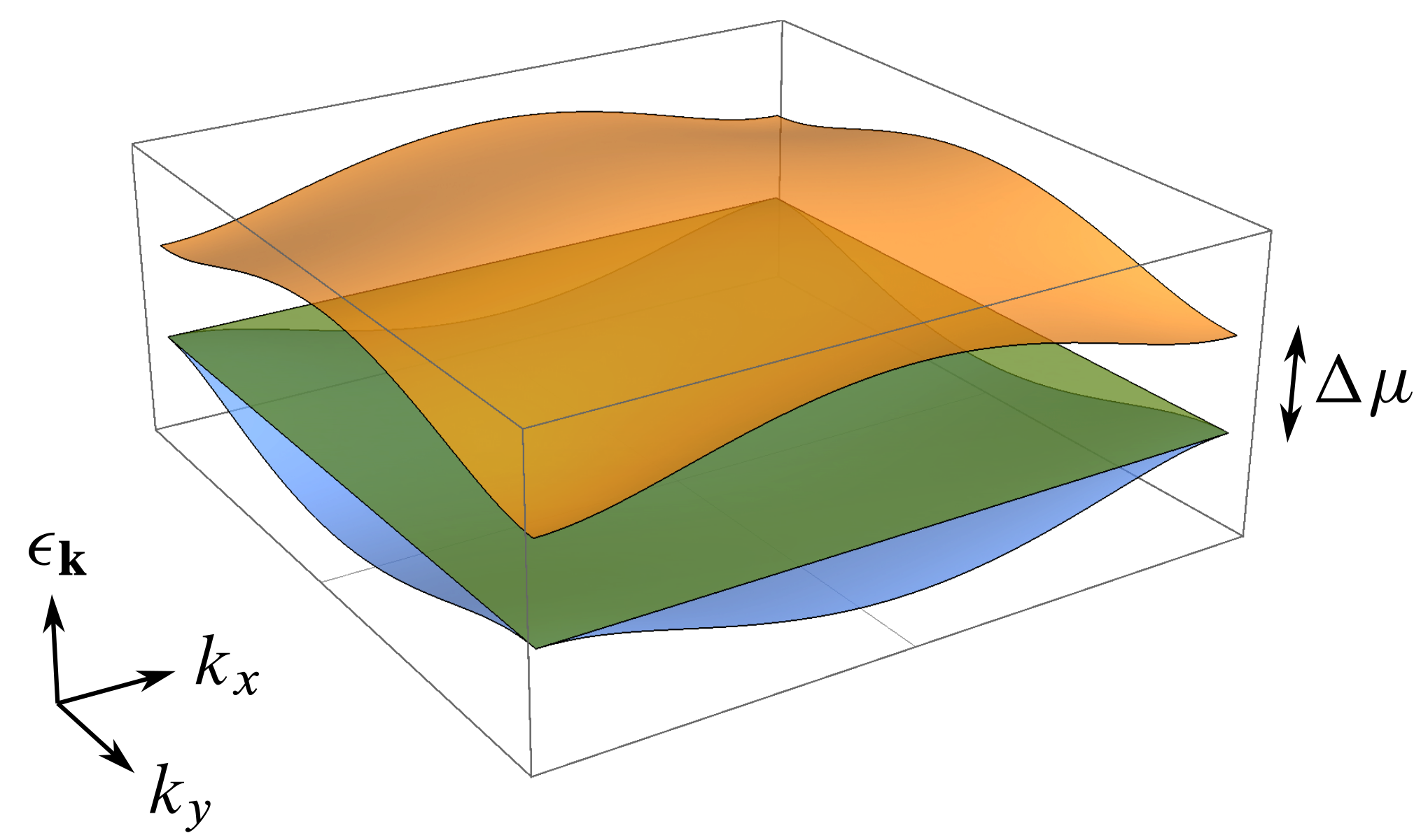}
  \caption{Doublon energy bands for the Lieb lattice. The effective
  local potential opens a gap between the upper band and the rest.
  The driving allows the band width to be reduced, 
  flattening the bands, while keeping the gap the same.}
  \label{fig:liebbands}
\end{figure}

To understand why, it is useful to
look at the effective Hamiltonian in momentum representation, which in the 
absence of an external magnetic flux adopts the same form for both lattices
$\Heff=\sum_\bk V^\dagger_\bk H_\bk V_\bk$, with
\begin{equation}
  H_\bk=\myvec{\Delta\mu & f_1(\bk) & f_2(\bk)\\f^*_1(\bk) & 0 & 0\\
  f^*_2(\bk) & 0 & 0}\,,\quad
  V_\bk=\myvec{d_{A\bk}\\d_{B_1\bk}\\d_{B_2\bk}}\,.
\end{equation}
Here, $d_{\alpha\bk}$ is the annihilation operator of a doublon with 
momentum $\bk$ in sublattice $\alpha\in\{A,B_1,B_2\}$. The 
eigenenergies and eigenstates are as follows:
\begin{gather}
  \epsilon^0_\bk=0\,,\quad
  \epsilon^\pm_\bk=\frac{1}{2} \left[\Delta\mu
  \pm\sqrt{4\abs{f_1(\bk)}^2+4\abs{f_2(\bk)}^2+\Delta\mu^2}\right]
  \,,\\
  \ket*{u^0_\bk}\propto\left[-\frac{f_2(\bk)}{f_1(\bk)}d^\dagger_{B_1\bk}
  +d^\dagger_{B_2\bk}\right]\ket{0}
  \,,\\
  \ket*{u^\pm_\bk}\propto\left[
  \frac{\epsilon^\pm_\bk}{f^*_2(\bk)}d^\dagger_{A\bk}
  +\frac{f^*_1(\bk)}{f^*_2(\bk)}d^\dagger_{B_1\bk}
  +d^\dagger_{B_2\bk}\right]\ket{0} \,.
\end{gather}
Note how the states of the flat band do not have weight on the $A$ 
sites of the lattice. The chemical potential difference 
$\Delta\mu=2J^2(z_A-z_B)/U$ produces a splitting between the upper band 
and the rest of the bands, see Fig.~\ref{fig:liebbands}. The functions 
$f_1$ and $f_2$ depend on the particular lattice geometry as shown in 
the table below,
\begin{center}
  \vspace*{1ex}
  \begin{tabular}{ccc}
    Lattice & $f_1(\bk)$ & $f_2(\bk)$ \\[0.5ex]
    \hline\hline
    &&\\[-2ex]
    $\mathcal{T}_3$ & $\jeff\left[
      e^{-i\left(k_x+\frac{k_y}{\sqrt{3}}\right)\frac{1}{2}}
     +e^{i\left(k_x-\frac{k_y}{\sqrt{3}}\right)\frac{1}{2}}
     +e^{i\frac{k_y}{\sqrt{3}}}\right]$
   & $f_2(\bk)=f^*_1(\bk)$\\
    &&\\[-2ex]
    Lieb & $2\jeff\cos\left(\frac{k_x}{2}\right)$
    & $2\jeff\cos\left(\frac{k_y}{2}\right)$\\
    &&\\[-2ex]
    \hline
  \end{tabular}
  \vspace*{1ex}
\end{center}
They are proportional to $\jeff$, which can be tuned by the ac 
driving. In particular, the relative weight on the $A$ sublattice
of the Bloch states corresponding to the upper (lower) band can 
be increased (reduced) by tuning the ac field parameters closer to 
a zero of the Bessel function.

\begin{figure}[!htb]
  \centering
  \includegraphics{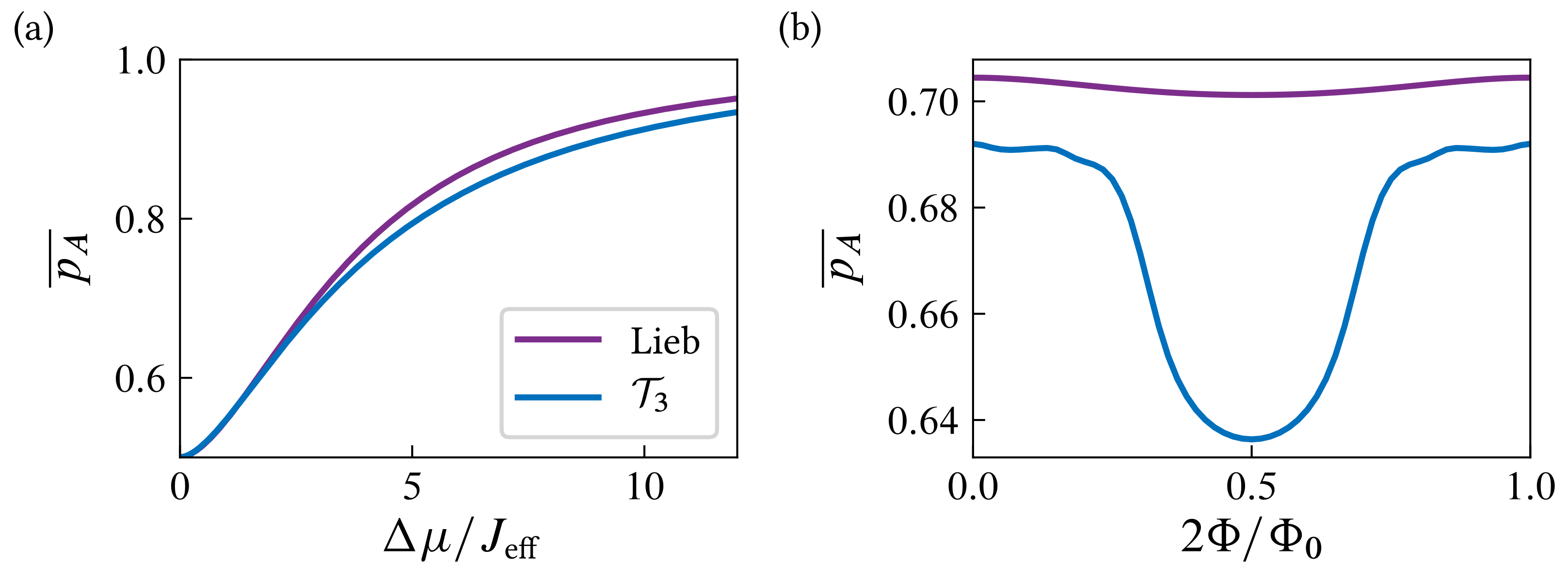}
  \caption{Probability to remain in sublattice $A$ as a function of the 
  ratio $\Delta\mu/\jeff$, for $\Phi=0$ (a), and as a function of 
  the magnetic flux per plaquette, for $\Delta\mu/\jeff=3$ (b).}
  \label{fig:pA}
\end{figure}

When studying \emph{quantum walks}~\cite{mulken2011}, i.e., the 
coherent evolution of particles in networks, it is natural to ask 
about the probability of finding a particle that was initially on 
site $i$ to be on site $j$ after a certain time $t$, that is, 
$p_{ij}(t)\equiv\abs{\bra{i}U(t)\ket{j}}^2=
\abs{\bra{i}e^{-iHt}\ket{j}}^2$. Using~\eqref{eq:Heff2D} as the 
effective single-particle Hamiltonian for the doublon, we define 
$p_A(t)\equiv\sum_{i,\,j\in A}p_{ij}(t)/N_A$, which is the probability for 
the doublon to remain in sublattice $A$ at time $t$; $N_A$ is the total 
number of sites that belong to sublattice $A$. To demonstrate sublattice 
confinement, we can compute the long-time average 
\begin{equation}
  \overbar{p_A}\equiv\lim_{t\to\infty}\frac{1}{t}\int_0^t
  \diff{t'}p_A(t')\,.
\end{equation}
According to the definition, the probability $p_A(t)$ is 
$\norm{U_A(t)}^2/N_A$, where $\norm{\cdot}$ denotes the 
Hilbert-Schmidt norm, and $U_A(t)=P_AU(t)P_A$ is the time evolution
operator projected on the subspace of the $A$ sublattice. Using
the spectral decomposition 
\begin{equation}
  U(t)=\summ_\bk\summ_{\alpha=0,\pm}e^{-i\epsilon^\alpha_\bk t}
  \ket*{u^\alpha_\bk}\bra*{u^\alpha_\bk}\,,
\end{equation}
we can express
\begin{align}
  \norm{U_A(t)}^2 & =
  \summ_\bk\abs*{\frac{e^{-i\epsilon^+_\bk t}}{1+g_+(\bk)}
  + \frac{e^{-i\epsilon^-_\bk t}}{1+g_-(\bk)}}^2
  \\
  & =\summ_{\bk,\,\alpha=\pm} \frac{1}{\left[1+g_\alpha(\bk)\right]^2}
  +\summ_\bk\frac{2\cos\left(\epsilon^+_\bk t - \epsilon^-_\bk t\right)}
  {\left[1+g_+(\bk)\right]\left[1+g_-(\bk)\right]}\,,
\end{align}
where we have defined $g_\pm(\bk)=
\left[\abs{f_1(\bk)}^2+\abs{f_2(\bk)}^2\right]
\left(\epsilon^\pm_\bk\right)^{-2}$. The time average is, thus, given by 
\begin{equation}
 \overbar{p_A}=\frac{1}{V}\int_\mathrm{FBZ} d^2\bk\,
  \left\{\frac{1}{\left[1+g_+(\bk)\right]^2}
  +\frac{1}{\left[1+g_-(\bk)\right]^2}\right\}\,.
\end{equation}
Here, we have taken the thermodynamic limit, replacing the sum over 
momenta by an integral on the first Brillouin zone (FBZ); $V$ 
stands for the FBZ area. As can be seen in Fig.~\ref{fig:pA}(a) the 
probability $\overbar{p_A}$ can be enhanced by tuning the ratio 
$\Delta\mu/\jeff$ to larger values, meaning that it is possible to confine 
the doublon's dynamics to a single sublattice by suitably changing the ac 
field parameters. We have also computed the dependence of $\overbar{p_A}$ 
with the magnetic flux threading the smallest plaquette (the smallest 
closed path in the lattice), see Fig.~\ref{fig:pA}(b); however, its 
variation turns out to be minor with $\overbar{p_A}$ gently increasing as 
the flux is tuned away from $2\Phi/\Phi_0=1/2$. A much stronger dependence 
is observed for the $\mathcal{T}_3$ than for the Lieb lattice. This is to 
be expected as Aharonov-Bohm phases have more dramatic effects in the 
$\mathcal{T}_3$ lattice, notably the caging effect that occurs for a 
magnetic flux $\Phi/\Phi_0=1/2$ in the singly-charged particle 
case~\cite{vidal1998}. 

\begin{figure}[!htb]
  \centering
  \includegraphics{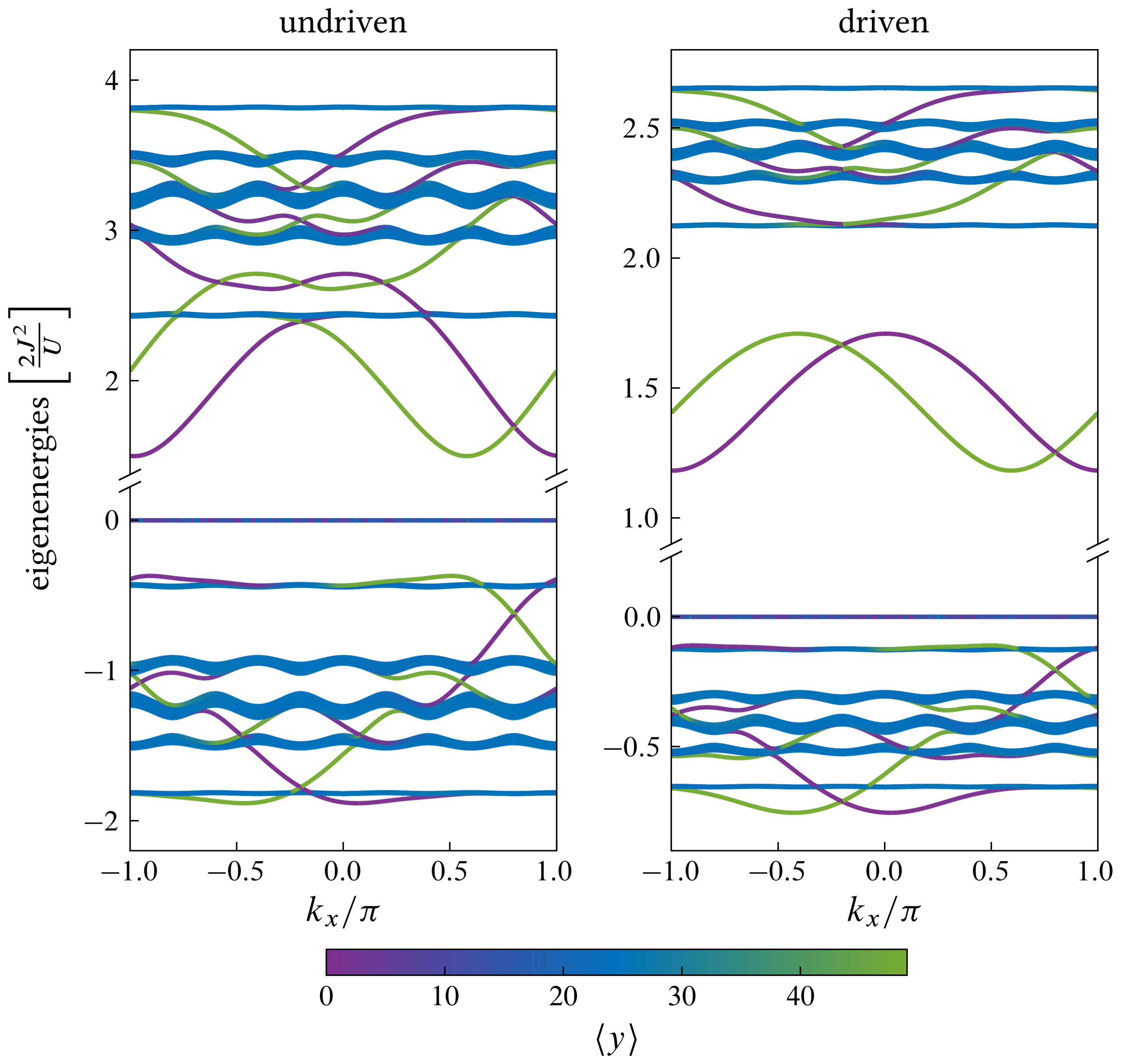}
  \caption{Energy spectrum of a finite ribbon of the Lieb lattice, 
  with $N_y=50$ unit cells along the $y$ direction in the presence of a 
  magnetic flux $\Phi/\Phi_0=1/10$. The ac field in the case shown
  in the right plot is such that $\bes{0}\left(2Ea/\omega\right)=1/2$. 
  The color indicates the localization of the state along the finite 
  dimension of the lattice: green and purple denote states localized 
  on each edge of the ribbon (topological and Shockley-like edge states), 
  while blue is used for bulk states that are not localized (minibands). 
  Topological edge states always connect different minibands, while 
  Shockley edge states do not necessarily do so.}
  \label{fig:LiebFlux}
\end{figure}

It is worth mentioning that a similar effect constrains the motion of 
doublons in any lattice with boundaries. The sites on the edges necessarily 
have fewer neighbors than those in the bulk and therefore have a smaller 
chemical potential. This produces eigenstates localized on the edges, which 
are of the Shockley or Tamm type. As a consequence, the doublon's dynamics 
can be confined to just the edges of the lattice. Furthermore, since 
having non-trivial topology in 2D does not require any symmetry, in the 
presence of a nonzero magnetic flux any lattice can support both 
Shockley-like edge states and topological chiral edge states at the same 
time. Let us be more specific. When comparing the 
effective model~\eqref{eq:Heff2D} with that of a Chern 
insulator~\cite{bernevig2013topo}, the only difference is the local chemical
potential term. It is well known that strong disorder potentials eventually 
destroy the topological properties of Chern insulators as they transition to
a trivial Anderson insulator by a mechanism known as ``levitation and 
annihilation'' of extended states~\cite{prodan2010,castro2015}. Nonetheless,
the chemical potential term in our Hamiltonian constitutes a very particular
form of disorder that does not affect the topology of the system. In 
Fig.~\ref{fig:LiebFlux} we show the energy spectrum of a ribbon of the Lieb 
lattice in the presence of a magnetic flux. There, we can observe 
topological edge states appearing in the minigaps opened by the magnetic 
flux that propagate in a fixed direction depending on their energy and the 
edge where they localize. We can also find Shockley-like edge states that 
can propagate in both directions along each edge. When reducing the 
effective hopping, these states are pulled further out of the bulk 
minibands, making them interfere less with the topological edge states. 
After this analysis we conclude that, depending on their energy, a doublon 
can propagate chirally or not along the edges of a lattice threaded by a 
magnetic flux.

\clearpage
\section{Doublon decay in dissipative systems} \label{sec:doublondecay}

The high controllability and isolation achieved in cold atom experiments 
make them a great platform for observing doublons. But can doublons be 
observed in other kind of systems? Nowadays, solid-state devices such as 
quantum dot arrays are being investigated as platforms for quantum 
simulation. However, phonons, nuclear spins, and fluctuating charges and 
currents make for a much noisier environment in these setups as compared 
with others~\cite{forster2014}. In this section we investigate how 
the coupling to the environment may affect the stability of doublons in QD 
arrays and give an estimate of their lifetime in current devices.

\begin{figure}[!htb]
  \centering
  \includegraphics{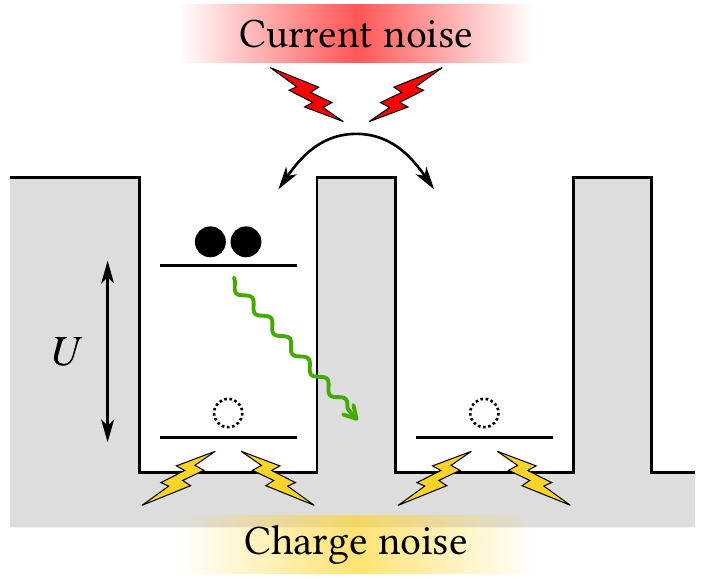}
  \caption{Schematic picture of the system under consideration. A QD array
  modelled as a tight-binding 1D lattice is occupied by two electrons. An 
  initial state with a doubly occupied site (doublon) may decay 
  dissipatively into a single-occupancy state with lower energy. The 
  released energy is of the order of the on-site interaction $U$ and will 
  be absorbed by the heat baths representing environmental charge and 
  current noise.}
  \label{fig:scheme_noise}
\end{figure}

We will analyze the case of a 1D array of $N$ quantum dots, see 
Fig.~\ref{fig:scheme_noise}. Electrons trapped in the QD array are modelled 
by the Hubbard Hamiltonian with an homogeneous hopping amplitude $J$ and 
interaction strength $U$. For the environment, we assume the chain is 
coupled to several independent baths of harmonic oscillators. The system 
and the environment can be modelled as a whole by the Hamiltonian 
$H=H_S+H_B+H_I$, with
\begin{align}
  H_S & = -J\summ_{j,\,\sigma}\left(c^\dagger_{j+1\sigma}c_{j\sigma}
  +\HC\right) + U\summ_jn_{j\up}n_{j\dn}\,,\\
  H_B & = \summ_{j,\,n}\omega_na^\dagger_{nj}a_{nj}\,,\\
  H_I & = \summ_j X_jB_j\,,\quad
  B_j=\summ_n g_n(a^\dagger_{nj}+a_{nj})\,.
\end{align}
Here, $B_j$ is the collective coordinate of the $j$th bath that couples to 
the system operator $X_j$, which will be specified below. Moreover, we 
assume that all baths are equal and statistically independent. 

In the Markovian regime, the time evolution of the system's density matrix 
$\rho$, can be suitably described by a Bloch-Redfield master equation of the
form~\cite{redfield1957,petruccione2002open} (see 
appendix~\ref{app:BlochRedfield})
\begin{align}
  \dot{\rho} & = -i[H_S,\rho]-\summ_j [X_j,[Q_j,\rho]]- 
  \summ_j [X_j,\{R_j,\rho\}] \nonumber\\
  & \equiv -i[H_S,\rho] + \mathcal{L}[\rho] \,,
  \label{eq:BlochRedfield}
\end{align}
with the anticommutator $\{A,B\}=AB+BA$ and
\begin{align}
  Q_j & = \frac{1}{\pi}\int_0^\infty \diff{\tau}\int_0^\infty \diff{\omega}
  \mathcal{S}(\omega)\tilde X_j(-\tau) \cos \omega\tau \,, \label{eq:Q}\\
  R_j &= \frac{-i}{\pi}\int_0^\infty \diff{\tau}\int_0^\infty \diff{\omega} 
  \mathcal{J}(\omega)\tilde X_j(-\tau) \sin \omega\tau \,. \label{eq:R}
\end{align}
The tilde denotes the interaction picture, 
$\tilde X_j(-\tau)={e^{-i H_S\tau}X_je^{i H_S\tau}}$, while the spectral
density of the baths is 
$\mathcal{J}(\omega)=\pi\sum_n |g_n|^2 \delta(\omega-\omega_n)$, and 
 ${\mathcal{S}(\omega)=\mathcal{J}(\omega)\coth(\beta\omega/2)}$ is the
Fourier transformed of the symmetrically ordered equilibrium
autocorrelation function $\mean{\{B_j(\tau),B_j(0)\}}/2$.
$\mathcal{J}(\omega)$ and $\mathcal{S}(\omega)$ are independent of the bath
subindex $j$ since all baths are identical. We will assume an ohmic spectral
density ${\mathcal{J}(\omega)=\pi\alpha\omega/2}$, where the dimensionless 
parameter $\alpha$ characterizes the dissipation strength.

\subsection{Charge noise \label{sec:charge}}

Fluctuations of the background charges in the substrate essentially act upon
the charge distribution of the chain. We model it by coupling the occupation
of each site to a heat bath, such that 
\begin{equation}
  H_I=\summ_{j,\,\sigma}n_{j\sigma}B_j\,,\quad
  X_j=n_{j\up}+n_{j\dn}\,.
\end{equation}
This fully specifies the master equation~\eqref{eq:BlochRedfield}.

\begin{figure}[h]
  \centering\includegraphics{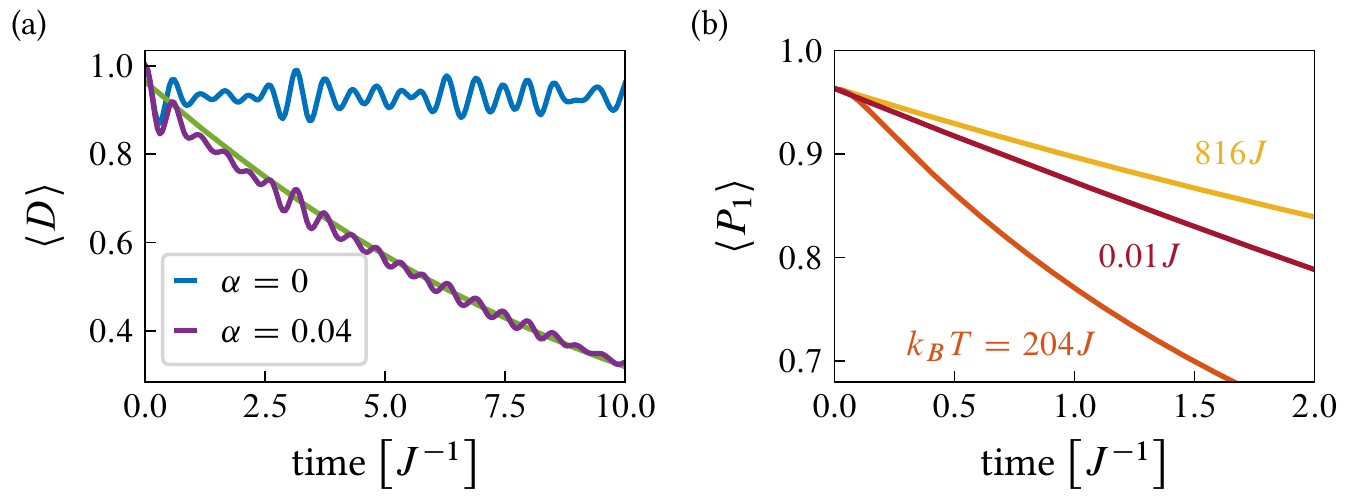}
  \caption{Time evolution of the double occupancy in a system with charge
  noise. The initial state consists of a doublon localized in a 
  particular site of a chain with periodic boundary conditions.  
  Parameters: $N=5$, $U=10J$ and $\alpha=0.04$. (a) Comparison between 
  free dynamics ($\alpha=0$) and dissipative dynamics ($\alpha\neq 0$). 
  Temperature is set to $k_B T=0.01J$. The green line corresponds to the 
  occupancy of the high-energy subspace for the case with $\alpha\neq 0$ 
  and illustrates the bound given in Eq.~\eqref{eq:bound}. (b) Decay of the 
  high-energy subspace occupancy for different temperatures ranging from  
  $0.01J$ to $1000J$. The slope of the curves at time $t=0$ is the same 
  in all cases and coincides with the value given by 
  Eq.~\eqref{eq:avergamma}.} 
  \label{fig:dynamics_chargenoise} 
\end{figure}

To get a qualitative understanding of the decay dynamics of a doublon, let 
us start by discussing the time evolution of the total double occupancy 
\begin{equation}
   D\equiv\summ_j n_{j\up}n_{j\dn}=\summ_j d^\dagger_jd_j\,,
\end{equation}
for an initial doublon state, shown in 
Fig.~\ref{fig:dynamics_chargenoise}(a). For 
$\alpha=0$, i.e., in the absence of dissipation, the two electrons will 
essentially remain together throughout time evolution. However, since the 
doublon states are not eigenstates of the system Hamiltonian, we observe 
some slight oscillations of the double occupancy. Still the time average of 
this quantity stays close to unity. On the contrary, if the system is 
coupled to a bath, doublons will be able to split, releasing energy into the
environment. Then the density operator eventually becomes the thermal state 
$\rho_\infty\propto e^{-\beta H_S}$. Depending on the temperature and the 
interaction strength, the corresponding asymptotic doublon occupancy 
$\mean{D}_\infty$ may still assume an appreciable value.

\begin{figure}[tb]
  \centering\includegraphics{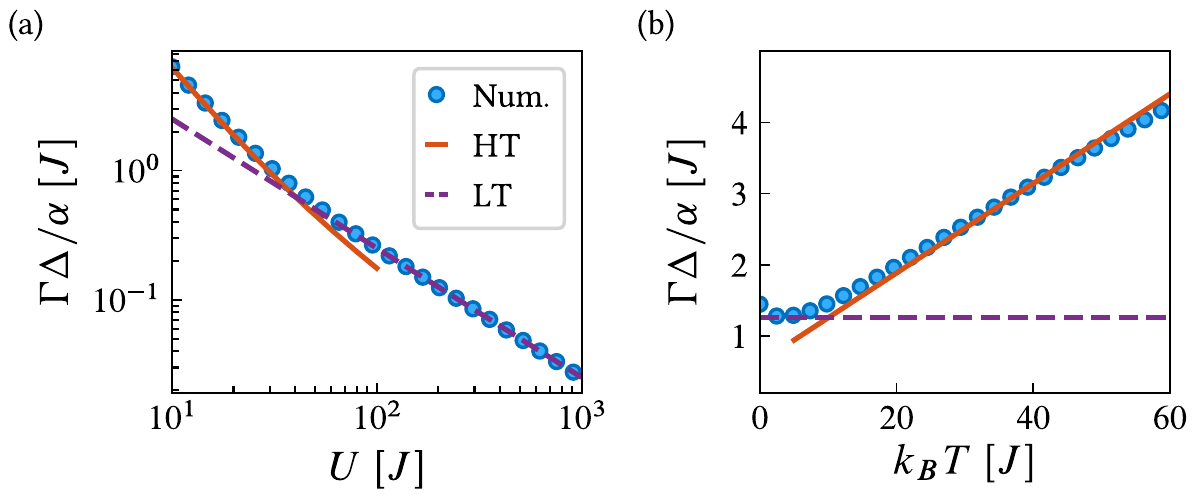} 
    \caption{Comparison between the numerically computed decay rate and the 
        analytic formulas \eqref{eq:GOHT} and \eqref{eq:GOLT} for a chain
        with $N=5$ sites and periodic boundary conditions in the case of 
        charge noise.  The dissipation strength is $\alpha=0.02$.
        (a) Dependence on the interaction strength for a fixed temperature 
        $k_B T=20J$. (b) Dependence on the temperature for a fixed 
        interaction strength $U=20J$.}
    \label{fig:analyticsOccup}
\end{figure}

To gain a quantitative insight, we decompose our master equation
\eqref{eq:BlochRedfield} into the system eigenbasis and obtain a form
convenient for numerical treatment (appendix~\ref{app:BlochRedfield}). 
A typical time evolution of the 
total double occupancy exhibits an almost monoexponential decay, such 
that the doublon life time $T_1$ can be defined as the time when
\begin{equation}
  \frac{\mean{D}_{T_1}-\mean{D}_\infty}{1-\mean{D}_\infty}=\frac{1}{e} \,.
  \label{eq:numericgamma}
\end{equation}
The corresponding decay rate $\Gamma=1/T_1$ is shown in 
Fig.~\ref{fig:analyticsOccup} as a function of the temperature and 
interaction strength for a fixed small $\alpha$. 

An analytical estimate for the decay rates can often be gained from the
behavior at the initial time $t=0$, i.e.\ from $\dot\rho_0$. In the present 
case, however, the calculation is hindered by the fast initial oscillations 
witnessed in Fig.~\ref{fig:dynamics_chargenoise}(a). To circumvent this 
problem, we focus instead on the occupancy of the high-energy subspace, 
$\langle P_1\rangle$, with $P_1$ being the projector onto the high energy 
subspace, shown in Fig.~\ref{fig:dynamics_chargenoise}(b). Using the 
Schrieffer-Wolff transformation derived in 
section~\ref{sec:whataredoublons} we can express it in terms of the 
projector onto the doublon subspace, $P_D$, as
\begin{equation}
  P_1 = P_D + \frac{1}{U}\left(H^+_J + H^-_J\right) + 
  O\left(U^{-2}\right)\,. \label{eq:projector}
\end{equation}
It turns out that this quantity evolves more smoothly while it decays also 
on the time scale $T_1$. To understand this similarity, notice that
\begin{equation}
  \abs*{\tr\left(P_1\rho\right)-\tr\left(D\rho\right)}
  \leq \sqrt{2}\norm{\rho}\sqrt{N-\tr\left(P_1P_D\right)} 
  \simeq \frac{2\sqrt{2N}J}{U} \,, \label{eq:bound}
\end{equation}
where we have used the Cauchy-Schwarz inequality for the inner product of 
operators, $(A,B)=\tr\left(A^\dagger B\right)$, and the perturbative 
expansion of $P_1$ mentioned before. The reason for its lack of fast 
oscillations is that the projector $P_1$ commutes with the system 
Hamiltonian, so that its expectation value is determined solely by 
dissipation.

Following our hypothesis of a monoexponential decay, we expect
\begin{equation}
 \mean{P_1}\simeq {\Delta e^{-\Gamma t} + \mean{P_1}_\infty}\,,
\end{equation}
therefore
\begin{equation}
  \Gamma\simeq - \frac{1}{\Delta}
  \left. \frac{d\mean{P_1}}{dt}\right|_{t=0}=
  -\frac{\tr\left(P_1\mathcal{L}[\rho_0]\right)}
  {\mean{P_1}_0-\mean{P_1}_\infty} \,.  \label{eq:gamma}
\end{equation}
This expression still depends slightly on the specific choice of the
initial doublon state. To obtain a more global picture, we consider an 
average over all doublon states, which can be performed 
analytically~\cite{storcz2005} (see appendix~\ref{app:Average}). 
Substituting the expression for the Liouvillian in Eq.~\eqref{eq:gamma}, we 
find the average decay rate
\begin{equation}
  \overbar{\Gamma}=\frac{1}{N\Delta}\summ_j 
  \tr\left(P_D[Q_j,[X_j,P_1]]\right)
  -\tr\left(P_D\{R_j,[X_j,P_1]\}\right) \,.  \label{eq:avergamma}
\end{equation}
For a further simplification, we have to evaluate the expressions
\eqref{eq:Q} and \eqref{eq:R} which is possible by approximating the
interaction picture coupling operator as $\tilde X_j(-\tau)\simeq X_j
-i\tau[H_S,X_j]$.  This is justified as long as the decay of the
environmental excitations is much faster than the typical system evolution,
i.e., in the high-temperature regime (HT). Inserting our approximation for
$\tilde X_j$ and neglecting the imaginary part of the integrals, we arrive at
\begin{gather} 
  Q_j \simeq \frac{1}{2}\lim_{\omega\rightarrow 0^+} 
  \mathcal{S}(\omega) X_j = \frac{\pi}{2}\alpha k_B T X_j\,, \\ 
  R_j \simeq - \frac{1}{2}\lim_{\omega\rightarrow 0^+} 
  \mathcal{J}'(\omega)[H_S,X_j]=\frac{\pi}{4}\alpha[H_S,X_j] \,.  
\end{gather} 
With these expressions, Eq.~\eqref{eq:avergamma} results in a 
temperature independent decay rate. Notice that any temperature dependence 
stems from the $Q_j$ in the first term of Eq.~\eqref{eq:avergamma}, 
which vanishes in the present case. While this observation agrees with the 
numerical findings in Fig.~\ref{fig:dynamics_chargenoise}(b) for very 
short times, it does not reflect the temperature dependent decay of 
$\langle P_1\rangle$ at the more relevant intermediate stage.

This particular behavior hints at the mechanism of the bath-induced doublon
decay. Let us remark that for the coupling to charge noise, $X_j$ commutes 
with $D$. Therefore, the initial state is robust against the influence of 
the bath. Only after mixing with the single-occupancy states due to the 
coherent dynamics, the system is no longer in an eigenstate of $X_j$, 
such that decoherence and dissipation become active. Thus, it is the 
combined action of the system's unitary evolution and the effect of the 
environment which leads to the doublon decay. An improved estimate of the 
decay rate, can be calculated by averaging the transition rate of states 
from the high-energy subspace to the low-energy subspace. Let us first 
focus on regime $k_BT\gtrsim U$ in which we can evaluate the operators $Q_j$
in the high-temperature limit. Then the average rate can be computed using 
expression \eqref{eq:avergamma} replacing $P_D$ by $P_1$. With the 
perturbative expansion of $P_1$ in Eq.~\eqref{eq:projector} we obtain to
leading order in $J/U$ the averaged rate
\begin{equation} 
  \overbar{\Gamma}_\mathrm{HT} \simeq 
  \frac{4\pi\alpha J^2}{U^2 \Delta}\left(2k_B T + U \right) \,,
  \label{eq:GOHT}
\end{equation}
valid for periodic boundary conditions. For open boundary conditions, the
rate acquires an additional factor $(N-1)/N$. Notice that we have
neglected back transitions via thermal excitations from singly occupied
states to doublon states. We will see that this leads to some deviations 
when the temperature becomes extremely large. Nevertheless, we refer to this
case as the high-temperature limit.

In the opposite limit, for temperatures ${k_B T < U}$, the decay rate 
saturates at a constant value. To evaluate $\overbar{\Gamma}$ in this limit,
it would be necessary to find an expression for $\tilde{X}_j(-\tau)$ dealing
properly with the $\tau$-dependence for evaluating the noise kernel, a
formidable task that may lead to rather involved expressions. However, 
one can make some progress by considering the transition of one initial 
doublon to one particular single-occupancy state. This corresponds to 
approximating our two-particle lattice model by a dissipative two-level 
system for which the decay rates in the Ohmic case can be taken from the 
literature~\cite{weiss1989,makhlin2001} (see appendix~\ref{app:TwoLevel}). 
Relating $J$ to the tunnel matrix element of the two-level system and $U$ to
the detuning, we obtain the temperature-independent expression
\begin{equation}
  \overbar{\Gamma}_\text{LT}\simeq\frac{8\pi\alpha J^2}{U\Delta}\,,
  \label{eq:GOLT}
\end{equation}
which formally corresponds to Eq.~\eqref{eq:GOHT} with the temperature set
to $k_B T=U/2$.

\subsection{Current noise}

Fluctuating background currents mainly couple to the tunnel matrix elements
of the system.  Then the system-bath interaction is given by
\begin{gather}
  H_I = \summ_{j,\,\sigma} \left(c^\dagger_{j+1\sigma}c_{j\sigma} 
  + c^\dagger_{j\sigma} c_{j+1\sigma}\right) B_j\,,\\
  X_j=c^\dagger_{j+1\sigma}c_{j\sigma} 
  + c^\dagger_{j\sigma} c_{j+1\sigma}\,.
\end{gather}
Depending on the boundary conditions, the sum may include the term 
connecting the first and last QD of the array. The main difference with 
respect to the case of charge noise is that now $H_I$ does not commute with 
the projector onto the doublon subspace and, thus, generally 
${\tr\left(D\mathcal{L}[\rho_0]\right)\neq 0}$. This allows doublons to 
decay without having to mix with single-occupancy states. Therefore, for the
same value of the dissipation strength $\alpha$, the decay may be much 
faster. 

\begin{figure}
  \centering\includegraphics{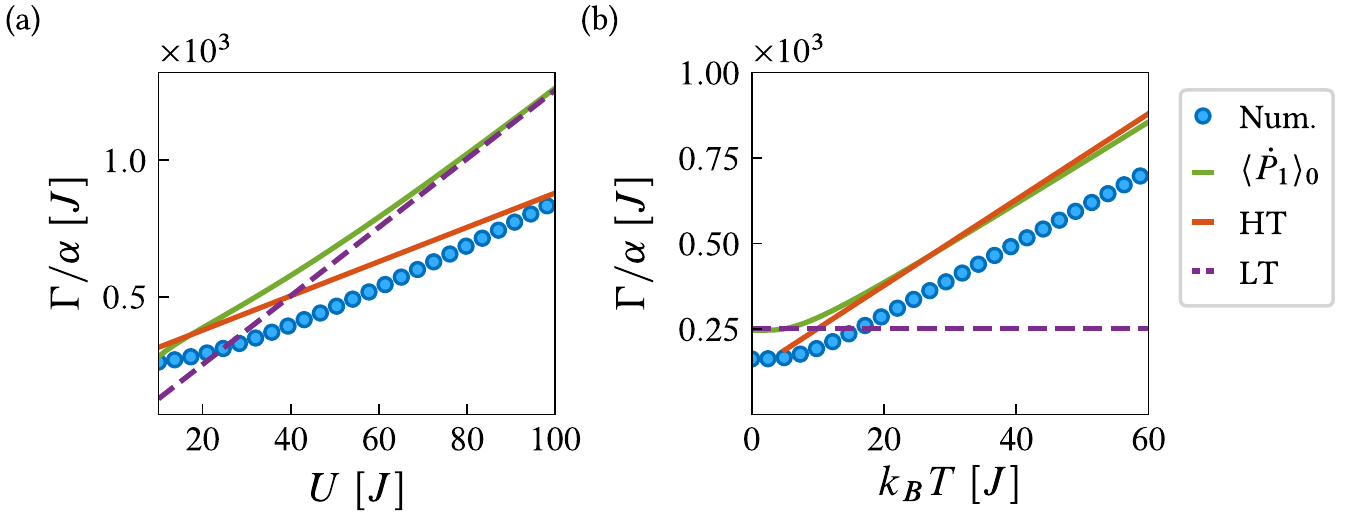}
  \caption{Numerically obtained decay rate in comparison with the
    approximations~\eqref{eq:decay_hopp}, \eqref{eq:GTHT} and 
    \eqref{eq:GTLT} for a chain with $N=5$ sites and periodic boundary 
    conditions in the case of current noise with strength $\alpha=0.02$.  
    The results are plotted as a function of (a) the interaction and the 
    temperature $k_B T=20J$ and (b) for a fixed interaction $U=20J$ as a 
    function of the temperature.}
  \label{fig:decay_current}
\end{figure}

As in the last section, we proceed by calculating analytical estimates for
the decay rates. However, the time evolution is no longer monoexponential.
In this case, we estimate the rate from the slope of the occupancy 
$\langle P_1\rangle$ at initial time,
\begin{equation}
    \Gamma \simeq - \left. \frac{d\mean{P_1}}{dt} \right|_{t=0} =
        - \tr\left(P_1\mathcal{L}[\rho_0]\right) \ .
    \label{eq:decay_hopp}
\end{equation}
We again perform the average over all doublon states for $\rho_0$ in the 
limits of high and low temperatures. For periodic boundary conditions, we 
obtain to lowest order in $J/U$ the high and low temperature rates 
\begin{align}
    \overbar{\Gamma}_\mathrm{HT} & = 2\pi\alpha \left(2k_B T+U \right) \,,
    \label{eq:GTHT}  \\
    \overbar{\Gamma}_\text{LT} & = 4\pi\alpha U \,,
    \label{eq:GTLT}
\end{align}
respectively, while open boundary conditions lead to the same expressions 
but with a correction factor $(N-1)/N$.
In Fig.~\ref{fig:decay_current}, we compare these results with the
numerically evaluated ones as a function of the interaction and the 
temperature. Both show that the analytical approach correctly predicts the 
(almost) linear behavior at large values of $U$ and $k_BT$, as well as the 
saturation for small values. However, the approximation slightly 
overestimates the influence of the bath.

While the rates reflect the decay at short times, it is worthwhile to
comment on the long time behavior under the influence of current
noise. As it turns out, the steady state is not unique. The reason for 
this is the existence of a doublon state
${\ket{\Phi}=\frac{1}{\sqrt{N}}\sum_{j=1}^N (-1)^j
c^\dagger_{j\up}c^\dagger_{j\dn}\ket{0}}$ which is an
eigenstate of $H_S$ without any admixture of single-occupancy states.
Since $X_j\ket{\Phi}=0$ for all $j$, current noise may affect the phase 
of ${\ket{\Phi}}$, but cannot induce its dissipative decay. For a closed 
chain with an odd number of sites, by contrast, the alternating phase of 
the coefficients of $\ket{\Phi}$ is incompatible with periodic boundary 
conditions, unless a flux threads the ring. As a consequence,
the state of the chain eventually becomes the thermal state 
$\rho_\infty\propto e^{-\beta H_S}$. The difference is manifest in the 
final value of the doublon occupancy at low temperatures. For closed 
chains with an odd number of sites, it will fully decay, while in the 
other cases, the population of $\ket{\Phi}$ will survive.

\subsection{Experimental implications}

A current experimental trend is the fabrication of larger QD 
arrays~\cite{puddy2015,zajac2016}, which triggered our question on the
feasibility of doublon experiments in solid-state systems. While the size
of these arrays would be sufficient for this purpose, their dissipative
parameters are not yet fully known. For an estimate we therefore consider
the values for GaAs/InGaAs quantum dots which have been determined recently
via Landau-Zener interference~\cite{forster2014}. Notice, that for the 
strength of the current noise, only an upper bound has been reported. We 
nevertheless use this value, but keep in mind that it leads to a 
conservative estimate. In contrast to the former sections, we now compute 
the decay for the simultaneous action of charge noise and current noise.

\begin{figure}
  \centering\includegraphics{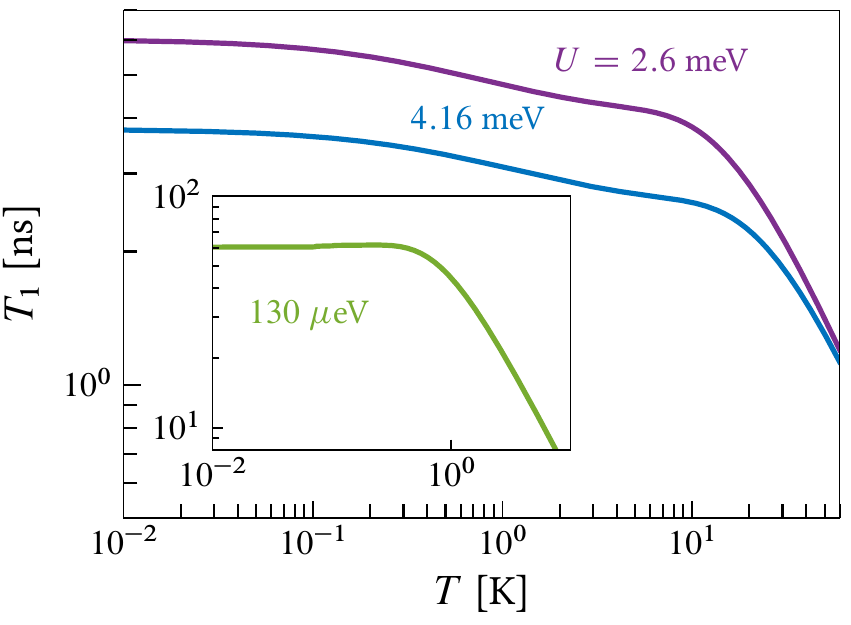}
  \caption{Doublon life time as a function of the temperature for
  different interaction strengths. The parameters are: 
  $\alpha_\mathrm{charge}=3\times 10^{-4}$,
  $\alpha_\mathrm{current}=5\times 10^{-6}$, and
  $J=13\,\mu$eV. The $T_1$ time has been measured for a doublon starting 
  in the middle of a chain with $N=5$ and open boundary conditions. 
  Inset: $T_1$ time for the optimized value of the interaction, 
  $U =10J =130\,\mu$eV and a current noise with 
  $\alpha_\mathrm{current}=2\times10^{-6}$.}
  \label{fig:lifetime_experiment}
\end{figure}

Figure~\ref{fig:lifetime_experiment} shows the $T_1$ times for two different
interaction strengths. It reveals that for low temperatures 
$T\lesssim J/k_BT$, the life time is essentially constant, while for larger 
temperatures, it decreases moderately until $k_BT$ comes close to the 
interaction $U$. For higher temperatures, $\Gamma$ starts to grow linearly. 
On a quantitative level, we expect life times of the order $T_1\sim 5$\,ns 
already for moderately low temperatures $T\lesssim 100$\,mK. Since we 
employed the value of the upper bound for the current noise, the life time 
might be even larger.

Considering the estimates for the decay rates at low temperatures, 
Eqs.~\eqref{eq:GOLT} and \eqref{eq:GTLT}, separately,
we conclude that for smaller values of $U$, current noise becomes
less important, while the impact of charge noise grows.  Therefore, a
strategy for reaching larger $T_1$ times is to design QD arrays with smaller
on-site interaction, such that the ratio $U/J$ becomes more favorable. The 
largest $T_1$ is expected in the case in which both low-temperature decay 
rates are equal, $\overbar{\Gamma}_\mathrm{LT}^\mathrm{charge}=
\overbar{\Gamma}_\mathrm{LT}^\mathrm{current}$, which for the present 
experimental parameters is found at $U\sim 10J$. 
%(while our data is for $U\sim 100 J$). 
This implies that in an optimized device, the doublon life times could be 
larger by one order of magnitude to reach values of $T_1\sim 50$\,ns, which 
is corroborated by the data in the inset of 
Fig.~\ref{fig:lifetime_experiment}.

\section{Summary}

For understanding the dynamics of doublons, we have derived an effective 
single-particle Hamiltonian taking into account also the effect of a 
periodic driving on the lattice. It contains two terms: one corresponding 
to an effective doublon hopping renormalized by the driving, and another 
one corresponding to an effective on-site chemical potential, with the 
peculiarity of being proportional to the number of neighbors of each 
site. Importantly, in the regime where the Hubbard interaction is larger 
than the frequency of the driving and these are both larger than any 
other energy scale of the system, the driving allows to tune the doublon 
hopping independently of the effective chemical potential. This produces 
several interesting phenomena regarding the doublons' motion:
\begin{enumerate}
  \item In any finite lattice doublons experience an effective chemical 
    potential difference between the sites on the edges and the rest of 
    the sites. This allows the generation of Shockley-like edge states 
    by reducing the doublon hopping with the driving.
  \item In the SSH-Hubbard model, the chemical potential difference 
    between the ending sites of the chain and the rest of the sites 
    causes the disappearance of the topological edge states for doublons. 
    This chemical potential difference breaks the chiral symmetry of the 
    SSH model, which is essential for having non-trivial topology. On the 
    other hand, for 2D lattices threaded by a magnetic flux, no symmetry 
    is required for having non-trivial topology, thus, they do support 
    topological edge states for doublons, which may coexist with 
    Shockley-like edge states induced by the driving.
  \item Both topological and Shockley-like edge states allow for the direct 
    long-range transfer of doublons between distant sites in the edges of 
    a finite lattice.
  \item In lattices with sites with different coordination numbers it is 
    possible to confine the doublon's motion to a single sublattice at the
    expense of slowing it down by reducing the doublon hopping with the 
    driving.
\end{enumerate}
We have also studied the stability of doublons in quantum dot arrays in the 
presence of charge noise and current noise. While the dependence on 
temperature of the doublon's lifetime $T_1$ is similar for both types of 
noise, the dependence with the interaction strength $U$ is very different. 
For charge noise $T_1\sim U$, whereas for current noise $T_1\sim U^{-1}$, in
the low temperature regime ($k_BT<U$). In current devices we predict a 
doublon lifetime of the order of 10 ns, although it can be improved up to 
one order of magnitude in devices specifically designed to that end.

\begin{subappendices}
  \section{Effective Hamiltonian for doublons \label{app:EffectiveHamiltonian}}
  We start from a Fermi-Hubbard model with an ac field that couples to the 
particle density and a magnetic flux that introduces complex phases in the 
hoppings. The Hamiltonian of the system is $H(t)=H_J+H_U+H_{AC}(t)$, with
\begin{equation}
  H_{AC}(t)=\summ_j V_j(t)(n_{j,\up}+n_{j,\dn})\,.
\end{equation}
For a time-periodic Hamiltonian, $H(t+T)=H(t)$, with $T=2\pi/\omega$, 
Floquet's theorem allows us to write the time-evolution operator as
\begin{equation}
  U(t_2,t_1)=e^{-iK(t_2)}e^{-iH_\mathrm{eff}(t_2-t_1)}e^{iK(t_1)}\,,
\end{equation}
where $H_\mathrm{eff}$ is a time-independent, effective Hamiltonian, and 
$K(t)$ is a $T$-periodic Hermitian operator. $H_\mathrm{eff}$ governs
the long-term dynamics, whereas $e^{-iK(t)}$, also known as the 
\emph{micromotion-operator}, accounts for the fast dynamics occurring
within a period. Following several perturbative 
methods~\cite{eckardt2015,mikami2016}, it is possible to find expressions 
for these operators as power series in $1/\omega$,
\begin{equation}
	\Heff=\summ\limits_{n=0}^\infty \frac{\Heff^{(n)}}{\omega^n} \,, \quad 
  K(t)=\summ\limits_{n=0}^\infty \frac{K^{(n)}(t)}{\omega^n} \,.
\end{equation}
These are known in the literature as high-frequency expansions (HFE).
The different terms in these expansions have a progressively more 
complicated dependence on the Fourier components of the original 
Hamiltonian, 
\begin{equation}
  H_q=\frac{1}{T}\int_0^T \diff{t}H(t)e^{i\omega q t}\,.
\end{equation}
The first three of them are:
\begin{align}
  \Heff^{(0)} & =H_0 \,,\\
  \Heff^{(1)} & = \summ\limits_{q\neq 0}\frac{H_{-q}H_q}{q} \,,\\
  \Heff^{(2)} & = \summ\limits_{q,\,p\neq 0} \left(\frac{H_{-q}H_{q-p}H_p}{q p} 
  -\frac{H_{-q}H_qH_0}{q^2}\right)\,.
\end{align}

Before deriving the effective Hamiltonian, it is convenient to transform 
the original Hamiltonian into the rotating frame with respect to both 
the interaction and the ac field~\cite{bukov2016},
\begin{gather}
  \tilde H(t)=\U^\dagger(t)H(t)\U(t)-i\U^\dagger(t)\partial_t\U(t)
  \,,\\
  \U(t)=\exp\left[-i H_U t - i\int \diff{t} H_{AC}(t)\right]\,.
\end{gather}
Noting that for fermions 
\begin{equation}
  e^{iH_Ut}c^\dagger_{i\sigma}c_{j\sigma}e^{-iH_Ut}=
  \left[1-n_{i\ol\sigma}\left(1-e^{iUt}\right)\right]
  c^\dagger_{i\sigma}c_{j\sigma}
  \left[1-n_{j\ol\sigma}\left(1-e^{-iUt}\right)\right]\,,
\end{equation}
the Hamiltonian in the rotating frame can be written as
\begin{align}
  \tilde H(t)&=-\summ_{i,\,j,\,\sigma}J_{ij}(t) \left(T^0_{ij\sigma} 
  + e^{iUt}T^+_{ij\sigma} + e^{-iUt}T^-_{ij\sigma}\right) \,,
\end{align}
with $J_{ij}(t)=J_{ij}e^{i\mbf A(t)\cdot \mbf d_{ij}}$. Note that this is a
different way of expressing the coupling to an electric field described 
by the vector potential $\mbf A(t)$. In the case of circular polarization, 
$\mbf A(t)=\left(\sin(\omega t),-\cos(\omega t)\right)E/\omega$;
$\mbf d_{ij}=\mbf r_i-\mbf r_j$ is the vector connecting sites $i$ and $j$. 
We have defined: 
\begin{gather}
	T^0_{ij\sigma}=n_{i\ol\sigma}c^\dagger_{i\sigma}c_{j\sigma}n_{j\ol\sigma} 
  +h_{i\ol\sigma}c^\dagger_{i\sigma}c_{j\sigma}h_{j\ol\sigma}
  \,,\\
	T^+_{ij\sigma} =n_{i\ol\sigma}c^\dagger_{i\sigma}c_{j\sigma}h_{j\ol\sigma} 
  \,, \quad T^-_{ij\sigma} =(T^+_{ji\sigma})^\dagger= 
	h_{i\ol\sigma}c^\dagger_{i\sigma}c_{j\sigma}n_{j\ol\sigma} 
  \,.
\end{gather}
The operators $T^0_{ij\sigma}$ involve hopping processes that conserve the 
total double occupancy, while $T^+_{ij\sigma}$ and $T^-_{ij\sigma}$ raise 
and lower the total double occupancy respectively. 

In order to apply the HFE method we need to find a common frequency. We will
consider first the resonant regime, $U=l\omega$, and then, by means of 
analytical continuation, obtain the \textit{strongly-interacting limit} 
($U\gg\omega>J$) and the \textit{ultrahigh-frequency limit} 
($\omega\gg U >J$). The Fourier components of $\tilde H(t)$ are
\begin{equation}
  \tilde H_q=-\summ_{i,\,j,\,\sigma} \left(J_{ij,q}T^0_{ij\sigma}+
  J_{ij,q+l}T^+_{ij\sigma}+J_{ij,q-l}T^-_{ij\sigma} \right)\,,
\end{equation}
with 
\begin{align}
  J_{ij,q} & =\frac{J_{ij}}{T}\int_0^T \diff{t} 
  e^{i\frac{E}{\omega}d^x_{ij}\sin(\omega t)}
  e^{-i\frac{E}{\omega}d^y_{ij}\cos(\omega t)}
  e^{iq\omega t}
  \\
  & = \frac{J_{ij}}{T}\int_0^T \diff{t} 
  \summ_{m,\,n}\bes{m}\left(\tfrac{Ed^x_{ij}}{\omega}\right)
  \bes{-n}\left(\tfrac{Ed^y_{ij}}{\omega}\right)
  i^n e^{i(m+n+q)\omega t}
  \\
  & = J_{ij}\summ_{n}
  \bes{n+q}\left(\tfrac{Ed^x_{ij}}{\omega}\right)
  \bes{n}\left(\tfrac{Ed^y_{ij}}{\omega}\right)
  e^{-in\pi/2}
  \\
  & = J_{ij} e^{-iq\alpha_{ij}}\bes{q}\left(\A\right) \,,
\end{align}
where $\alpha_{ij}=\arg\left(d^x_{ij}+id^y_{ij}\right)$, and $\bes{q}$ 
stands for the Bessel function of first kind of order $q$. To go from the 
first to the second line we have used the Jacobi--Anger expansion (the 
sums run over all positive and negative integers), and we have used Graf's 
addition theorem to derive the last expression. Note that 
$J_{ij,q}=J^*_{ji,-q}$.

Now, the zeroth-order approximation in the HFE is given by:
\begin{equation}
  \tilde H_\mathrm{eff}^{(0)} =- J\summ_{i,\,j,\,\sigma} J_{ij}\left[
	\bes{0}\left(\A\right) T^0_{ij\sigma} 
  % \right.\\ \left.
  +e^{-il\alpha_{ij}}\bes{l}\left(\A\right)T^+_{ij\sigma}
  +e^{il\alpha_{ij}}\bes{-l}\left(\A\right)T^-_{ij\sigma}\right]\,.
\end{equation}
In contrast to the undriven case, the total double occupancy is not 
necessarily an approximate conserved quantity in the regime $U\gg J$. There 
are terms proportional to 
$\bes{l}\left(E\abs{\mbf d_{ij}}/\omega\right)$ that correspond to 
the formation and dissociation of doublons assisted by the ac field
($l$-photon resonance). 
However, for low driving amplitudes ($E\abs{\mbf d_{ij}}/\omega<l$) 
the probability for these processes to occur is very small and we 
can neglect them. It is in this low amplitude regime where it makes 
sense to consider an effective Hamiltonian for doublons. We neglect the 
terms that go with $T^0_{ij\sigma}$ because they act non-trivially only on 
states with some single-occupancy. 

In the next order of the HFE, there are more terms that do not conserve the 
total double occupancy, which we neglect, and from those which do conserve 
it, we only keep the ones that act non-trivially on the doublon's subspace:
\begin{multline}
  \frac{\tilde H_\mathrm{eff}^{(1)}}{\omega} 
  \reseq \summ_{i,\,j,\,\sigma}\left[ \summ_{q\neq 0} 
  \frac{\bes{-q+l}^2\left(\A\right)\abs{J_{ij}}^2
  T^+_{ij\sigma}T^-_{ji\sigma}}{q\omega}\right. \\ 
  + \left.\summ_{q\neq 0}\frac{\bes{-q+l}\left(\A\right)
  \bes{q-l}\left(\A\right)J_{ij}^2
  T^+_{ij\sigma}T^-_{ij\ol\sigma}}{q\omega}\right] \,.
\end{multline}
Here, the first term is equal to
\begin{equation}
  \abs{J_{ij}}^2\summ_{p\neq -l} \frac{\bes{p}^2\left(\A\right)}
  {(l-p)\omega}T^+_{ij\sigma}T^-_{ji\sigma}= 
  \frac{\abs{J_{ij}}^2}{U}\summ_{p\neq -l} 
  \frac{\bes{p}^2\left(\A\right)}{1-p\omega/U}
	\left(n_{i\ol\sigma}n_{i\sigma}
  -n_{i\ol\sigma}n_{i\sigma}n_{j\sigma}n_{j\ol\sigma}\right) \,,
\end{equation}
and the second term is equal to
\begin{multline}
  J_{ij}^2\summ_{p\neq -l} 
  \frac{\bes{p}\left(\A\right)\bes{-p}\left(\A\right)}{(l-p)\omega}
  T^+_{ij\sigma}T^-_{ij\ol\sigma}=\\
  \frac{J_{ij}^2}{U}\summ_{p\neq -l} 
  \frac{\bes{p}\left(\A\right)\bes{-p}\left(\A\right)}{1-p\omega/U}
	c^\dagger_{i\sigma}c^{\dagger}_{i\ol\sigma}c_{j\ol\sigma} c_{j\sigma} \,.
\end{multline}
In the limit $U\gg\omega >J$, $p\omega/U\ll 1$ and we can approximate 
all the denominators in the above expressions as 1. Note that for fixed 
argument $\alpha$, the Bessel functions $\bes{q}(\alpha)$ decay for 
increasing order $\abs{q}$. Also, when analytically continuing the formulas 
for values of $U$ other than multiples of $\omega$, we may safely neglect 
the restriction $p\neq -l$. Finally, using the identities 
$\sum_q\bes{q}^2(\alpha)=1$ and 
$\sum_q\bes{q}(\alpha)\bes{k-q}(\beta)=\bes{k}(\alpha+\beta)$,
we arrive at
\begin{gather}
  \Heff^{U\gg\omega}=\summ_{i,\,j}\Jeff_{ij}d^\dagger_id_j 
  + \summ_i \mu_i d^\dagger_id_i
  -\summ_{i,\,j} \frac{2\abs{J_{ij}}^2}{U}d^\dagger_id_id^\dagger_jd_j  
  \,, \label{effective}\\
  \Jeff_{ij}\equiv\frac{2J_{ij}^2}{U}\bes{0}
  \left(\tfrac{2E\abs{\mbf d_{ij}}}{\omega}\right)\,, \quad
  \mu_i\equiv\summ_j \frac{2\abs{J_{ij}}^2}{U} \ . 
\end{gather}

For completeness we give also the result in the other limit: 
$\omega\gg U >J$. Now $p\omega/U$ is very large and all the terms in the 
sums are very small except those for $p=0$. The effective Hamiltonian in 
this case would be:
\begin{gather}
  \Heff^{\omega\gg U}=\summ_{i,\,j}\Jeff_{ij}d^\dagger_id_j 
  + \summ_i \mu_i d^\dagger_id_i 
  -\summ_{i,\,j} \abs{\Jeff_{ij}}d^\dagger_id_id^\dagger_jd_j\,,\\
  \Jeff_{ij}\equiv\frac{2J_{ij}^2}{U}\bes{0}^2\left(\A\right)\,, \quad 
  \mu_i\equiv \summ_j\abs{\Jeff_{ij}} \,.
\end{gather}
It is worth mentioning that these results could also be obtained by 
applying the HFE sequentially, integrating first the fast varying terms 
corresponding to the leading energy scale in the system. We also note that 
higher order corrections will include complex next-nearest-neighbor hoppings
that break the time-reversal symmetry, even in systems without any external 
magnetic flux.
 
  \section{Bloch-Redfield master equation \label{app:BlochRedfield}}
  Expanding the integrand of Eq.~\eqref{eq:BornMarkovSpicture} we get
\begin{equation}
  \begin{split}
    \tr_B[H_I,[\tilde{H_I}(-\tau),\rho_S\otimes\rho_B]] & =  
    \summ_{j,\,k}\big[C_{jk}(\tau)X_j\tilde X_k(-\tau)\rho_S 
    -C_{kj}(-\tau)X_j\rho_S\tilde X_k(-\tau)
    \\
    &\qquad -C_{jk}(\tau)\tilde X_k(-\tau)\rho_SX_j
    +C_{kj}(-\tau)\rho_S\tilde X_k(-\tau)X_j\big]\,.
  \end{split}
  \label{eq:integrandBM}
\end{equation}
Here, we have defined 
$\mean{\tilde B_j(t)\tilde B_k(t')}\equiv C_{jk}(t-t')$. If bath operators 
are independent from each other $C_{jk}(\tau)=C_j(\tau)\delta_{jk}$. Then, 
splitting the correlation functions into symmetric and antisymmetric parts
$C_j(\tau)=S_j(\tau)+A_j(\tau)$, 
\begin{equation}
  S_j(\tau)=\frac{C_j(\tau)+C_j(-\tau)}{2}\,, \
  A_j(\tau)=\frac{C_j(\tau)-C_j(-\tau)}{2}\,,
\end{equation}
Eq.~\eqref{eq:integrandBM} can be rewritten as
\begin{multline}
  \tr_B[H_I,[\tilde{H_I}(-\tau),\rho_S\otimes\rho_B]]=\\
  \summ_jS_j(\tau)[X_j,[\tilde X_j(-\tau),\rho_S]]
  +\summ_jA_j(\tau)[X_j,\{\tilde X_j(-\tau),\rho_S\}]\,.
\end{multline}
Putting everything together we get
\begin{equation}
  \dot\rho=-i[H_S,\rho]-\summ_j[X_j,[Q_j,\rho]]
  -\summ_j[X_j,\{R_j,\rho\}]\,,
\end{equation}
where we have dropped the subindex ``$S$'' of the system's reduced density 
matrix, and we have defined
\begin{equation}
  Q_j\equiv\int_0^\infty \diff{\tau} S_j(\tau)\tilde X_j(-\tau)\,,\
  R_j\equiv\int_0^\infty \diff{\tau} A_j(\tau)\tilde X_j(-\tau)\,.
\end{equation}

So far the derivation remained rather general, we will now particularize 
to the case where the environment is composed of several independent baths 
of harmonic oscillators $H_B=\sum_{j,n}\omega_n a^\dagger_{nj}a_{nj}$, that 
couple to the system via $B_j=\sum_ng_na_{nj}+\mathrm{H.c}$.  We consider
that these baths are identical and statistically independent. They are all 
characterized by the same spectral density
${\mathcal{J}(\omega)\equiv\pi\sum_n |g_n|^2 \delta(\omega-\omega_n)}$, 
which describes how the system couples to the different modes of a bath.
Both $S(\tau)$ and $A(\tau)$ can be expressed in terms of this spectral 
density as
\begin{gather}
  C(\tau) =\frac{1}{\pi}\int_0^\infty \diff{\omega} \mathcal{J}(\omega)
   \left\{e^{i\omega\tau}n_B(\omega)+e^{-i\omega\tau}[1+n_B(\omega)]\right\}
   \,, \label{autocorrelation}\\
  S(\tau) =\frac{1}{\pi}\int_0^\infty \diff{\omega} \mathcal{J}(\omega) 
  \coth(\beta\omega/2)\cos\omega\tau \,, \\	
  A(\tau) =\frac{-i}{\pi}\int_0^\infty \diff{\omega} \mathcal{J}(\omega)
  \sin\omega\tau\,.
\end{gather} 
Here, $n_B(\omega)=\left(e^{\beta\omega}-1\right)^{-1}$ is the bosonic 
thermal ocupation number. Thus,
\begin{gather}
  Q_j = \frac{1}{\pi}\int_0^\infty \diff{\tau}\int_0^\infty \diff{\omega} 
  \mathcal{S}(\omega)\tilde X_j(-\tau)\cos\omega\tau \,,\\
  R_j = \frac{-i}{\pi}\int_0^\infty \diff{\tau}\int_0^\infty \diff{\omega} 
  \mathcal{J}(\omega)\tilde X_j(-\tau)\sin\omega\tau \,,
\end{gather}
with $\mathcal{S}(\omega)\equiv\mathcal{J}(\omega)\coth(\beta\omega/2)$.

However, with the aim of doing numerical calculations, it is better to 
work directly with Eq.~\eqref{eq:BornMarkovSpicture} expressed in 
the system eigenbasis, $\{\ket{\phi_\alpha}\}$, fulfilling 
$H_S\ket{\phi_\alpha}=\epsilon_\alpha\ket{\phi_\alpha}$. Introducing the 
identity $\sum_\alpha \ket{\phi_\alpha} \bra{\phi_\alpha}$, noting that 
\begin{equation}
  \bra{\phi_\alpha}\tilde X_j(-\tau)\ket{\phi_\beta}=
  e^{-i(\epsilon_\alpha-\epsilon_\beta)\tau}\bra{\phi_\alpha}X_j 
  \ket{\phi_\beta}\,,
\end{equation}
and using the short-hand notation 
$X\supi{j}_{\alpha\beta}\equiv\bra{\phi_\alpha}X_j\ket{\phi_\beta}$, 
and $\rho_{\alpha\beta}\equiv\bra{\phi_\alpha}\rho\ket{\phi_\beta}$,
we can write:
\begin{align}
  - &\bra{\phi_\alpha}\mathcal{L}[\rho]\ket{\phi_\beta}= 
  \summ_{j,\,\alpha',\,\beta'} \int_0^\infty \diff{\tau}\nonumber\\
  \Big[
    & C(\tau)e^{i(\epsilon_{\alpha'}-\epsilon_{\beta'})\tau} X\supi{j}_{\alpha\beta'}X\supi{j}_{\beta'\alpha'}\rho_{\alpha'\beta} 
     -C(-\tau) e^{-i(\epsilon_{\beta'}-\epsilon_\beta)\tau}X\supi{j}_{\alpha\alpha'}X\supi{j}_{\beta'\beta}\rho_{\alpha'\beta'} 
     \nonumber\\
    &-C(\tau) e^{i(\epsilon_{\alpha'}-\epsilon_\alpha)\tau} X\supi{j}_{\alpha\alpha'}X\supi{j}_{\beta'\beta}\rho_{\alpha'\beta'}
     +C(-\tau) e^{-i(\epsilon_{\beta'}-\epsilon_{\alpha'})\tau} X\supi{j}_{\beta'\alpha'}X\supi{j}_{\alpha'\beta}\rho_{\alpha\beta'}
  \Big] \,. \label{eq:tocho}
\end{align}
% \begin{multline}
%   - \bra{\phi_\alpha}\mathcal{L}[\rho]\ket{\phi_\beta}=  
%   \summ_{j,\,\alpha',\,\beta'} \int_0^\infty \diff{\tau} \left[C(\tau) 
%   e^{i(\epsilon_{\alpha'}-\epsilon_{\beta'})\tau} X\supi{j}_{\alpha\beta'}
%   X\supi{j}_{\beta'\alpha'}\rho_{\alpha'\beta} \right. \\
%   -C(-\tau) e^{-i(\epsilon_{\beta'}-\epsilon_\beta)\tau}
%   X\supi{j}_{\alpha\alpha'}X\supi{j}_{\beta'\beta}\rho_{\alpha'\beta'} 
%   -C(\tau) e^{i(\epsilon_{\alpha'}-\epsilon_\alpha)\tau} 
%   X\supi{j}_{\alpha\alpha'}X\supi{j}_{\beta'\beta}\rho_{\alpha'\beta'}
%   \\\vphantom{\int_0^\infty}
%   \left.+C(-\tau) e^{-i(\epsilon_{\beta'}-\epsilon_{\alpha'})\tau} 
%   X\supi{j}_{\beta'\alpha'}X\supi{j}_{\alpha'\beta}\rho_{\alpha\beta'}
%   \right] \,. \label{eq:tocho}
% \end{multline}
Now, we define
\begin{equation}
  \Gamma(\omega)\equiv\int_0^\infty \diff{\tau} C(\tau)e^{i\omega\tau}=
  \begin{cases}
    \mathcal{J}(\omega)\left[1+n_B(\omega)\right]\,, & \omega> 0 \\
    \mathcal{J}(-\omega)n_B(-\omega)\,, & \omega<0
  \end{cases} \,,
\end{equation}
and $\Gamma_{\alpha\beta}\equiv\Gamma(\epsilon_\alpha-\epsilon_\beta)$.
We have neglected the imaginary part of the integral, i.e., the Lamb-Shift, 
since it only affects the coherent part of the dynamics.
The bath autocorrelation function satisfies $C(-\tau)=C(\tau)^*$, so 
we can express Eq.~\eqref{eq:tocho}, as:
\begin{multline}
  \bra{\phi_\alpha}\mathcal{L}[\rho]\ket{\phi_\beta}=  
  \summ_{j,\,\alpha',\,\beta'} \Bigg[
    (\Gamma^*_{\beta'\beta}+\Gamma_{\alpha'\alpha})
  X\supi{j}_{\alpha\alpha'}X\supi{j}_{\beta'\beta}
  \\
  -\delta_{\beta\beta'}\summ_{\beta''}\Gamma_{\alpha'\beta''}
  X\supi{j}_{\alpha\beta''}X\supi{j}_{\beta''\alpha'}
  -\delta_{\alpha\alpha'}\summ_{\alpha'' }\Gamma^*_{\beta'\alpha''}
  X\supi{j}_{\beta'\alpha''}X\supi{j}_{\alpha''\beta}\Bigg]
\rho_{\alpha'\beta'} \,.
\end{multline}
We have rearranged a bit the indices so that we readily identify the matrix 
form of the Liouvillian superoperator 
$\mathcal{L}$, $\bra{\phi_\alpha}\mathcal{L}
[\rho]\ket{\phi_\beta}=\sum_{\alpha'\beta'}
\mathcal{L}_{\alpha\beta,\alpha'\beta'}\rho_{\alpha'\beta'}$. 
Together with the coherent part of the evolution, we have the following set 
of first-oder differential equations for the matrix elements of the density 
matrix:
\begin{equation}
  \dot{\rho}_{\alpha\beta}=
  -i(\epsilon_\alpha-\epsilon_\beta)\rho_{\alpha\beta} + 
  \summ_{\alpha',\,\beta'}\mathcal{L}_{\alpha\beta,\alpha'\beta'}
  \rho_{\alpha'\beta'} \,.
\end{equation}
% From \eqref{autocorrelation} we see that:
% \begin{equation}
%   \Gamma(\omega)=\begin{cases}
%     \mathcal{J}(\omega)\left[1+n_B(\omega)\right]\,, & \omega> 0\,. \\
%     \mathcal{J}(-\omega)n_B(-\omega)\,, & \omega<0\,.
%   \end{cases} 
%   % = \frac{J(|\omega|)}{2}\left[\mathrm{sign}(\omega)+\coth\left(\frac{\beta|\omega|}{2}\right)\right]
% \end{equation}

  % \newpage
  \section{Average over pure initial states \label{app:Average}}
  As an ensemble of pure states, we consider all normalized linear 
combinations $\ket{\psi}=\sum_{n=1}^N c_n \ket{n}$ of orthonormal 
basis states $\ket{n}$, $n=1,\dots,N$. For the probability 
distribution of the coefficients $c_n$, we request invariance under 
unitary transformations, which leads to
\begin{equation}
  P(c_1,\cdots,c_N)=\frac{(N-1)!}{\pi^N}\delta\left(1-r^2\right) \,,
\end{equation}
where $r^2=\sum_{n=1}^N |c_n|^2$. This corresponds to an homogeneous 
distribution on the surface of a $2N$-dimensional unit sphere, while 
averages of the kind
\begin{gather}
 \overbar{c_n c^*_m}=\frac{1}{N}\delta_{nm} \,,\\
 \overbar{c_n c^*_m c_{n'} c^*_{m'}}=\frac{1}{N(N+1)}
  (\delta_{nm}\delta_{n'm'}+\delta_{nm'}\delta_{n'm}) \,,
\end{gather}
follow from integrals of polynomials over its $(2N-1)$-dimensional 
surface~\cite{folland2001}. Consequently, we find the ensemble averages
\begin{gather}
  \overbar{\tr(\rho A)}=\frac{1}{N}\tr(A) \,, \\
  \overbar{\tr(\rho A \rho B)}=\frac{\tr(A)\tr(B)+\tr(AB)}{N(N+1)} \,.
\end{gather}

To compute averages for pure states belonging to a particular subspace of
dimension $N_D$, we have to replace $N$ by $N_D$ and the operators $A$
and $B$ by their projections onto that subspace, $P_DAP_D$ and $P_DBP_D$. 

  % \newpage
  \section{Two-level system decay rates \label{app:TwoLevel}}
  For completeness, we summarize the Bloch-Redfield result for the decay
rates of the two-level system coupled to an Ohmic 
bath~\cite{weiss1989,makhlin2001}. Following the notation used in the main 
text, its Hamiltonian is defined by 
\begin{equation}
  H=\frac{\Delta}{2}\sigma_x+\frac{\epsilon}{2}\sigma_z+\frac{1}{2}XB 
  \,, \label{eq:tls}
\end{equation}
with the tunnel matrix element $\Delta$ and the detuning $\epsilon$.  The 
bath coupling is specified by (i) $X=\sigma_z$ for charge noise and (ii)
$X=\sigma_x$ for current noise, respectively. To establish a relation to 
our Hubbard chain, we identify the detuning by the interaction, 
$\epsilon\sim U$, and the tunnel coupling by $\Delta\sim J$. Note that 
replacing charge noise by current noise corresponds to changing 
$\epsilon\to-\Delta$ and $\Delta\to\epsilon$. Therefore, we can restrict 
the derivation of the decay rate to case (ii).

It is straightforward to transform the Hamiltonian into the eigenbasis of
the two-level system, where it reads
\begin{equation}
 H' = \frac{E}{2}\sigma_z+X' B \,, 
\end{equation}
with $E=\sqrt{\epsilon^2 + \Delta^2}$, while the system-bath coupling 
becomes
\begin{equation}
	X'= \frac{\epsilon}{2E}\sigma_x + \frac{\Delta}{2E}\sigma_z \,.
\end{equation}
In the interaction picture, it is
\begin{equation}
  \tilde{X}(-\tau) = \frac{1}{2E}\left[\epsilon\sigma_x\cos(E\tau)
  + \epsilon\sigma_y\sin(E\tau) + \Delta\sigma_z\right] \,.
\end{equation}
Again ignoring the imaginary part of the integral in Eq.~\eqref{eq:Q}, the
noise kernel can be written as
\begin{equation}
 Q = \frac{\epsilon}{2E}\frac{\mathcal{S}(E)}{2}\sigma_x + 
  \frac{\Delta}{2E}\frac{\mathcal{S}(0)}{2}\sigma_z \,.
\end{equation}
The projector to the high-energy state is $P_1=(\sigma_0 + \sigma_z)/2$, so
that the decay rate can be found as
\begin{equation}
  \Gamma_\mathrm{ii} = \tr\left(P_1[Q,[X,P_1]]\right)
    = \left(\frac{\epsilon}{2E}\right)^2\mathcal{S}(E) \,.
\end{equation}
Accordingly, we find for case (i) the rate
\begin{equation}
    \Gamma_\mathrm{i}
    = \left(\frac{\Delta}{2E}\right)^2\mathcal{S}(E) \,.
\end{equation}
As it turns out, the low-temperature limit for an Ohmic spectral density 
\begin{equation}
  \mathcal{S}(E)\propto E \coth\left(\beta E/2\right) \,,
\end{equation}
coincides with the high-temperature limit at $k_BT=E/2$, since 
$\coth(x)\sim x^{-1}$ in the limit $x\to 0$.

\end{subappendices}

\chapter{Topological quantum optics \label{chap:topologicalQED}}
The spectacular progress in recent years in the field of topological matter 
has motivated the application of topological ideas to the field of quantum 
optics. The starting impulse was the observation that topological bands 
also appear with electromagnetic waves~\cite{haldane2008}. Soon after that, 
many experimental realizations followed~\cite{ozawa2019}. 
Nowadays, topological photonics is a burgeoning field with many 
experimental and theoretical developments. Among them, one of the current 
frontiers of the field is the exploration of the interplay between 
topological photons and quantum 
emitters~\cite{perczel2017,bettles2017,barik2018}.   

In this chapter we analyze what happens when quantum emitters interact
with a topological wave-guide QED bath, namely a photonic analogue of the 
SSH model, and show that it causes a number of unexpected 
phenomena~\cite{bello2019} such as 
the emergence of chiral photon bound states and unconventional scattering. 
Furthermore, we show how these properties can be harnessed to simulate 
exotic many-body Hamiltonians.

\section{Quantum emitter dynamics \label{sec:QEdynamics}}

\begin{figure}[!htb]
  \centering
  \includegraphics{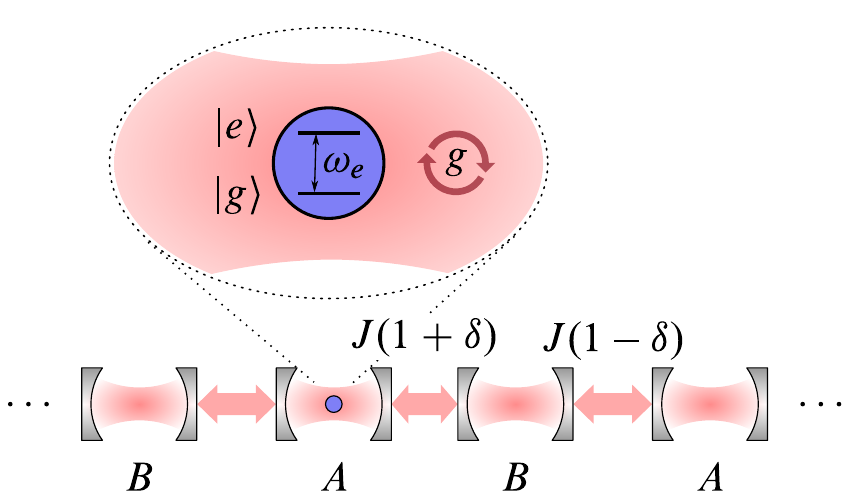}
  \caption{Schematic picture of the system under consideration: One or 
  many two-level quantum emitters (in blue) interact with the photonic 
  analogue of the SSH model. The interaction with photons (in red) induces 
  non-trivial dynamics between them.}
  \label{fig:schematics_photonicSSH}
\end{figure}

The system under consideration is depicted schematically in 
Fig.~\ref{fig:schematics_photonicSSH}, $N_e$ quantum emitters (QEs) interact
with a common bath, which behaves as the photonic analogue of the SSH 
model. This bath model consists of two interspersed photonic lattices $A/B$ 
with alternating nearest neighbour hoppings $J(1\pm\delta)$. We assume that 
the $A/B$ modes have the same energy $\omega_c$, that from now on we take as 
the reference energy of the problem, i.e., $\omega_c=0$. Since we have 
already analyzed this model in section~\ref{sec:SSH} we do not give any 
further details here. For the QEs, we consider they all have a single 
optical transition $g$-$e$ with a detuning $\omega_e$ respect to $\omega_c$, 
and they couple to the bath locally. Thus, the QEs and the bath are jointly 
described by the Hamiltonian ${H=H_S+H_B+H_I}$, with
\begin{equation}
  H_S=\omega_e\summ_{n=1}^{N_e}\sigma^n_{ee}\,,
\end{equation}
\begin{gather}
  H_B=-J\summ_{j=1}^N\left[(1+\delta)c^\dagger_{jA}c_{jB}
  +(1-\delta)c^\dagger_{j+1A}c_{jB}+\HC\right]\,, \\
  H_I=g\summ_{n=1}^{N_e}\left(\sigma^n_{eg}c_{x_n\alpha_n}+\HC\right)\,.
  \label{eq:HintQuantumOptics}
\end{gather}
Here, $c_{jA}$ ($c_{jB}$) annihilates a photon at the $j$th unit cell in
the $A$ ($B$) sublattice; $x_n$ and $\alpha_n$ denote the unit cell and 
sublattice to which the $n$th QE is coupled. We use the notation 
$\sigma^n_{\mu\nu}=\ket{\mu}_n\bra{\nu}$, $\mu,\nu\in\{e,g\}$ for the 
$n$th QE operator. The interaction is treated within the rotating-wave 
approximation such that only number-conserving terms appear in $H_{I}$.

To compute the dynamics of the system we can use two different approaches. 
In the weak-coupling regime the bath can be effectively traced out, such 
that the evolution of the QE reduced density matrix $\rho$ is described by 
a Markovian master equation~\cite{petruccione2002open} 
(see appendix~\ref{app:QuantumOpticalME}):
\begin{equation}
  \dot{\rho} = i[\rho,H_S] + 
  i\summ_{m,\,n} J_{mn}^{\alpha\beta} 
  \left[\rho, \sigma_{eg}^m\sigma_{ge}^{n}\right] 
  + \summ_{m,\,n} \frac{\Gamma^{\alpha\beta}_{mn}}{2}
  \left[2\sigma_{ge}^n\rho\sigma_{eg}^m - \sigma_{eg}^m\sigma_{ge}^n\rho 
  - \rho\sigma_{eg}^m\sigma_{ge}^n\right]\,.
  \label{eq:optical_mastereq}
\end{equation}
The functions $J_{mn}^{\alpha\beta}$ and $\Gamma^{\alpha\beta}_{mn}$, which 
control the QE coherent and dissipative dynamics, are the real and imaginary 
parts of the collective self-energy 
${\Sigma_{mn}^{\alpha\beta}(\omega_e+i0^+)}=
J_{mn}^{\alpha\beta}-i\Gamma^{\alpha\beta}_{mn}/2$.
This collective self-energy depends on the sublattices $\alpha,\beta\in\{A,B\}$ 
to which the $m$th and $n$th QE couple respectively, as well as on their 
relative position $x_{mn}=x_n-x_m$. Remarkably, for our model they can be
calculated analytically in the thermodynamic limit ($N\rightarrow\infty$) 
yielding (see appendix~\ref{app:SelfEnergies}):
\begin{gather}
\Sigma^{AA/BB}_{mn}(z)=-\frac{g^2 z\left[y^{|x_{mn}|}_+\Theta(1-|y_+|)
	-y^{|x_{mn}|}_-\Theta(|y_{+}|-1)\right]}
{\sqrt{z^4-4J^2(1+\delta^2)z^2+16J^4\delta^2}} \,, \label{eq:SigmaAA}\\
\Sigma^{AB}_{mn}(z)=\frac{g^2J\left[F_{x_{mn}}(y_+)\Theta(1-|y_+|)-
	F_{x_{mn}}(y_-)\Theta(|y_+|-1)\right]}
{\sqrt{z^4-4J^2(1+\delta^2)z^2+16J^4\delta^2}}\label{eq:SigmaAB} \,,
\end{gather}
where $F_n(z)=(1+\delta)z^{|n|}+(1-\delta)z^{|n+1|}$, $\Theta(z)$ is 
Heaviside's step function, and 
\begin{equation}
  y_\pm=\frac{z^2-2J^2(1+\delta^2) 
  \pm\sqrt{z^4-4J^2(1+\delta^2)z^2+16J^4\delta^2}}{2J^2(1-\delta^2)}\,.
  \label{eq:ypm}
\end{equation} 

When the transition frequency of the emitters lays in one of the bath's
energy bands, generally $\Gamma^{\alpha\beta}_{mn}\neq 0$, so the Markovian 
approximation predicts the decay of those emitters that are excited, 
emitting a photon into the bath. 
% In fact, the decay rate of a single 
% emitter is the one given by Fermi's golden rule.
On the other hand, if the transition frequency lays in one of the band gaps,
$\Gamma^{\alpha\beta}_{mn}=0$, so the emitters will not decay, but they 
will interact with each other through the emission and absorption of virtual
photons in the bath, that is, the bath mediates dipolar interactions between
the emitters. However, since we have a highly structured bath, this 
perturbative description will not be valid in certain regimes, e.g., close 
to band-edges, and we will use resolvent operator 
techniques~\cite{cohen1992atom} to solve the problem exactly for infinite 
bath sizes (see section \ref{sec:resolvent} for a brief introduction to the 
resolvent operator formalism).

\subsection{Single emitter dynamics \label{sec:SingleQEdynamics}}

Let us begin analyzing the dynamics of a single QE coupled to the bath. 
If the QE is initially excited, the wavefunction of the system at time 
$t=0$ is given by $\ket{\psi(0)}=\ket{e}\ket{\vac}$ ($\ket{\vac}$ denotes 
the vacuum state of the bath). Since the Hamiltonian conserves the number of 
excitations, the wavefunction of the system at any later time has the form:
\begin{equation}
  \ket{\psi(t)}=\left[\psi_e(t)\sigma_{eg}
  +\summ_j\summ_{\alpha=A,B}\psi_{j\alpha}(t)c^\dagger_{j\alpha}\right]
  \ket{g}\ket{\vac}\,.
  \label{eq:psi}
\end{equation}
The probability amplitude $\psi_e(t)$ can be computed as the Fourier 
transform of the corresponding matrix element of the resolvent, i.e., 
the emitter's Green's function 
$G_e=\left[z-\omega_e-\Sigma_e(z)\right]^{-1}$,
\begin{equation}
  \psi_e(t)=\frac{-1}{2\pi i}\int_{-\infty}^\infty\diffE
  G_e(E+i0^+)e^{-iEt}\,, \label{eq:integral_green}
\end{equation}
which depends on the emitter self-energy $\Sigma_e$,
\begin{equation}
  \Sigma_e(z)=\frac{g^2z\sign(|y_+|-1)}
  {\sqrt{z^4-4J^2(1+\delta^2)z^2+16J^4\delta^2}}\,,
  \label{eq:selfe3}
\end{equation}
obtained from Eq.~\eqref{eq:SigmaAA} defining 
$\Sigma_e(z)\equiv\Sigma^{AA}_{nn}(z)$. As we can see, $\Sigma_e$ 
does not depend on the sign of $\delta$. Thus, the single emitter 
dynamics is insensitive to the bath's topology.

\begin{figure}[!htb]
  \centering
  \includegraphics{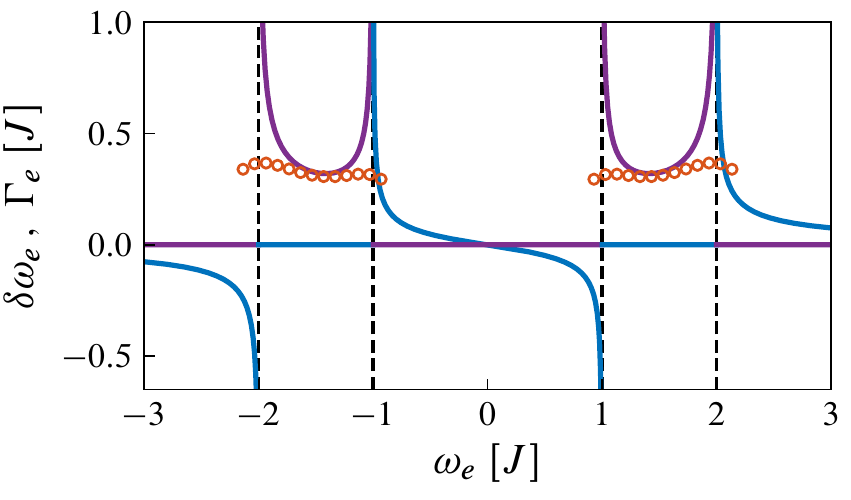}
  \caption{Comparison between the exact decay rate given by the 
  imaginary part of the complex poles of $G_e$ (orange circles) and 
  the Markovian decay rate (purple line) as a function of the 
  emitter's bare frequency. We also plot the Markovian Lamb shift 
  (blue line). The dashed vertical lines mark the position of the 
  bath's band edges. The parameters of the system are $\delta=0.5$, 
  $g=0.4J$.}
  \label{fig:selfenergy}
\end{figure}

To compute the integral in Eq.~\eqref{eq:integral_green}, we can use residue 
integration closing the contour of integration in the lower half of the 
complex plane. 
As a first approximation, one can assume that $\Sigma_e$ is small and 
varies little in the neighbourhood of $\omega_e$, and substitute $z$ by 
$\omega_e$ in its argument. This is essentially the same as the Markovian 
approximation, since the Green's function has then a single pole at 
$\omega_e+\Sigma_e(\omega_e+i0^+)$ and the evolution is given by 
$\psi_e(t)=e^{-i[\omega_e+\Sigma_e(\omega_e+i0^+)]t}$. One can readily 
identify $\delta\omega_e\equiv\re\Sigma_e(\omega_e+i0^+)$ as a shift of the 
emitter's frequency, known as the Lamb shift, and 
$\Gamma_e\equiv-2\im\Sigma_e(\omega_e+i0^+)$ as the decay rate of the 
emitter's excited state, which can be expressed in terms of the bath's 
density of modes $D(E)$ at emitter frequency as 
$\Gamma_e=\pi g^2D(\omega_e)$ (Fermi's golden rule). They are plotted in 
Fig.~\ref{fig:selfenergy}. Note that $\Gamma_e$ is nonzero only inside 
the bath's band regions, while the $\delta\omega_e$ is nonzero only outside 
them. This result corresponds to the basic expectation that if the 
frequency of the emitter is within the bath's energy bands of allowed modes, 
the emitter will decay exponentially with a decay rate given by Fermi's 
golden rule. On the contrary, if the emitter's frequency lays outside the 
bath's energy bands, it will remain excited.

\begin{figure}[!htb]
  \centering
  \includegraphics{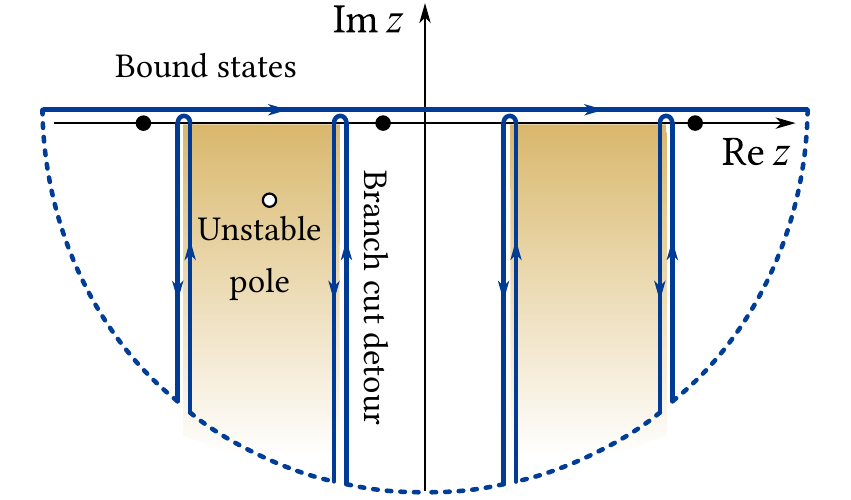}
  \caption{Integration path to compute the dynamics (in blue). Since the 
  Green's function has branch cuts along the real axis, it is necessary to 
  take a detour to the second Riemann sheet of the function (shaded areas). 
  The dynamics can be split in the contribution from the real poles of the 
  function (black dots), the complex poles (white dot) and the detours taken 
  at the band edges. $G^I_e$ has no complex poles in the lower half complex 
  plane thus we only have to consider them in the band regions where we go 
  to the second Riemann sheet. Real poles only appear in the band gap 
  regions.}
  \label{fig:contour_integration}
\end{figure}

As we will see next, an exact calculation of the integral yields somewhat 
different results. It requires choosing a proper contour of integration 
due to the branch cuts that the Green's function has along the real axis
in the regions where the bands of the bath are defined. A way to do it is 
to take a detour at the band edges to other Riemann sheets of the function, 
see Fig.~\ref{fig:contour_integration}. The formula for the Green's function 
in the first Riemann sheet $G^I_e$ is the one we have provided already. Its 
analytical continuation to the second Riemann sheet $G^{I\!I}_e$ can be 
obtained changing the sign of the square root in the denominator of 
$\Sigma_e$. 

According to this procedure the dynamics can be split in contributions of 
three different kinds:
\begin{equation}
  \psi_e(t)=\summ_{z_\mathrm{BS}}R(z_\mathrm{BS})e^{-iz_\mathrm{BS}t}
  + \summ_{z_\mathrm{UP}}R(z_\mathrm{UP})e^{-iz_\mathrm{UP}t}
  + \summ_j \psi_{\mathrm{BC},j}(t)\,.
\end{equation}
The first term accounts for the contribution of real poles of $G_e$, the 
so-called \emph{bound states} (BS). These are non-decaying solutions of the 
Schr\"odinger equation. The second term accounts for the contribution of 
\emph{unstable poles} (UP), i.e., complex poles of $G_e$. These are solutions 
that decay exponentially. The residue at both the real and complex poles 
can be computed as $R(z_0)=[1-\Sigma'_e(z_0)]^{-1}$, where $\Sigma'_e(z_0)$ 
denotes the first derivative of the appropriate function $\Sigma^I_e(z)$ 
or $\Sigma^{I\!I}_e(z)$. It can be interpreted as the overlap between the 
initial wavefunction and these solutions. Finally, we should 
subtract the detours taken due to the branch cuts. Their contribution can 
be computed as
\begin{equation}
  \psi_{\mathrm{BC},j}(t)=\frac{\pm 1}{2\pi}\int_0^\infty\diff{y}
  \left[G^I_e(x_j-iy)-G^{I\!I}_e(x_j-iy)\right]e^{-i(x_j-iy)t}\,,
\end{equation}
with $x_j\in\{\pm 2J,\pm 2|\delta|J\}$. The sign has to be chosen positive 
if when going from $x_j+0^+$ to $x_j-0^+$ the integration goes from the 
first to the second Riemann sheet, and negative if it is the other way 
around.

We can now point out several differences between the exact dynamics and the 
dynamics within the Markovian approximation. To begin with, the actual decay 
rate does not diverge at the band edges, but acquires a finite value, 
contrary to the Markovian prediction, see Fig.~\ref{fig:selfenergy}. 
Furthermore, the actual decay is not purely exponential, as the BC 
contributions decay algebraically $\sim t^{-3}$~\cite{sanchezburillo2017}
(see appendix~\ref{app:AlgebraicDecay}). In 
Fig.~\ref{fig:singleQE_dynamics}(a) it is shown an example of this 
subexponential decay when $\omega_e$ is placed at the lower band edge 
of the bath's spectrum.  
Another difference between the Markovian and the exact result is the 
fractional decay that the emitter experiences when its frequency lays 
outside the bath's energy bands . This is due to the emergence of
photon bound states which localize the photon around the 
emitter~\cite{bykov1975,john1990,kurizki1990}. 
For example, in Fig~\ref{fig:singleQE_dynamics}(b) we show the dynamics of 
an emitter with frequency in the middle of the band gap ($\omega_e=0$). It 
remains in the excited state with a long time limit given by the residue at 
$z_\mathrm{BS}=0$, $\lim_{t\to\infty}|\psi_e(t)|^2=|R(0)|^2=
\left[1+g^2/(4J^2\abs{\delta})\right]^{-2}$.
% \begin{equation}
%   \lim_{t\to\infty}|\psi_e(t)|^2=|R(0)|^2
%   =\left(1+\frac{g^2}{4J^2|\delta|}\right)^{-2}\,.
% \end{equation}

\begin{figure}[!htb]
  \centering
  \includegraphics{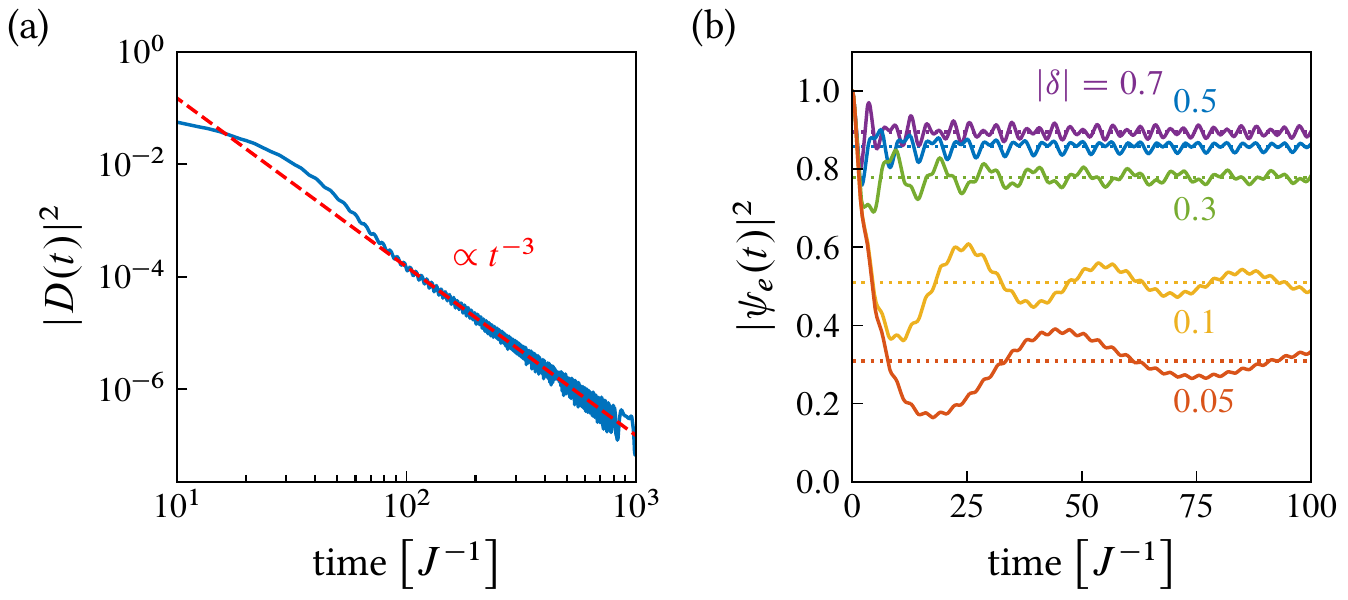}
  \caption{(a) Sub-exponential decay for a system with parameters:
  $\omega_e=-2J$, $\delta=0.5$ and $g=0.2J$. We plot the squared absolute
  value of the decaying part of the dynamics 
  $D(t)=\psi_e(t)-\sum_{z_\mathrm{BS}}R(z_\mathrm{BS})e^{-iz_\mathrm{BS}t}$.
  (b) Fractional decay for 
  different values of the dimerization parameter. The rest of parameters 
  of the system are $\omega_e=0$ and $g=0.4J$. As the band gap closes 
  ($\delta\to 0$) the decay becomes stronger. The dashed lines mark the 
  value of $|R(0)|^2$}
  \label{fig:singleQE_dynamics}
\end{figure}

These photon bound states are not unique to this particular 
bath~\cite{john1994}.
However, the BSs appearing in the present topological waveguide bath 
have some distinctive features with no analogue in other systems, and 
deserve special attention. As we will see later, they play a crucial 
role in the coherent evolution of many emitters. 
We can find their energy and wavefunction solving the secular equation 
$H\ket{\Psi_\mathrm{BS}}=\EBS\ket{\Psi_\mathrm{BS}}$, with 
$\EBS$ outside the band regions and $\ket{\Psi_\mathrm{BS}}$ 
in the form of Eq.~\eqref{eq:psi} with time-independent coefficients. 
Without loss of generality we assume that the emitter couples to 
sublattice $A$ at the $j=0$ cell. After some algebra, one finds 
that the energy of the BS is given by the pole equation 
$\EBS=\omega_e+\Sigma_e(\EBS)$. Irrespective of the values of $\omega_e$ or 
$g$, there are always three solutions to this equation. This is 
because the self-energy diverges in all band edges (see 
Fig.~\ref{fig:selfenergy}), which guarantees finding a BS in each of the 
band gaps. The wavefunction amplitudes can be obtained as
\begin{gather}
  \psi_{jA}=\frac{g\EBS\psi_e}{2\pi}
  \int_{-\pi}^\pi\diff{k}\frac{e^{ikj}}{\EBS^2-\omega_k^2}\,,
  \label{eq:BSwavefunA}\\
  \psi_{jB}=\frac{g\psi_e}{2\pi}
  \int_{-\pi}^\pi\diff{k}\frac{\omega_ke^{i(kj-\phi_k)}}{\EBS^2-\omega_k^2}\,,
  \label{eq:BSwavefunB}
\end{gather}
where $\psi_e$ is a constant obtained from the normalization 
condition that is directly related with the long-time population
of the excited state in spontaneous emission.
\begin{figure}[!htb]
  \centering
  \includegraphics{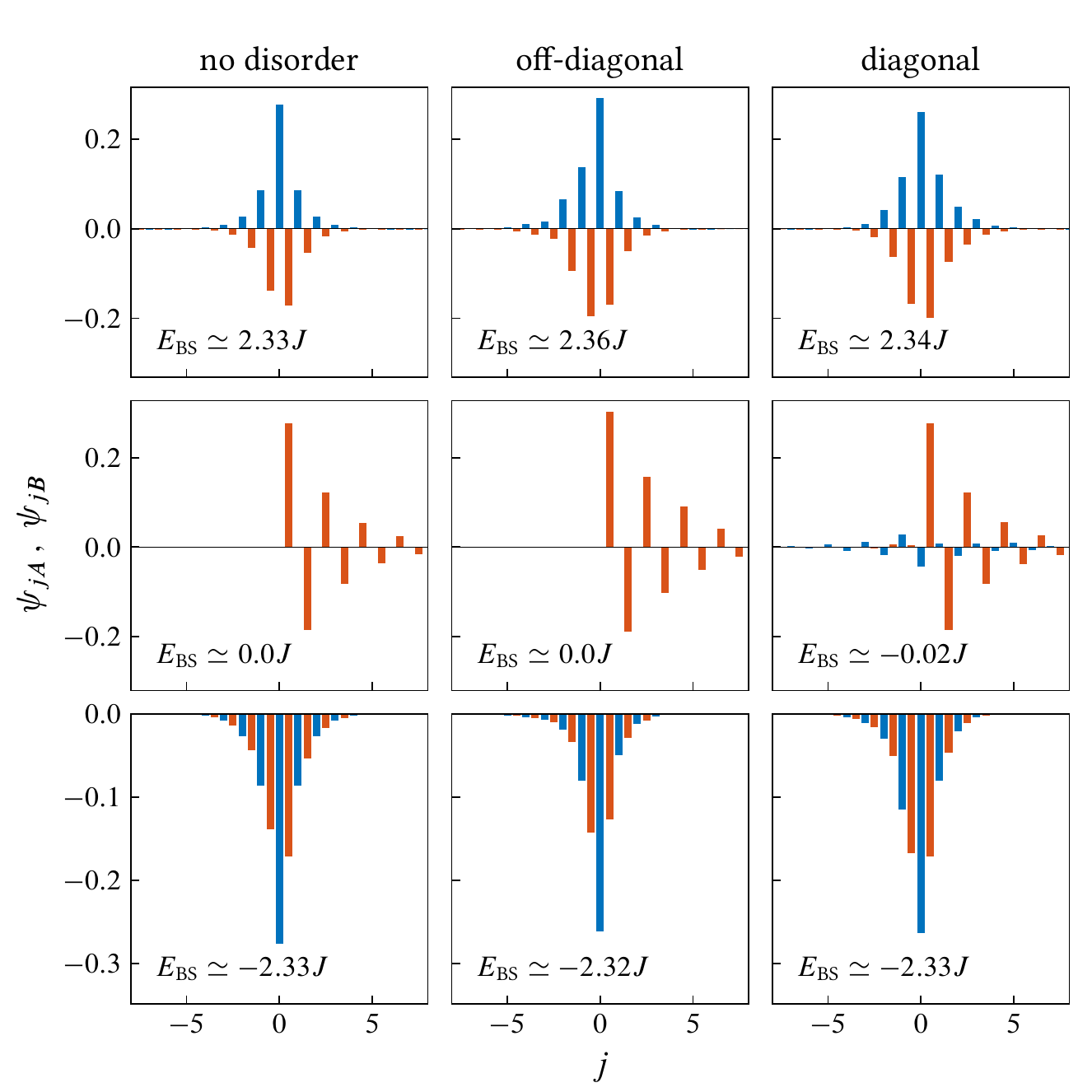}
  \caption{Single-photon bound states for a single emitter with transition 
  frequencies $\omega_e=2.2J$ (upper row), $\omega_e=0$ (middle row) and 
  $\omega_e=-2.2J$ (bottom row). Only the dominant bound state is plotted for
  each case. Rest of parameters: $g=0.4J$, $\delta=0.2$, and $w=0.6J$ for the 
  disordered cases}
  \label{fig:boundstates}
\end{figure}
They are plotted in the left column of Fig.~\ref{fig:boundstates}.
From Eqs.~\eqref{eq:BSwavefunA} and \eqref{eq:BSwavefunB} we can extract 
several properties of the spatial wavefunction distribution. On the 
one hand, above or below the bands ($|\EBS|>2J$, upper and lower band gaps) 
the largest contribution to the integrals is that of $k=0$. Thus, the 
amplitude of the wavefunction in any sublattice $\psi_{j\alpha}$ has the 
same sign regardless the unit cell $j$. In the lower (upper) band-gap, 
the amplitude on the different sublattices has the same (opposite) sign.
On the other hand, in the inner band gap ($|\EBS|<2|\delta|J$), the main 
contribution to the integrals is that of $k=\pi$. This gives an extra 
factor $(-1)^j$ to the coefficients $\psi_{j\alpha}$. Furthermore, in any 
band gap, the amplitudes on the sublattice to which the QE couples are 
symmetric with respect to the position of the QE, whereas they are 
asymmetric in the other sublattice, that is, the BSs are chiral. 
Changing $\delta$ from positive to negative results in a spatial 
inversion of the BS wavefunction. The asymmetry of the BS 
wavefunction is more extreme in the middle of the inner band-gap, for
$\omega_e=0$. At this point the BS has energy $\EBS=0$. If $\delta>0$, 
its wavefunction is given by $\psi_{jA}=0$ and 
\begin{equation}
  \psi_{jB}=\begin{dcases}
    \frac{g\psi_e(-1)^j}{J(1+\delta)}
    \left(\frac{1-\delta}{1+\delta}\right)^j\,, & j\geq 0\\
    0\,, & j< 0\\
  \end{dcases}\,,
\end{equation}
whereas for $\delta<0$ the wavefunction decays for $j<0$, 
and is strictly zero for $j\geq 0$. Its decay length diverges as 
$\lambda_\mathrm{BS}\sim 1/(2|\delta|)$ when the gap closes. Away from this 
point, the BS decay length shows the usual behavior for 1D baths 
$\lambda_\mathrm{BS}\sim |\Delta_\mathrm{edge}|^{-1/2}$,
% $\lambda_\mathrm{BS}\sim 1/\sqrt{|\Delta_\mathrm{edge}|}$, 
with $\Delta_\mathrm{edge}$ being the smallest detuning between 
the QE frequency and the band-edges. 

The physical intuition behind the appearance of such chiral BS at 
$\EBS=0$ is that the QE with $\omega_e=0$ acts as an 
\emph{effective edge} in the middle of the chain, or equivalently, 
as a boundary between two semi-infinite chains with different 
topology. In fact, one can show that this chiral BS has the same 
properties as the edge-state that appears in a semi-infinite SSH 
chain in the topologically non-trivial phase, for example, 
inheriting its robustness to disorder. To illustrate it, we study 
the effect of two types of disorder: one that appears in the 
cavities' bare frequencies, and another one that appears 
in the tunneling amplitudes. The former 
corresponds to the addition of random diagonal terms to the bath's 
Hamiltonian and breaks the chiral symmetry of the original model, 
\begin{equation}
  H_B\to H_B+ \summ_{j}\summ_{\alpha=A,B}
  \epsilon_{j\alpha}c^\dagger_{j\alpha}c_{j\alpha}\,,\\
\end{equation}
while the latter corresponds to the addition of off-diagonal random 
terms and preserves it,
\begin{equation}
H_B\to H_B + \summ_j\left(\epsilon_{j1}c^\dagger_{jB}c_{jA} +
\epsilon_{j2}c^\dagger_{j+1A}c_{jB} + \HC\right)\,.
\end{equation}
We take the coefficients $\epsilon_{j\nu}$, $\nu\in\{A,B,1,2\}$, 
from a uniform distribution within the range $[-w/2,w/2]$. To prevent 
changing the sign of the coupling amplitudes between the cavities, $w$ 
is restricted to $w/2 < J(1-|\delta|)$ in the case of off-diagonal 
disorder.

In the middle (right) column of Fig.~\ref{fig:boundstates} we plot the 
shape of the three BS appearing in our problem for a situation with 
off-diagonal (diagonal) disorder with $w=0.6J$. There, we observe that while
the upper and lower BS get modified for both types of disorder, the chiral 
BS has the same protection against off-diagonal disorder as a regular SSH 
edge-state: its energy is pinned at $E_\mathrm{BS}=0$ as well as keeping its
shape with no amplitude in the sublattice to which the QE couples to. On the
contrary, for diagonal disorder the middle BS is not protected any more and 
may have weight in both sublattices.  

\begin{figure}[!htb]
  \centering
  \includegraphics{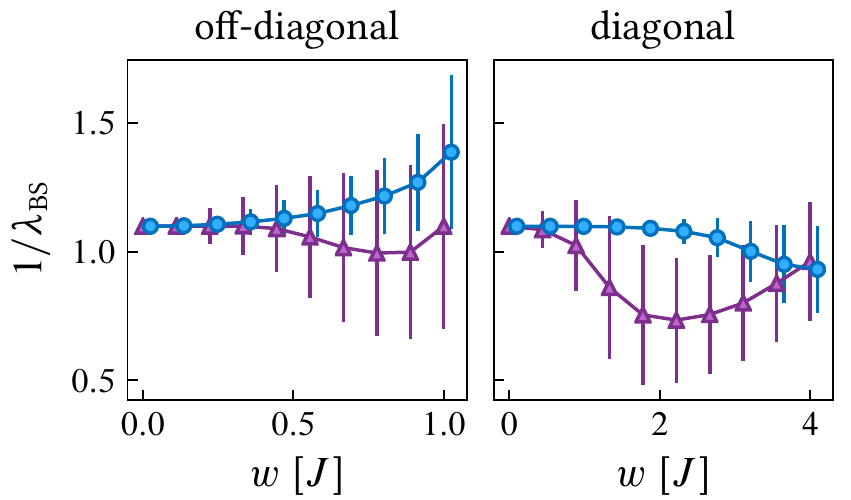}
  \caption{Inverse BS localization length as a function of the 
  disorder strength $w$ for both diagonal and off-diagonal disorder.
  The markers correspond to the average value computed with a total of $10^4$
  instances of disorder, and the error bars mark the value of one standard 
  deviation above and below the average value. The two sets of points are 
  slightly offset along the $x$ axis for better visibility. The two cases 
  shown correspond to $\omega_e\simeq 2.06J$ (purple triangles) and 
  $\omega_e=0$ (blue circles), which in the case without disorder have the 
  same decay length. The rest of parameters are: $g=0.4J$ and $\delta=0.5$.}
  \label{fig:BS_decay_length}
\end{figure}

Finally, to make more explicit the different behavior with disorder of the
middle BS compared to the other ones, we compute their localization length
$\lambda_{\mathrm{BS}}$ as a function of the disorder strength $w$ averaging
for many realizations. In Fig.~\ref{fig:BS_decay_length} we plot both the 
average value (markers) of $\lambda_\mathrm{BS}^{-1}$ and its standard 
deviation (bars) for the cases of the middle (blue circles) and upper 
(purple triangles) BSs. Generally, one expects that for weak disorder, 
states outside the band regions tend to delocalize, while for strong 
disorder all eigenstates become localized (see, for example, 
Ref.~\cite{perezgonzalez2019a}). In fact, this is the behavior we observe for the 
upper BS for both types of disorder. However, the numerical results suggest 
that for off-diagonal disorder the chiral BS never delocalizes (on average).
Furthermore, the chiral BS localization length is less sensitive to the 
disorder strength $w$ manifested in both the large initial plateau 
as well as the smaller standard deviations compared to the upper BS results.

\subsection{Two emitter dynamics}

The next simplest case we can study is that of two QE coupled to the bath.
From the master equation~\eqref{eq:optical_mastereq}, we can see that if the
QEs frequency is in one of the band gaps, the interaction with the bath 
leads to an effective unitary dynamics governed by the following 
Hamiltonian:
\begin{equation}
  H_\mathrm{dd}=J^{\alpha\beta}_{12}
  \left(\sigma^1_{eg}\sigma^2_{ge}+\HC\right)\,.
\end{equation}
That is, the bath mediates dipole-dipole interactions between the emitters.
One way to understand the origin of these interactions is that the emitters
exchange virtual photons through the bath. In fact, these virtual photons 
are nothing but the photon BS that we have studied in the previous section. 
Thus, these interactions $J^{\alpha\beta}_{mn}$ inherit many properties of 
the BSs. For example, the interactions are exponentially localized in space, 
with a localization length that can be tuned and made large by setting 
$\omega_e$ close to the band-edges, or fixing $\omega_e=0$ and letting the 
middle band gap close, $\delta\to 0$. Moreover, one can also change 
qualitatively the interactions by moving $\omega_e$ to different band gaps: 
for $|\omega_e|>2J$ all the $J^{\alpha\beta}_{mn}$ have the same sign, while 
for $|\omega_e|<2|\delta|J$ they alternate sign as $x_{mn}$ increases. Also,
changing $\omega_e$ from positive to negative changes the sign of 
$J^{AA/BB}_{mn}$, but leaves unaltered $J^{AB/BA}_{mn}$. Furthermore, while
$J^{AA/BB}_{mn}$ are insensitive to the bath's topology, the 
$J^{AB/BA}_{mn}$ mimic the dimerization of the underlying bath, but allowing
for longer range couplings. The most striking regime is reached for 
$\omega_e=0$. In that case $J^{AA/BB}_{mn}$ identically vanish, so
the QEs only interact if they are coupled to different sublattices. 
Furthermore, in such a situation the interactions have a strong directional 
character, i.e., the QEs only interact if they are in some particular order.
Assuming that the first QE at $x_1$ couples to sublattice $A$, and the 
second one at $x_2$ couples to $B$, we have
\begin{equation}
  J^{AB}_{12}=\begin{cases}
    \sign(\delta)\frac{g^2(-1)^{x_{12}}}{J(1+\delta)}
    \left(\frac{1-\delta}{1+\delta}\right)^{x_{12}} \,, 
    & \delta\cdot x_{12}>0\\
    0\,, & \delta\cdot x_{12}<0\\
    \Theta(\delta)\frac{g^2}{J(1+\delta)}\,, & x_{12}=0
  \end{cases}\,.
  \label{eq:JABmarkov}
\end{equation}

\begin{figure}[!htb]
  \centering
  \includegraphics{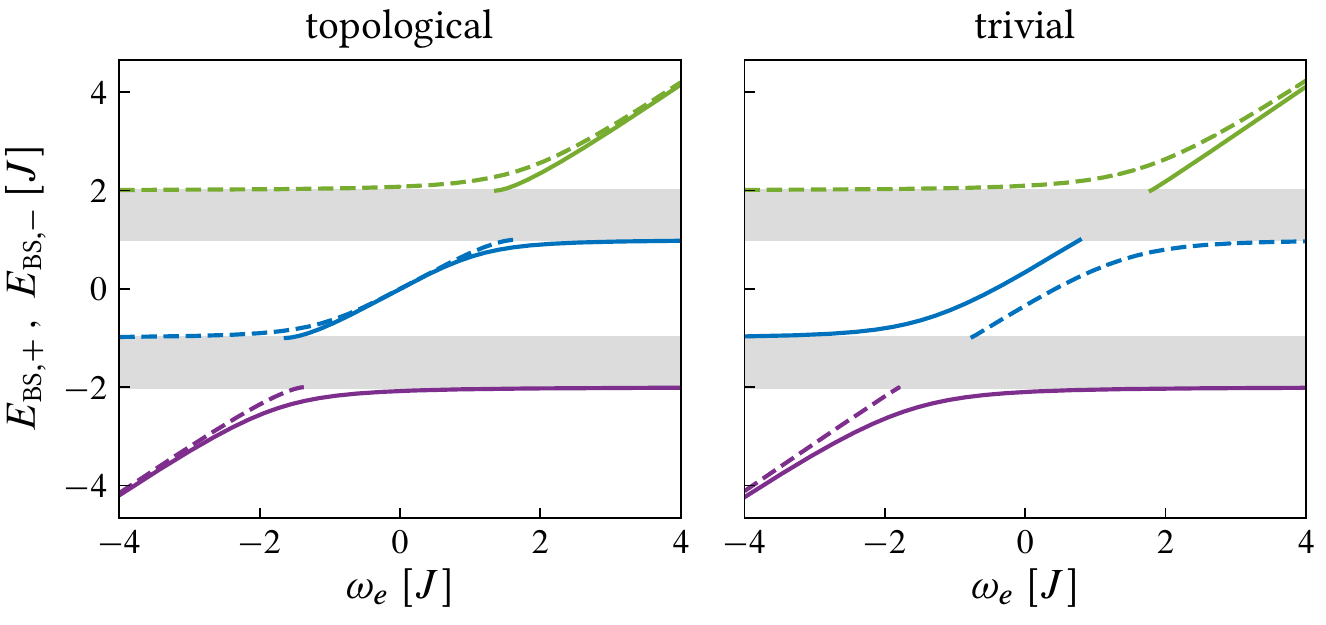}
  \caption{Exact energies for the symmetric (continuous line) and 
  antisymmetric (dashed line) bound states for a system with parameters: 
  $g=0.8J$, and $\delta=0.5$ (right) or $\delta=-0.5$ (left). The
  two emitters are in the $AB$ configuration coupled to the same unit cell,
  $x_{12}=0$. The grey areas mark the span of the bath's energy bands.}
  \label{fig:disappearance}
\end{figure}

We can also apply resolvent operator techniques to compute the dynamics of 
two emitters exactly. It can be shown that the symmetric and antisymmetric 
combinations
$\sigma^\dagger_\pm=\left(\sigma^1_{eg}\pm\sigma^2_{eg}\right)/\sqrt{2}$ 
evolve independently as they couple to orthogonal bath 
modes~\cite{gonzaleztudela2017} (see appendix \ref{app:SelfEnergies}). Thus, 
the two-emitter problem is equivalent to two independent single-emitter 
problems. Remarkably the self-energies adopt the simple form 
$\Sigma^{\alpha\beta}_\pm=\Sigma_e\pm\Sigma^{\alpha\beta}_{mn}$.
We can now compute the two-emitter BSs' energies solving the pole equations 
$\EBSsub{\pm}=\omega_e+\Sigma^{\alpha\beta}_\pm(\EBSsub{\pm})$. However, 
there are some subtleties which differentiate this problem from the 
single-emitter problem. Now, the cancellation of divergences in $\Sigma_\pm$ 
at the band edges results in critical values for the emitter transition 
frequency above (or below) which some bound states cease to exist. For 
example, for two emitters in the $AB$ configuration, in the symmetric 
subspace we have that the lower bound state ($\EBSsub{+}<-2J$) always 
exists, while the upper bound state ($\EBSsub{+}>2J$) exists only for 
$\omega_e>\omega_\mathrm{crit}$,
\begin{equation}
  \omega_\mathrm{crit}=2J-\frac{g^2(2x_{12}+1-\delta)}{2J(1-\delta^2)}\,.
\end{equation}
For the middle bound state, there are two possibilities: either the 
divergence vanishes at $-2|\delta|J$, in which case the bound state will 
exist for $\omega_e>\omega_\mathrm{crit}$, or the divergence vanishes at 
$2|\delta|J$, then the middle bound state exists for 
$\omega_e<\omega_\mathrm{crit}$. In both cases $\omega_\mathrm{crit}$ takes 
the same form
\begin{equation}
  \omega_\mathrm{crit}=(-1)^{x_{12}}\left\{
    2\delta J+\frac{g^2[(2x_{12}+1)\delta-1]}{2J(1-\delta^2)}\right\}\,.
\end{equation}
The situation in the antisymmetric subspace can be readily understood 
realizing that 
$\re\Sigma^{\alpha\beta}_-(z)=-\re\Sigma^{\alpha\beta}_+(-z)$, 
which implies that if $\EBSsub{+}$ is a solution of the pole equation for 
$\Sigma^{\alpha\beta}_+$ for a particular value of $\omega_e$, then 
$\EBSsub{-}=-\EBSsub{+}$ is a solution of the pole equation for 
$\Sigma^{\alpha\beta}_-$ for the opposite value of $\omega_e$. 
Fig.~\ref{fig:disappearance} summarizes at a glance the different 
possibilities and the dependence on the bath's topology. 

\begin{figure}[!htb]
  \centering
  \includegraphics{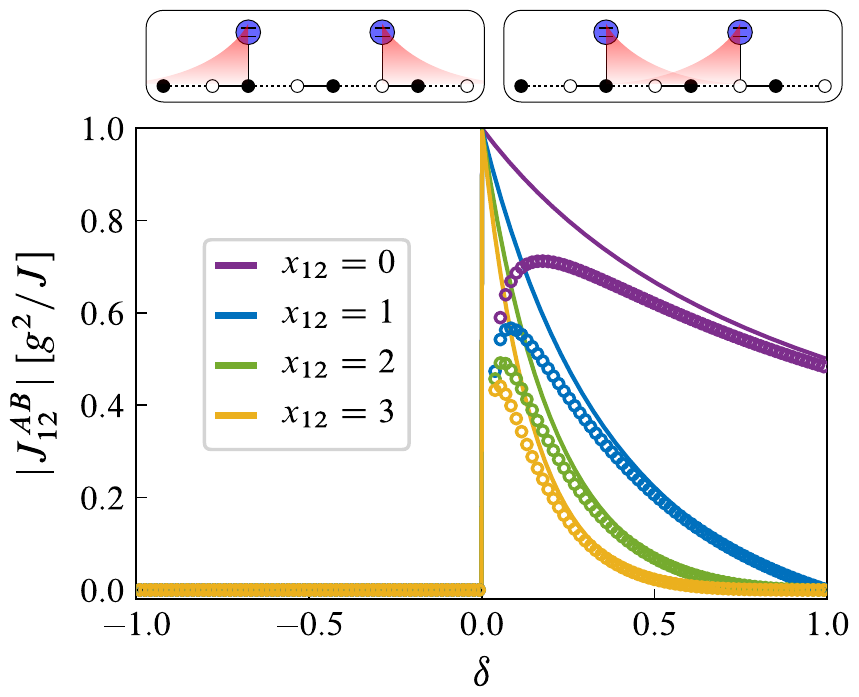}
  \caption{Effective dipolar coupling as given by the BSs energy difference 
  (dots), and the Markovian approximation~\eqref{eq:JABmarkov} (lines) as 
  a function of the dimerization constant. The rest of parameters are 
  $\omega_e=0$ and $g=0.4J$. The schematics above show the shape of the 
  bound states in the topological (right) and trivial (left) phases. The 
  situation for the $BA$ configuration is the same, reversing the role of 
  $\delta$.}
  \label{fig:dipolecoupling}
\end{figure}

The physical intuition behind this phenomenon is the following: when 
bringing close together two emitters, their respective single-emitter bound 
states hybridize forming (anti)symmetric superpositions which have 
energies above and below the single emitter bound state energy. Back in the 
localized basis the splitting of the bound state energies corresponds to 
the ``hopping'' of the photon, i.e., an effective dipole-dipole interaction 
between the emitters, $J^{\alpha\beta}_{12}=(\EBSsub{+}-\EBSsub{-})/2$. 
If this splitting is very strong, it may happen that one of the two bound 
states merges into the bulk bands. Then it is not possible to rewrite the 
low-energy Hamiltonian as an effective dipole-dipole 
interaction~\cite{shi2018}.
In Fig.~\ref{fig:dipolecoupling} we show the exact value of the interaction 
constant and compare it with the Markovian result \eqref{eq:JABmarkov}. 
Apart from small deviations when the gap closes for $\delta\to 0$, it is 
important to highlight that the directional character agrees perfectly in 
both cases.

When the QEs frequency is resonant with one of the bath's bands, the bath 
typically induces non-unitary dynamics in the emitters. However, when many 
QEs couple to the bath there are situations in which the interference 
between their emission may enhance or suppress (even completely) the decay 
of certain states. This phenomenon is known as 
super/subradiance~\cite{dicke1954}, respectively. Let us illustrate this 
effect with two QEs: In that case, the decay rate of a 
symmetric/antisymmetric combination of excitations is 
$\Gamma_e\pm\Gamma^{\alpha\beta}_{12}$. When 
$\Gamma^{\alpha\beta}_{12}=\pm\Gamma_e$, these states decay at a rate that 
is either twice the individual one or zero. In this latter case they are 
called perfect subradiant or dark states.  

In standard one-dimensional baths 
$\Gamma_{12}(\omega_e)=\Gamma_e(\omega_e)\cos\big(k(\omega_e)|x_{mn}|\big)$,
so the dark states are such that the wavelength of the photons involved, 
$k(\omega_e)$, allows for the formation of a standing wave between the QEs 
when both try to decay, i.e., when $k(\omega_e)|x_{mn}|=n\pi$, with 
$n\in\mathbb{N}$. Thus, the emergence of perfect super/subradiant states 
solely depends on the QE frequency $\omega_e$, bath energy dispersion 
$\omega_k$, and their relative position $x_{mn}$, which is the common 
intuition for this phenomenon.  
This common wisdom gets modified in the topological bath that we have 
considered, where we find situations in which, for the same values of 
$x_{mn}$, $\omega_k$ and $\omega_e$, the induced dynamics is very different 
depending on the sign of $\delta$. In particular, when two QEs couple to the
$A$ and $B$ sublattice respectively, the collective decay reads: 
\begin{equation}
  \Gamma^{AB}_{12}(\omega_e)=
  \Gamma_e\sign(\omega_e)\cos\big(k(\omega_e)x_{12}-\phi(\omega_e)\big)\,, 
  \label{eq:colltop}
\end{equation}
which depends both on the photon wavelength mediating the interaction
\begin{equation}
  k(\omega_e)=\arccos \left(\frac{\omega_e^2-2J^2(1+\delta^2)}
  {2J^2(1-\delta^2)}\right) \,,
\end{equation}
an even function of $\delta$, and on the phase 
$\phi(\omega_e)\equiv\phi(k(\omega_e))$, which is sensitive to the sign 
of $\delta$. This 
$\phi$-dependence enters through the system-bath coupling when rewriting 
the interaction Hamiltonian $H_I$ [Eq.~\eqref{eq:HintQuantumOptics}] in 
terms of the eigenoperators $u_k,l_k$ (see appendix~\ref{app:SelfEnergies}). 
Thus, even though the sign of $\delta$ does not play a role in the 
properties of an infinite bath, when the QEs couple to it, the bath embedded 
between them is different for $\delta\gtrless 0$, making the two situations 
inequivalent.  

Using Eq.~\eqref{eq:colltop}, we find that the detunings at which 
perfect super/subradiant states appear satisfy
$k(\omega_s)x_{12}-\phi(\omega_s)=n\pi$, $n\in\mathbb{N}$. They come in 
pairs: If $\omega_s$ corresponds to a superradiant (subradiant) state in 
the upper band, $-\omega_s$ corresponds to a subradiant (superradiant) 
state in the lower band. In particular, it can be shown that when 
$\delta<0$, the previous equation has solutions for $n=0,\dots,x_{12}$, 
while if $\delta>0$, the equation has solutions for $n=0,\dots,x_{12}+1$. 
Besides, the detunings, $\omega_s$ at which the subradiant states appear 
also satisfy that $J_{12}^{AB}(\omega_s)\equiv 0$, which guarantees that 
these subradiant states survive even in the non-Markovian regime with a 
correction due to retardation which is small as long as 
$x_{12}\Gamma_e(\omega_e)/(2|v_g(\omega_e)|)\ll 1$ [$v_g(\omega_e)$ is 
the group velocity of the photons in the bath at frequency 
$\omega_e$]~\cite{gonzaleztudela2017}. Apart from inducing different decay 
dynamics, these different conditions for super/subradiance at fixed 
$\omega_e$ also translate in different reflection/transmission 
coefficients when probing the system through photon scattering, as we show 
in the next section.

\section{Single-photon scattering}

Here, we study the case when the emitter frequencies are resonant with one 
of the bands. We will see how for a single emitter coupled to both the $A$ 
and $B$ cavities inside a unit cell there is a $\delta$-dependent Lamb-Shift
that can be detected in single-photon scattering experiments. Also, we will 
see how the different super/subradiant states for $\pm\delta$ lead to 
different behavior when a single photon scatters off two QEs.

\subsection{Scattering formalism}

\begin{figure}[!htb]
  \centering
  \includegraphics{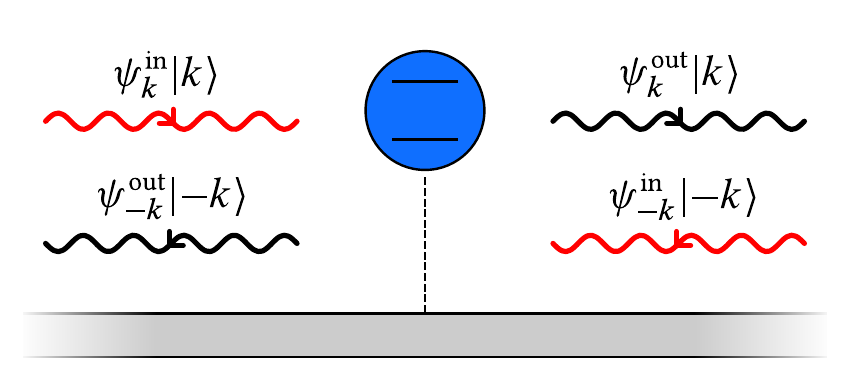}
  \caption{Schematic picture of a scattering eigenstate and the different 
  amplitudes involved. The QE divides the space in left and right regions. 
  Incoming modes are those that propagate towards the emitter (red), 
  while outgoing modes are those that propagate away from the emitter 
  (black).}
  \label{fig:scatteringeigenstate}
\end{figure}

The scattering properties of a single photon impinging into one or several 
QEs in the ground state can be obtained from the scattering eigenstates,
which are solutions of the secular equation
$H\ket{\Psi_k}=\pm\omega_k\ket{\Psi_k}$ (the sign depends on the band we 
are probing). First, let us assume that a single emitter couples to the 
$A$ sublattice at the $x_1$ unit cell. We use the ansatz
\begin{equation}
  \lvert\Psi_k\rangle=\begin{cases}
    \psi^\inn_k\ket{k}+\psi^\out_{-k}\ket{-k}\,, & j< x_1\\
    \psi^\out_k\ket{k}+\psi^\inn_{-k}\ket{-k}\,, & j\geq x_1
  \end{cases}\,,
\end{equation}
where $\ket{\pm k}=u^\dagger_{\pm k}\ket{\vac}$ or 
$\ket{\pm k}=l^\dagger_{\pm k}\ket{\vac}$, depending on the band we are 
probing ($u_k$ and $l_k$ are the eigenmodes of the SSH model, see 
section~\ref{sec:SSH}). For a schematic representation of $\ket{\Psi_k}$ see
Fig.~\ref{fig:scatteringeigenstate}. 

The coefficients of the scattering eigenstate in position representation 
are then
\begin{align}
  \psi_{jA}&=\pm\begin{cases}
    \psi_{k}^\inn e^{i(kj+\phi_k)}+\psi_{-k}^\out e^{-i(kj+\phi_k)}\,,
    & j\leq x_1 \\
    \psi_{k}^\out e^{i(kj+\phi_k)}+\psi_{-k}^\inn e^{-i(kj+\phi_k)}\,,
    & j\geq x_1 
  \end{cases}\,, \\
  \psi_{jB}&=\begin{cases}
    \psi_{k}^\inn e^{ikj}+\psi_{-k}^\out e^{-ikj}\,,
    & j<x_1 \\
    \psi_{k}^\out e^{ikj}+\psi_{-k}^\inn e^{-ikj}\,,
    & j\geq x_1 
  \end{cases}\,.
\end{align}                           
The matching condition at the emitter position
\begin{equation}
  \psi_{k}^\inn e^{i(kx_1+\phi_k)}+\psi_{-k}^\out e^{i(kx_1+\phi_k)} =
  \psi_{k}^\out e^{i(kx_1+\phi_k)}+\psi_{-k}^\inn e^{i(kx_1+\phi_k)}\,,
\end{equation}
together with the secular equation
\begin{gather}
  \begin{cases}
    \omega_e\psi_e + g\psi_{x_1A}=\pm\omega_k\psi_e\,,\\
    g\psi_e + -J(1+\delta)\psi_{x_1B}-J(1-\delta)\psi_{x_1-1B}
    =\pm\omega_k\psi_{x_1A}\,,
  \end{cases}\\
  \Rightarrow
  -J(1+\delta)\psi_{x_1B}-J(1-\delta)\psi_{x_1-1B}
  =\left(\pm\omega_k-\frac{g^2}{\Delta_k}\right)\psi_{x_1A}\,,
\end{gather}
where we have defined $\Delta_k\equiv \pm\omega_k - \omega_e$,
allow us to write a linear relation between the wave
amplitudes on the left of the emitter and those on the right as
\begin{equation}
  \myvec{\psi_{k}^\out\\\psi_{-k}^\inn}=
  T\myvec{\psi_{k}^\inn\\\psi_{-k}^\out}\,,
\end{equation}
where the \emph{transfer matrix} $T$ is
\begin{equation}
  \mbox{\small\(
  T_A=\myvec{1\pm\dfrac{g^2}{i2\Delta_k J(1-\delta)\sin(k+\phi_k)}&
  \dfrac{\pm g^2 e^{-i2(kx_1+\phi_k)}}{i2\Delta_k J(1-\delta)\sin(k+\phi_k)} \\[1em]
  \dfrac{\mp g^2 e^{i2(kx_1+\phi_k)}}{i2\Delta_k J(1-\delta)\sin(k+\phi_k)} &
  1\mp\dfrac{g^2}{i2\Delta_k J(1-\delta)\sin(k+\phi_k)}}\,.
  \)}
\end{equation}
Similarly, if the emitter couples to the $B$ sublattice at the $x_1$ 
unit cell we have
\begin{equation}
  T_B=\myvec{1\mp\dfrac{g^2}{i2\Delta_k J(1+\delta)\sin(\phi_k)} &
  \dfrac{\mp g^2 e^{-i2kx_1}}{i2\Delta_k J(1+\delta)\sin(\phi_k)} \\[1em]
  \dfrac{\pm g^2 e^{i2kx_1}}{i2\Delta_k J(1+\delta)\sin(\phi_k)} &
  1\pm\dfrac{g^2}{i2\Delta_k J(1+\delta)\sin(\phi_k)}}\,.
\end{equation}
From the transfer matrix we can compute the scattering matrix $S$, 
which relates the asymptotic incoming modes with the outgoing modes:
\begin{equation}
  \myvec{\psi_{k}^\out\\\psi_{-k}^\out}=
  S\myvec{\psi_{k}^\inn\\\psi_{-k}^\inn}\,, 
\end{equation}
with
\begin{equation}
  S=\myvec{t_{11}-\dfrac{t_{12}t_{21}}{t_{22}} & \dfrac{t_{12}}{t_{22}}\\[1em]
    -\dfrac{t_{21}}{t_{22}} & \dfrac{1}{t_{22}}}
  \equiv\myvec{t_L & r_R \\ r_L & t_R}\,.
\end{equation}
Here, $t_{ij}$ denote the matrix elements of the transfer matrix, while 
$t_{L/R}$ and $r_{L/R}$ denote the matrix elements of the scattering matrix.
They correspond to the transmission and reflection probability amplitudes 
for a wave coming from the left/right.

If evolution is unitary, that is, there are no photon losses,
$S^\dagger S= SS^\dagger=I$, which implies 
$|t_L|^2 + |r_L|^2 = |t_R|^2 + |r_R|^2 = 1$ and
$|t_L|^2 + |r_R|^2 = |t_R|^2 + |r_L|^2 = 1$. Therefore, $|t_L|=|t_R|$ and 
$|r_L|=|r_R|$. Furthermore if the system is time-reversal symmetric ($H$ is 
real), as is the case in our model, the scattering is reciprocal, i.e., 
$t_L=t_R\equiv t$. To see this, let us consider the scattering eigenstate 
with amplitudes
$(\psi_{k}^\inn,\psi_{-k}^\inn,\psi_{k}^\out,\psi_{-k}^\out)=(1,0,t_L,r_L)$,
and call it $\lvert\Psi_{k,L}\rangle$. Then, its complex conjugate is also a
scattering eigenstate with the same energy and so is the linear combination 
$(1/t_L^*)\ket{\Psi_{k,L}}^*-(r_L^*/t_L^*)\ket{\Psi_{k,L}}$,
which has coefficients $\left(0, 1, -r_L^*t_L/t_L^*,t_L\right)$, 
but this must be the scattering eigenstate with coefficients $(0, 1, r_R, t_R)$.

The scattering coefficients for the many-emitter case can be readily 
obtained noting that if we label the emitters with an increasing index
from left to right, the fields on the right of the $m$th emitter are those
on the left of the $(m + 1)$th emitter. Thus, the transfer matrix of the 
entire system can be written as the product of single-emitter transfer 
matrices $T=T_{N_e}T_{N_e-1}\dots T_1$ ($N_e$ is the number of emitters) and
from it one can compute the scattering matrix of the entire system. 

\subsection{Scattering off one and two emitters}

\begin{figure}[!htb]
  \centering
  \includegraphics{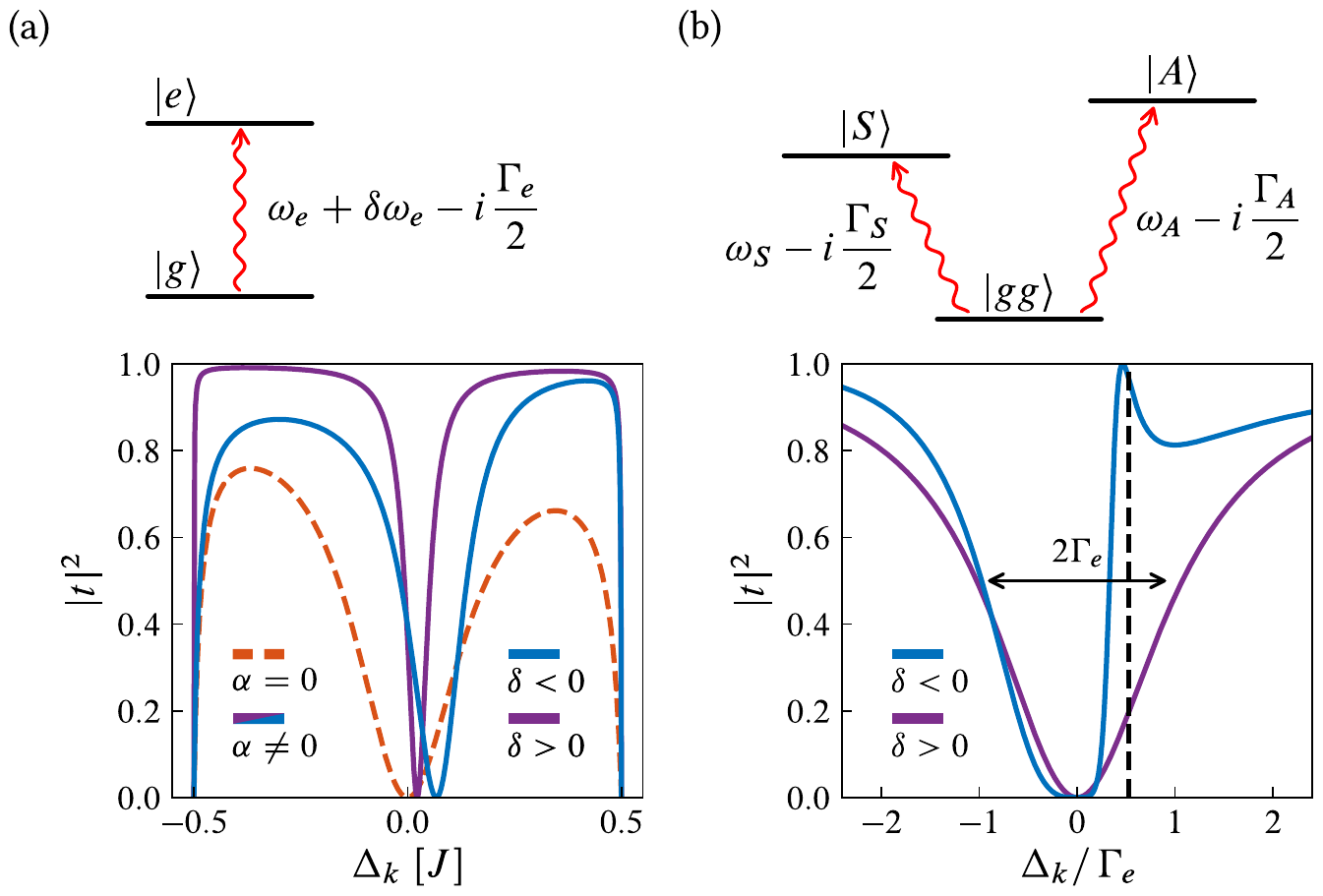}
  \caption{Relevant level structure and transmission probability for a 
  single photon scattering off (a) one QE coupled to both $A$ and $B$ 
  cavities inside a unit cell, (b) two QEs in an $AB$ configuration 
  separated a distance of $x_{12}=2$ unit cells; $\ket{gg}=\ket{g}_1\ket{g}_2$
  denotes the common ground state, while $\ket{S/A}=\left(
  \ket{e}_1\ket{g}_2\pm\ket{g}_1\ket{e}_2\right)/\sqrt{2}$ denotes the 
  (anti)symmetric excited state combination of the two QEs.
  The parameters in (a) are 
  $g=0.4J$, $\delta=\pm 0.5$, $\omega_e=1.5J$, and $\alpha=0$
  or $\alpha=0.3$. The dashed orange line corresponds to the case where the 
  emitter couples to a single sublattice ($\alpha=0,1$), it does not depend
  on the sign of $\delta$. The parameters in (b) are $g=0.1J$, 
  $\delta=\pm 0.5$, and $\omega_e\simeq 1.65J$, for which the two QEs are 
  in a subradiant configuration if $\delta>0$. The black dashed line 
  marks the value of $\omega_\mathrm{EIT}$.}
  \label{fig:scattering}
\end{figure}

For a single emitter, we find the same transmission coefficient regardless 
the sublattice to which the emitter is coupled
\begin{equation}
  t=\frac{2J^2(1-\delta^2)\Delta_k\sin(k)}
  {2J^2(1-\delta^2)\Delta_k\sin(k)\mp i g^2\omega_k}\,.
\end{equation}
A well-known feature for this type of system is the perfect reflection 
($\abs{r}^2=1\Leftrightarrow\abs{t}^2=0$) when the frequency of the 
incident photon matches exactly that of the QE~\cite{zhou2008}. This can be 
seen in Fig.~\ref{fig:scattering}(a) as a full dip in the transmission 
probability at $\Delta_k=0$. The dip has a bandwidth determined by the 
individual decay rate $\Gamma_e$. Besides, it also shows the vanishing of 
the transmission at the band edges. Since there is no dependence on the 
sign of $\delta$, the scattering in this configuration is insensitive to 
the bath's topology.

A more interesting situation occurs when a single emitter couples to both
the $A$ and $B$ cavities in a single cell. We choose the coupling constants
$g\alpha$ and $g(1-\alpha)$, such that we can interpolate between the cases
where the QE couples only to sublattice $A$ ($\alpha=1$) or $B$ 
($\alpha=0$). Using the same ansatz as in the previous case, we find
\begin{equation}
  t=\frac{2iJ(1-\delta)\sin(k)\left[J(1+\delta)\Delta_k-g^2\alpha(1-\alpha)\right]}
  {2iJ^2(1-\delta^2)\Delta_k\sin(k)+g^2\omega_k\left[2\alpha(1-\alpha)(e^{-i\phi_k}\mp 1)
  \pm 1\right]}\,.
  \label{eq:transmisiontwo}
\end{equation}
Now, for $0<\alpha<1$, the transmission is different for $\pm\delta$. 
In Fig.~\ref{fig:scattering}(a) we plot this formula for $\delta=\pm 0.3$, 
and show that the transmission dip gets shifted. This is due to a 
$\delta$-dependent Lamb-Shift 
$\delta\omega_e=g^2\alpha(1-\alpha)/\left[J(1+\delta)\right]$.
Notice that Eq.~\eqref{eq:transmisiontwo} is invariant under the 
transformation $\alpha\to 1-\alpha$.

For two emitters coupled equally to the bath at unit cells $x_1$ and $x_2$, 
in the $AB$ configuration, we find
\begin{equation}
  t=\frac{\left[2J^2(1-\delta^2)\Delta_k\sin(k)\right]^2}
  {g^4\omega^2_k e^{i2\left(kx_{12}-\phi_k\right)}-
  \left[g^2\omega_k\pm 2iJ^2(1-\delta^2)\Delta_k\sin(k)\right]^2}\,,
\end{equation}
whose squared absolute value is plotted in Fig.~\ref{fig:scattering}(b) for 
$\delta=\pm 0.5$. The difference between bath in the topological and trivial
phases is more pronounced than in the single-emitter case, since now the 
transmission is qualitatively different in each case: While the case with 
$\delta>0$ features a single transmission dip at the QEs frequency, for 
$\delta<0$, the transmission dip is followed by a window of frequencies with
perfect photon transmission, i.e., $\abs{t}^2=1$. We can understand this 
behavior realizing that a single photon only probes the (anti)symmetric 
states in the single excitation subspace $\ket{S}/\ket{A}$, with the 
following energies renormalized by the bath, 
$\omega_{S/A}=\omega_e\pm J^{AB}_{12}$, and linewidths 
$\Gamma_{S/A}=\Gamma_e\pm\Gamma_{12}$. For the parameters chosen, it can be
shown that for $\delta>0$ the QEs are in a perfect super/subradiant 
configuration in which one of the states decouples while the other has a
$2\Gamma_e$ decay rate. Thus, at this configuration, the two QEs
behave like a single two-level system with an increased linewidth. On the 
other hand, when $\delta<0$, both the (anti)symmetric states are coupled 
to the bath, such that the system is analogous to a V-type system where
perfect transmission occurs for an incident frequency 
$\pm\omega_\mathrm{EIT}=\left(\omega_S\Gamma_A-\omega_A\Gamma_S\right)/
(\Gamma_A-\Gamma_S)$~\cite{witthaut2010}

\section{Many emitters: effective spin models}

One of the main interests of having a platform with BS-mediated interactions
is to investigate spin models with long-range 
interactions~\cite{douglas2015,gonzaleztudela2015}. The study of these 
models has become an attractive avenue in quantum simulation because 
long-range interactions are the source of non-trivial many-body 
phases~\cite{hauke2010} and dynamics~\cite{richerme2014}, and are also very 
hard to treat classically.  

Let us now investigate how the shape of the QE interactions inherited from 
the topological bath translate into different many-body phases at zero 
temperature as compared to those produced by long-range interactions 
appearing in other setups such as trapped 
ions~\cite{hauke2010,richerme2014}, or standard waveguide setups. For that, 
we consider having $N_e$ emitters equally spaced and alternatively coupled 
to the $A/B$ lattice sites. After eliminating the bath, and adding a 
collective field with amplitude $\mu$ to control the number of spin 
excitations, the dynamics of the emitters (spins) is effectively given by: 
\begin{equation}
  H_\mathrm{spin} = \summ_{m,\,n} J^{AB}_{mn}
  \left(\sigma_{eg}^{m,A}\sigma_{ge}^{n,B}+\mathrm{H.c.}\right)
  - \frac{\mu}{2} \summ_n\left(\sigma_{z}^{n,A} + \sigma_{z}^{n,B}\right)
  \,, \label{eq:Hspinmiddle}
\end{equation}
denoting by $\sigma_{\nu}^{n,\alpha}$, $\nu=x,y,z$, the corresponding Pauli
matrix acting on the $\alpha\in\{A, B\}$ site in the $n$th unit cell. The
$J^{\alpha\beta}_{mn}$ are the spin-spin interactions derived in the 
previous subsection, whose localization length, denoted by $\xi$, and 
functional form can be tuned through system parameters such as $\omega_e$.

For example, when the lower (upper) BS mediates the interaction, the 
$J^{\alpha\beta}_{mn}$ has negative (alternating) sign for all sites, 
similar to the ones appearing in standard waveguide setups. When the range 
of the interactions is short (nearest neighbor), the physics is well 
described by the ferromagnetic XY model with a transverse 
field~\cite{katsura1962}, which goes from a fully polarized phase when 
$|\mu|$ dominates to a superfluid one in which spins start flipping as 
$|\mu|$ decreases. In the case where the interactions are long-ranged the 
physics is similar to that explained in Ref.~\cite{hauke2010} for power-law 
interactions ($\propto 1/r^3$). The longer range of the interactions tends 
to break the symmetry between the ferro/antiferromagnetic situations and 
leads to frustrated many-body phases. Since similar interactions also 
appear in other scenarios (standard waveguides or trapped ions), we now 
focus on the more different situation where the middle BS at $\omega_e=0$ 
mediates the interactions, such that the coefficients $J^{AB}_{mn}$ have
the form of Eq.~\eqref{eq:JABmarkov}.  

\begin{figure}[!htb]
  \centering
  \includegraphics{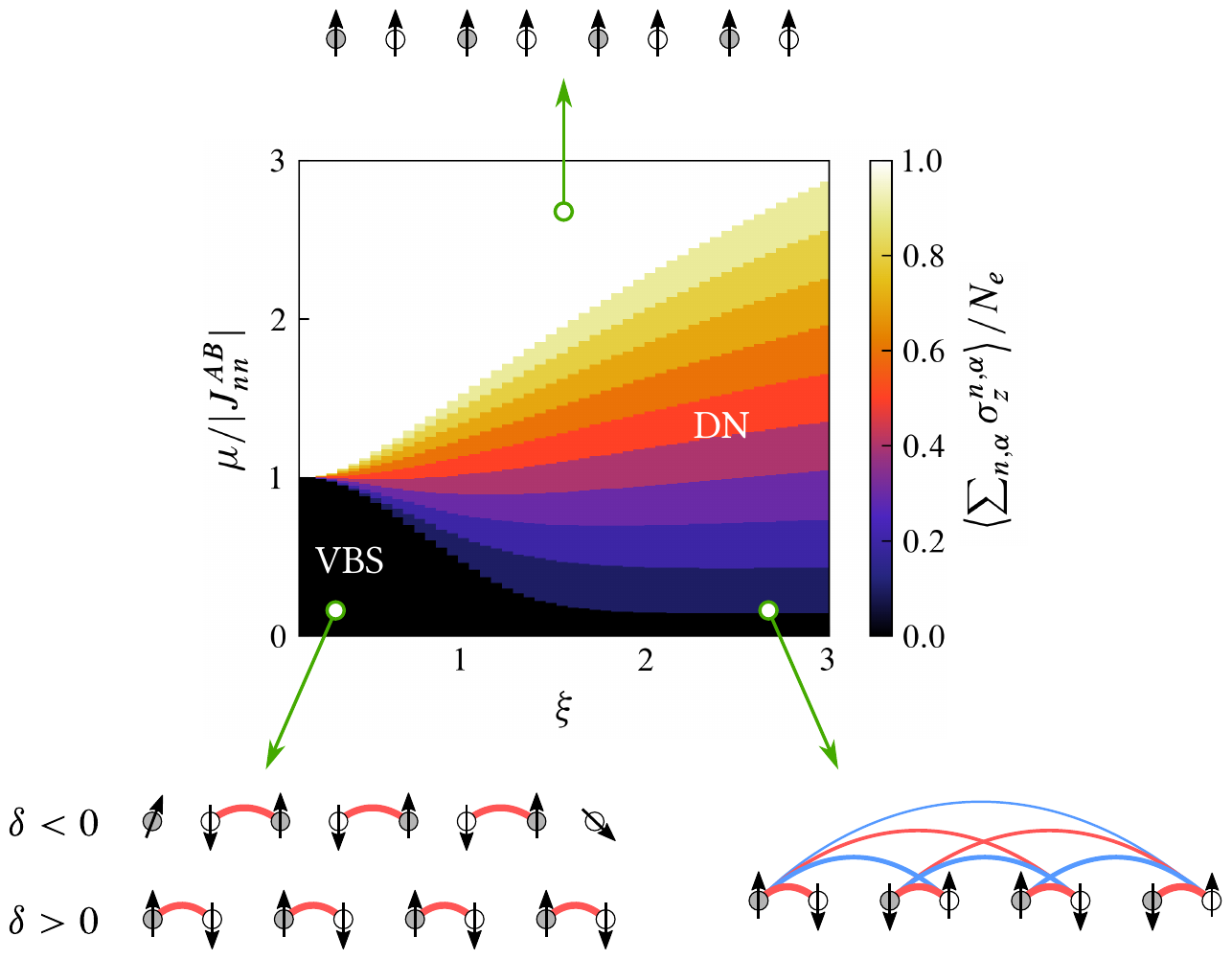}
  \caption{Ground state average polarization obtained by exact 
  diagonalization for a chain with $N_e=20$ emitters with frequency tuned to
  $\omega_e=0$ as a function of the chemical potential $\mu$ and the decay 
  length of the interactions $\xi$. The different phases discussed in the 
  text, a Valence-Bond Solid (VBS) and a Double N\'eel ordered phase (DN) 
  are shown schematically below, on the left and right respectively. 
  Interactions of different sign are marked with links of different color.}
  \label{fig:phasediag}
\end{figure}

\begin{figure}[!htb]
  \centering
  \includegraphics{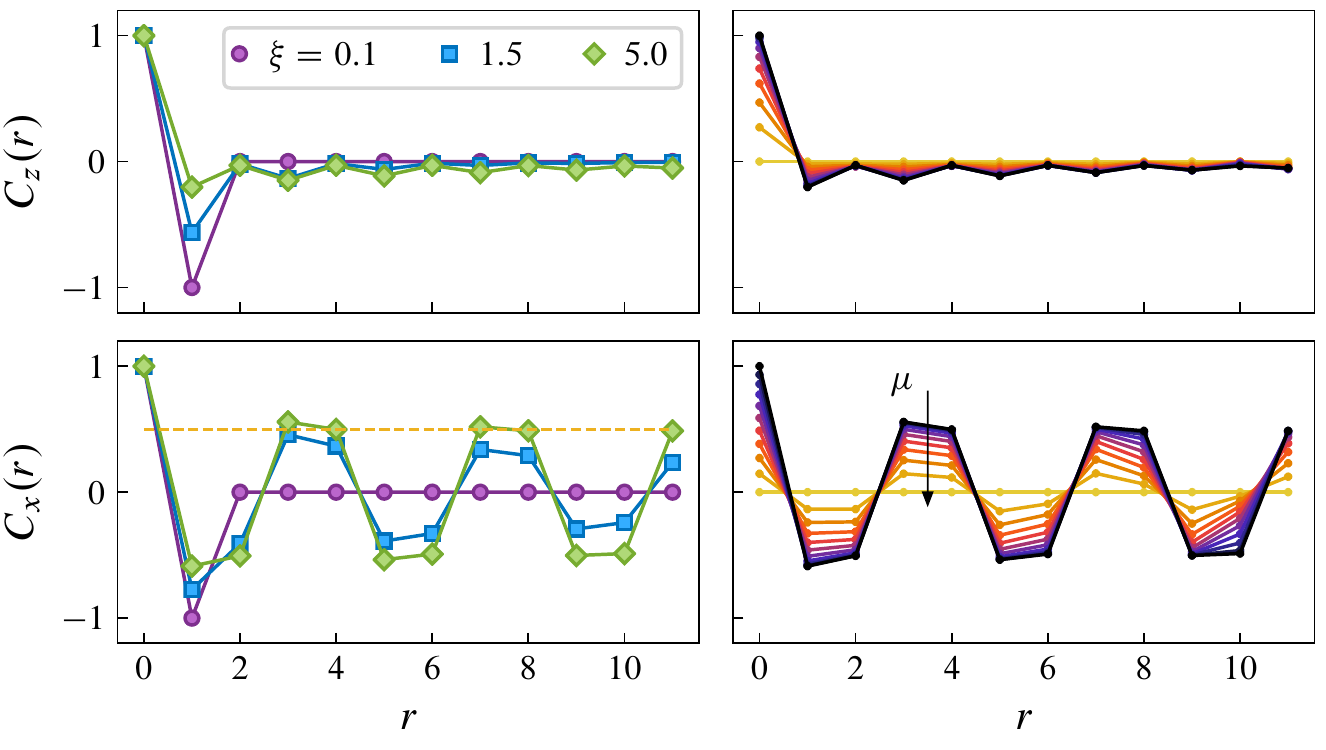}
  \caption{Correlations ${C_\nu(r) = \mean{\sigma^{9}_\nu
  \sigma^{9+r}_\nu} - \mean{\sigma^{9}_\nu}\mean{\sigma^{9+r}_\nu}}$,
  $\nu=x,y,z$, [$C_x(r)=C_y(r)$] for the same system as in 
  Fig.~\ref{fig:phasediag} for different interaction lengths, fixing $\mu=0$  
  (left column). Correlations for different chemical potentials fixing 
  $\xi=5$, darker colors correspond to lower chemical potentials (right 
  column). Note we have defined a single index $r$ that combines the unit 
  cell position and the sublattice index. The yellow dashed line marks the 
  value of $1/2$ expected when the interactions are of infinite range.}
  \label{fig:correlations}
\end{figure}

In that case, the Hamiltonian $H_\mathrm{spin}$ of 
Eq.~\eqref{eq:Hspinmiddle} is very unusual: i) spins only interact if they 
are in different sublattices, i.e., the system is bipartite ii) the 
interaction is chiral in the sense that they interact only in case they are 
properly sorted, i.e., the one in lattice $A$ to the left/right of that in 
lattice $B$, depending on the sign of $\delta$. Note that $\delta$ also 
controls the interaction length $\xi$. In particular, for $|\delta|=1$ the 
interaction only occurs between nearest neighbors, whereas for $\delta\to0$,
the interactions become of infinite range. These interactions translate into
a rich phase diagram as a function of $\xi$ and $\mu$, which we plot in 
Fig.~\ref{fig:phasediag} for a small chain with $N_e=20$ emitters (obtained 
with exact diagonalization). Let us guide the reader into the different 
parts: 
\begin{enumerate}
	\item The region with maximum average magnetization (in white) corresponds
    to the regimes where $\mu$ dominates such that all spins are aligned 
    upwards.  
	\item Now, if we decrease $\mu$ from this fully polarized phase in a 
    region where the localization length is short, i.e., $\xi\approx 0.1$, 
    we observe a transition into a state with zero average magnetization. 
    This behavior can be understood because in that short-range limit 
    $J^{AB}_{mn}$ only couples nearest neighbor $AB$ sites, but not $BA$ 
    sites as shown in the scheme of the lower part of the diagram for 
    $\delta>0$ (the opposite is true for $\delta<0$). Thus, the ground state
    is a product of nearest neighbor singlets (for $J>0$) or triplets (for 
    $J<0$). This state is usually referred to as Valence-Bond Solid in the 
    condensed matter literature~\cite{auerbach1994}. 
    Note, the difference between $\delta \gtrless 0$ is the presence (or 
    not) of uncoupled spins at the edges.  
	\item However, when the bath allows for longer range interactions
    ($\xi>1$), the transition from the fully polarized phase to the
    phase of zero magnetization does not occur abruptly but passing through 
    all possible intermediate values of the magnetization. Besides, we also 
    plot in Fig.~\ref{fig:correlations} the spin-spin correlations along the    
    $x$ and $z$ directions (note the symmetry in the $xy$ plane) for the 
    case of $\mu=0$ to evidence that a qualitatively different order 
    appears as $\xi$ increases. In particular, we show that the spins align 
    along the $x$ direction with a double periodicity, which we can 
    pictorially represent by $\ket{\up\up\dn\dn\up\up\dots}_x$, and that we 
    call double N\'eel order states. Such orders have been predicted as 
    a consequence of frustration in classical and quantum spin chains with 
    competing nearest and next-nearest neighbour 
    interactions~\cite{morita1972,sen1989,sen1992}, introduced to describe 
    complex solid state systems such as multiferroic 
    materials~\cite{qi2016}. In our case, this order emerges in a system 
    which has long-range interactions but no frustration as the system is 
    always bipartite regardless the interaction length.    
 \end{enumerate}
  
To gain analytical intuition of this regime, we take the limit 
$\xi\rightarrow\infty$, where the Hamiltonian~\eqref{eq:Hspinmiddle} reduces
to 
\begin{equation}
  H'_\mathrm{spin}=UH_\mathrm{spin}U^\dagger \simeq 
  J(S^+_AS^-_B + \mathrm{H.c.}) \,,
  \label{eq:Hinfinite}
\end{equation}
where $S^+_{A/B}=\sum_n \sigma^{n,A/B}_{eg}$, and the
unitary transformation $U=\prod_{n\in\mathbb{Z}_\mathrm{odd}}
\sigma^{n,A}_z\sigma^{n,B}_z$ is used to cancel the alternating signs of 
$J^{AB}_{mn}$. Equality in Eq.~\eqref{eq:Hinfinite} occurs for a system with
periodic boundary conditions, while for finite systems with open boundary 
conditions some corrections have to be taken into account due to the fact 
that not all spins in one sublattice couple to all spins in the other but 
only to those to their right/left depending on the sign of $\delta$. 
The ground state is symmetric under (independent) permutations in $A$ and 
$B$. In the thermodynamic limit we can apply mean field theory, which 
predicts symmetry breaking in the spin $xy$ plane. For instance, if $J<0$ 
and the symmetry is broken along the spin direction $x$, the spins will 
align so that 
$\mean{(S^x_A)^2}=\mean{(S^x_B)^2}=\mean{S^x_AS^x_B}=(N_e/2)^2$, and 
$\mean{S^x_A}^2 = \mean{S^x_B}^2 = (N_e/2)^2$\,.

Since $N_e$ is finite in our case, the symmetry is not broken, but it is
still reflected in the correlations, so that 
\begin{equation}
  \mean{\sigma^{m,A}_\nu \sigma^{n,A}_\nu} \simeq \mean{\sigma^{m,A}_\nu
	\sigma^{n,B}_\nu} \simeq 1/2 \,,\quad\nu=x,y\,.
\end{equation}
In the original picture with respect to $U$, we obtain the
double N\'eel order observed in Fig.~\ref{fig:correlations}. As can be
understood, the alternating nature of the interactions is crucial for 
obtaining this type of ordering. Finally, let us mention that the topology 
of the bath translates into the topology of the spin chain in a 
straightforward manner: regardless the range of the effective interactions, 
the ending spins of the chain will be uncoupled to the rest of spins if the 
bath is topologically non-trivial. 

This discussion shows the potential of the present setup to act as a quantum
simulator of exotic many-body phases not possible to simulate with other 
known setups.    

\section{Summary}

We have analyzed the dynamics of a set of quantum emitters (two-level 
systems) whose ground state-excited state transition couples to the modes of
a photonic lattice, which acts as a collective structured bath. For this, we
have computed analytically the collective self-energies, which allowed us to
use master equations and resolvent operator techniques to study the dynamics
of quantum emitters in the Markovian and non-Markovian regimes respectively.
The behavior depends fundamentally on whether the transition frequency of 
the emitters lays in a range of allowed bath modes or a band gap. If the 
transition frequency is tuned to a band gap we observe the following 
phenomena:
\begin{enumerate}
  \item Emergence of chiral bound states, that is, bound states that are 
    mostly localized on the left or right of the emitter depending on the 
    sign of the dimerization constant $\delta$ of the photonic bath. 
    Specifically, when the transition frequency of an emitter lays in the 
    middle of the inner band gap, it acts as a boundary between two photonic
    lattices with different topology. The resulting bound state has the same
    properties as a regular topological edge state of the SSH model 
    including the protection against certain types of disorder. 
  \item An exact calculation reveals that the existence conditions of the 
    bound states for two emitters are different depending on the sign of 
    $\delta$.
  \item These bound states give rise to dipolar interactions between the 
    emitters which depend on the topology of the underlying bath if the 
    emitters are coupled to different sublattices. In particular, when the 
    transition frequency of the emitters lays in the middle of the inner 
    band gap, the interactions can be toggled on and off by changing the 
    sign of $\delta$.
\end{enumerate}
When the emitters' frequency is tuned to the band, the topology of the bath 
reflects itself in:
\begin{enumerate}
  \item Different super/subradiant conditions depending on the sign of 
    $\delta$.
  \item A $\delta$-dependent Lamb Shift for a single emitter coupled 
    simultaneously to both $A$ and $B$ sublattices. This can be detected as
    a shift of the transmission dip in single-photon scattering experiments.
  \item Different scattering properties for two emitters in an $AB$ 
    configuration depending on the sign of $\delta$. This is a consequence 
    of the different super/subradiant conditions in each phase. 
\end{enumerate}
Last, we analyze the zero temperature phases of the effective spin 
Hamiltonians that can be generated in the many-emitter case after tracing 
out the bath degrees of freedom. We find that for short-range interactions 
the emitters realize a valence-bond solid. On the other hand, for long-range
interactions, the system becomes gapless and a double N\'eel order emerges.

\begin{subappendices}
  \section{Calculation of the self-energies \label{app:SelfEnergies}}
  To obtain analytical expressions for the self-energies, it is convenient
to first express $H_I$ in the bath eigenbasis. For this we just have to
invert~\eqref{eq:uandl} to obtain expresions for the local operators 
$c_{j\alpha}$ in terms of $u_k$ and $l_k$. Substituting 
in~\eqref{eq:HintQuantumOptics} we obtain
\begin{equation}
  \begin{split}
    H_I ={}& \frac{g}{\sqrt{2N}}\summ_{n\in S_A}\summ_k
    e^{i(kx_n+\phi_k)}(u_k + l_k)\sigma^n_{eg} \\
    & + \frac{g}{\sqrt{2N}}\summ_{n\in S_B}\summ_k
    e^{ikx_n}(u_k - l_k)\sigma^n_{eg} + \mathrm{H.c.} \,,
    \label{eq:Hint_eigenmodes}
  \end{split}
\end{equation}
where $S_A$ ($S_B$) denotes the set of emitters coupled to the $A$ ($B$)
sublattice.

Identifying $H_0=H_S$ and $V=H_B+H_I$, using 
$P=\ket{e}\ket{\vac}\bra{\vac}\bra{e}$, from Eq.~\eqref{eq:level-shift} we 
obtain
\begin{equation}
  \Sigma_e(z)\equiv\bra{\vac}\bra{e}R(z)\ket{e}\ket{\vac}
  =\frac{g^2}{2N}\summ_k\left(\frac{1}{z-\omega_k}
  +\frac{1}{z+\omega_k}\right)
  \,.
\end{equation}
In the thermodynamic limit ($N\to\infty$), the sum can be replaced by an 
integral, which can be computed easily with the change of variable 
$e^{ik}=y$ yielding the result shown in Eq.~\eqref{eq:selfe3}; the functions
$y_\pm(z)$ appearing in the expression are the roots of the polynomial 
\begin{equation}
  p(y)=y^2+\left[\frac{2J^2(1+\delta^2)-z^2}{J^2(1-\delta^2)}\right]y + 1\,.
\end{equation}

For the two emitter case, we will first show that the (anti)symmetric 
combinations 
$\sigma^\dagger_\pm=\left(\sigma^1_{eg}\pm\sigma^2_{eg}\right)/\sqrt{2}$
couple to orthogonal bath modes~\cite{gonzaleztudela2017}. Substituting 
$\sigma^n_{eq}$, $n=1,2$, in terms of $\sigma^\dagger_\pm$ in
Eq.~\eqref{eq:Hint_eigenmodes}, and pairing the terms with opposite
momentum, we obtain for the case where the two QEs couple to sublattice $A$
% \begin{gather}
%   H^{AA}_I = \frac{g}{\sqrt{N}}\summ_{k>0}\summ_{\beta=\pm}\sqrt{1+\beta\cos(kx_{12})}
%   (\tilde{u}_{k\beta} + \tilde{l}_{k\beta})\sigma^\dagger_\beta + \mathrm{H.c.} 
%   \,, \label{eq:HIAA_ini}\\
%   \tilde{u}_{k\pm} = \frac{\left[e^{i(kx_1+\phi_k)}\pm e^{i(kx_2+\phi_k)} \right]u_k 
%   + \left[e^{-i(kx_1+\phi_k)}\pm e^{-i(kx_2+\phi_k)} \right]u_{-k}}{2\sqrt{1\pm\cos(kx_{12})}}
%   \,, \\
%   \tilde{l}_{k\pm} = \frac{\left[e^{i(kx_1+\phi_k)}\pm e^{i(kx_2+\phi_k)} \right]l_k
%   + \left[e^{-i(kx_1+\phi_k)}\pm e^{-i(kx_2+\phi_k)} \right]l_{-k}}{2\sqrt{1\pm\cos(kx_{12})}}
%   \,,
%   \label{eq:HIAA_end}
% \end{gather}
\begin{gather}
  H^{AA}_I = \frac{g}{\sqrt{N}}\summ_{k>0}\summ_{\beta=\pm}
  \sqrt{1+\beta\cos(kx_{12})} (\tilde{u}_{k\beta} 
  + \tilde{l}_{k\beta})\sigma^\dagger_\beta + \mathrm{H.c.} 
  \,, \label{eq:HIAA_ini}\\
  \tilde{u}_{k\pm} = 
  \frac{\left[e^{i(kx_1+\phi_k)}\pm e^{i(kx_2+\phi_k)} \right]u_k 
  + \left[e^{-i(kx_1+\phi_k)}\pm e^{-i(kx_2+\phi_k)} \right]u_{-k}}
  {2\sqrt{1\pm\cos(kx_{12})}}
  \,, \\
  \tilde{l}_{k\pm} = 
  \frac{\left[e^{i(kx_1+\phi_k)}\pm e^{i(kx_2+\phi_k)} \right]l_k
  + \left[e^{-i(kx_1+\phi_k)}\pm e^{-i(kx_2+\phi_k)} \right]l_{-k}}
  {2\sqrt{1\pm\cos(kx_{12})}}
  \,.
\end{gather}
Here, $x_{12}=x_2-x_1$ is the signed distance between the two emitters. 
For the case where the two QEs are on a different sublattice
\begin{gather}
  \begin{split}
  H^{AB}_I = \frac{g}{\sqrt{N}}\summ_{k>0}\summ_{\beta=\pm}
    \Big[&\sqrt{1+\beta\cos(kx_{12}-\phi_k)}\, 
    \tilde{u}_{k\beta}\sigma^\dagger_\beta \\
  &\qquad+\sqrt{1-\beta\cos(kx_{12}-\phi_k)}\,
  \tilde{l}_{k\beta}\sigma^\dagger_\beta \Big] + \mathrm{H.c.} \,, 
  \label{eq:HIAB_ini}
  \end{split}
  \\
  \tilde{u}_{k\pm}=\frac{\left[e^{i(kx_1+\phi_k)}\pm e^{ikx_2} \right]u_k
  + \left[e^{-i(kx_1+\phi_k)}\pm e^{-ikx_2} \right]u_{-k}}
  {2\sqrt{1\pm\cos(kx_{12}-\phi_k)}}
  \,, \\
  \tilde{l}_{k\pm}=\frac{\left[e^{i(kx_1+\phi_k)}\mp e^{ikx_2} \right]l_k
  + \left[e^{-i(kx_1+\phi_k)}\mp e^{-ikx_2} \right]l_{-k}}
  {2\sqrt{1\mp\cos(kx_{12}-\phi_k)}}
  \,.
  \label{eq:HIAB_end}
\end{gather}
The denominators in the definition of $\tilde u_{k\pm}$ and $\tilde l_{k\pm}$ 
come from normalization. Importantly, these modes are orthogonal, 
they satisfy
\begin{equation}
  \left[\tilde u_{k\alpha},\tilde u^\dagger_{k'\alpha'}\right]
  =\left[\tilde l_{k\alpha},\tilde l^\dagger_{k'\alpha'}\right]
  =\delta_{kk'}\delta_{\alpha\alpha'}\,.
\end{equation}
Since $\omega_k=\omega_{-k}$, we have that the bath Hamiltonian is 
also diagonal in this new basis. The two other configurations, 
can be analyzed analogously. From these expressions for the 
interaction Hamiltonian, it is possible to obtain the self-energy 
for the (anti)symmetric states of the two QE, 
\begin{gather}
  \Sigma^{AA/BB}_\pm=\frac{g^2}{N}\summ_{k>0}
  \left[\frac{1\pm\cos(k x_{12})}{z-\omega_k}
  +\frac{1\pm\cos(k x_{12})}{z+\omega_k}\right]\,,\\
  \Sigma^{AB}_\pm=\frac{g^2}{N}\summ_{k>0}
  \left[\frac{1\pm\cos(k x_{12}-\phi_k)}{z-\omega_k}
  +\frac{1\mp\cos(k x_{12}-\phi_k)}{z+\omega_k}\right]\,.
\end{gather}
As it turns out, they can be cast in the form 
$\Sigma^{\alpha\beta}_\pm=\Sigma_e\pm\Sigma^{\alpha\beta}_{12}$, with
\begin{gather}
  \Sigma^{AA/BB}_{mn}(z;x_{mn})=\frac{g^2}{N}\summ_k
  \frac{ze^{ikx_{mn}}}{z^2-\omega^2_k}\,, \label{eq:selfeAABB}\\
  \Sigma^{AB}_{mn}(z;x_{mn})=\frac{g^2}{N}\summ_k
  \frac{\omega_ke^{i(kx_{mn}-\phi_k)}}{z^2-\omega^2_k} \,, 
  \label{eq:selfeAB}
\end{gather}
where $x_{mn}=x_n-x_m$. It can be shown that 
\begin{equation}
\Sigma^{BA}_{mn}(z; \delta, x_{mn}) = \Sigma^{AB}_{nm}(z; \delta, -x_{mn}) 
  = \Sigma^{AB}_{mn}(z; -\delta, x_{mn} - 1)\,.
\end{equation}
Again, these expressions in the thermodynamic limit can be evaluated
substituting the sum by an integral, which can be computed easily with
the change of variable $y=\exp(i\sign(x_{12})k)$, giving the results shown 
in Eqs.~\eqref{eq:SigmaAA} and \eqref{eq:SigmaAB}.

  \section{Quantum optical master equation \label{app:QuantumOpticalME}}
  From Eq.~\eqref{eq:Hint_eigenmodes}, we can readily obtain the expression 
of $H_I$ in the interaction picture
\begin{equation}
  \tilde{H_I}(t) = \summ_n e^{i\omega_e t}\sigma^n_{eg}\otimes 
  \tilde{B_n}(t) + \mathrm{H.c.}\,, \label{eq:Hintint}
\end{equation}
where 
\begin{equation}
  \tilde{B_n}(t)=\begin{cases} \frac{g}{\sqrt{2N}}\summ_k
    e^{i(kx_n+\phi_k)}\left(e^{-i\omega_kt}u_k + e^{i\omega_kt}l_k\right)
    &\text{ if }n\in S_A \,.\\
    \frac{g}{\sqrt{2N}}\summ_k
    e^{ikx_n}\left(e^{-i\omega_kt}u_k - e^{i\omega_kt}l_k\right)
    &\text{ if }n\in S_B \,.
  \end{cases}
\end{equation}
Now, expanding the integrand in Eq.~\eqref{eq:BornMarkov}, neglecting 
the fast-rotating terms $\propto e^{\pm i2\omega_e t}$,  we arrive at
\begin{equation}
\begin{split}
  \dot{\tilde{\rho_S}}(t) &=
  -\summ_{m,\,n}\sigma^m_{eg}\sigma^n_{ge}\tilde{\rho_S}(t)
  \int_0^\infty \diff{s}e^{i\omega_e s}\mean*{B_m(t)B^\dagger_n(t-s)}\\
  &\phantom{{}={}}\negmedspace
  +\summ_{m,\,n}\sigma^n_{ge}\tilde{\rho_S}(t)\sigma^m_{eg}
  \int_0^\infty \diff{s}e^{-i\omega_e s}\mean*{B_m(t-s)B^\dagger_n(t)}\\
  &\phantom{{}={}}\negmedspace
  +\summ_{m,\,n}\sigma^n_{ge}\tilde{\rho_S}(t)\sigma^m_{eg}
  \int_0^\infty \diff{s}e^{i\omega_e s}\mean*{B_m(t)B^\dagger_n(t-s)}\\
  &\phantom{{}={}}\negmedspace
  -\summ_{m,\,n}\tilde{\rho_S}(t)\sigma^m_{eg}\sigma^n_{ge}
  \int_0^\infty \diff{s}e^{-i\omega_e s}\mean*{B_m(t-s)B^\dagger_n(t)} \,.
\end{split}
\end{equation}
Let us compute the bath correlations assuming 
that the bath is in the vacuum state. If both $m,n\in S_B$, 
\begin{align}
  &\int_0^\infty \diff{s}e^{i\omega_e s}\mean*{B_m(t)B^\dagger_n(t-s)}
  \nonumber\\ 
  &\quad=\frac{g^2}{2N}\summ_k e^{ik(x_m-x_n)}\int_0^\infty \diff{s}
  e^{i\omega_e s} \left(e^{-i\omega_k s}+e^{i\omega_k s}\right) \\
  &\quad=i\frac{g^2}{2N}\summ_k e^{ik(x_m-x_n)}\left(\frac{1}{\omega_e+i0^+-\omega_k}
  +\frac{1}{\omega_e+i0^++\omega_k}\right)\\
  &\quad=i\Sigma^{BB}_{mn}(\omega_e+i0^+)\,.
\end{align}
In the last equality we have substituted the deffinition of the collective 
self energy for the $BB$ configuration, Eq.~\eqref{eq:selfeAABB}. Similarly,
the other correlator can be computed noting that 
$\big\langle B_m(t)B^\dagger_n(t')\big\rangle$ only depends on the time 
difference $t-t'$, so the integral is the same changing $s\to -s$. 
\begin{equation}
  \int_0^\infty \diff{s}e^{-i\omega_e s}\mean*{B_m(t-s)B^\dagger_n(t)}= 
  -i\Sigma^{BB}_{mn}(\omega_e-i0^+)\,.
\end{equation}
Analogously if $m\in S_A$ and $n\in S_B$, 
\begin{align}
  &\int_0^\infty \diff{s}e^{i\omega_e s}\mean*{B_m(t)B^\dagger_n(t-s)}
  \nonumber\\
  &\quad=\frac{g^2}{2N}\summ_k e^{i[k(x_m-x_n)+\phi_k]}\int_0^\infty \diff{s}
  e^{i\omega_e s} \left(e^{-i\omega_k s}-e^{i\omega_k s}\right) \\
  &\quad=i\frac{g^2}{2N}\summ_k e^{i[k(x_m-x_n)+\phi_k]}\left(\frac{1}{\omega_e+i0^+-\omega_k}
  -\frac{1}{\omega_e+i0^++\omega_k}\right)\\
  &\quad= i\Sigma^{AB}_{mn}(\omega_e+i0^+) \,,
\end{align}
and
\begin{equation}
  \int_0^\infty \diff{s}e^{-i\omega_e s}\mean*{B_m(t-s)B^\dagger_n(t)}
  = -i\Sigma^{AB}_{mn}(\omega_e-i0^+) \,.
\end{equation}
So in general, we can replace 
\begin{gather}
  \int_0^\infty \diff{s}e^{i\omega_e s}\mean*{B_m(t)B^\dagger_n(t-s)} = 
  i\Sigma^{\alpha\beta}_{mn}(\omega_e+i0^+) \,, \\
  \int_0^\infty \diff{s}e^{-i\omega_e s}\mean*{B_m(t-s)B^\dagger_n(t)} =
  -i\Sigma^{\alpha\beta}_{mn}(\omega_e-i0^+) \,.
\end{gather}
Finally, splitting the self-energies in their real and imaginary parts, 
$\Sigma^{\alpha\beta}_{mn}(\omega_e\pm i0^+)=J^{\alpha\beta}_{mn}\mp i
\Gamma^{\alpha\beta}_{mn}/2$, gathering the terms that go with 
$J^{\alpha\beta}_{mn}$ and those that go with $\Gamma^{\alpha\beta}_{mn}$,
\begin{equation}
  \dot{\tilde{\rho_S}} = -i\summ_{m,\,n}J^{\alpha\beta}_{mn}
  [\sigma^m_{eg}\sigma^n_{ge},\tilde{\rho_S}]
  +\summ_{m,\,n}\frac{\Gamma^{\alpha\beta}_{mn}}{2}
  \left(2\sigma^n_{ge}\tilde{\rho_S}\sigma^m_{eg}
  -\sigma^m_{eg}\sigma^n_{ge}\tilde{\rho_S} 
  -\tilde{\rho_S}\sigma^m_{eg}\sigma^n_{ge}\right)\,.
\end{equation}
Back to the Schr\"odinger picture we have
\begin{multline}
  \dot{\rho_S} = -i[H_S,\rho_S]
  -i\summ_{m,\,n}J^{\alpha\beta}_{mn}[\sigma^m_{eg}\sigma^n_{ge},\rho_S] \\
  +\summ_{m,\,n}\frac{\Gamma^{\alpha\beta}_{mn}}{2}
  \left(2\sigma^n_{ge}\rho_S\sigma^m_{eg} -\sigma^m_{eg}\sigma^n_{ge}\rho_S 
  -\rho_S\sigma^m_{eg}\sigma^n_{ge}\right)\,.
\end{multline}

  % \newpage
  \section{Algebraic decay \label{app:AlgebraicDecay}}
  The fractional decay of the emitter can be better seen when the 
emitter's frequency is precisely at any of the band edges. There, the 
contribution of the branch cuts on the dynamics is larger. Defining
\begin{equation}
   D(t)\equiv\psi_e(t)-\summ_{z_\mathrm{BS}}R(z_\mathrm{BS})
   e^{-iz_\mathrm{BS}t} \,,
\end{equation}
at long times we have
\begin{equation}
  \lim_{t\to\infty}D(t)
  \simeq\summ_j \psi_{\mathrm{BC},j}(t) = \summ_j K_j(t)e^{-ix_jt} \,,
\end{equation}
with
\begin{equation}
  K_j(t)=\frac{\pm 1}{2\pi}\int_0^\infty \diff{y} 
  \frac{2\Sigma_e(x_j-iy)e^{-yt}}
  {(x_j-iy-\omega_e)^2-\Sigma^2_e(x_j-iy)}\,. \label{eq:Kintegrand}
\end{equation}
The long-time average of the decaying part of the dynamics can be computed 
as
\begin{equation}
  \overline{|D(t)|^2}\equiv\lim_{t\to\infty}\frac{1}{t}
  \int_0^t \diff{t'} |D(t')|^2=\summ_j|K_j(t)|^2 \,.
\end{equation}
If the emitter's transition frequency is close to one of the band edges, 
$\omega_e\simeq x_0$, then $\overline{|D(t)|^2}\simeq|K_0(t)|^2$. In the 
long-time limit, we can expand the integrand of \eqref{eq:Kintegrand} in 
power series around $y=0$,
\begin{align} 
  K_0(t)&=\frac{\pm1}{2\pi}\int_0^\infty \diff{y}\left[
    \frac{4}{g^2}\sqrt{\frac{i(2-x^2_0+2\delta^2)}{x_0}} 
    + \mathcal{O}(y)\right]
  y^{1/2}e^{-yt}\\
  &\simeq \frac{\pm1}{\sqrt{\pi g^2}}\sqrt{\frac{i(2-x^2_0+2\delta^2)}{x_0}}
  t^{-3/2}+\mathcal{O}(t^{-5/2})\,.
\end{align}
Therefore, to leading order $\overline{|D(t)|^2}\sim t^{-3}$. 

\end{subappendices}

\chapter[Conclusions and outlook]{Conclusions and outlook/\\\textcolor{gray}{Conclusiones y perspectiva} \label{chap:conclu}}
In this thesis we have studied problems that generalize the physics of 
topological insulators. In the first part, we analyze the dynamics of 
doublons in 1D and 2D topological lattices. On the second part, we 
investigate the dynamics of quantum emitters interacting with a common 
topological waveguide QED bath, namely, a photonic analogue of the SSH 
model. %We summarize hereunder the main achievements of our investigation.

To understand the dynamics of doublons, we have derived an effective 
single-particle Hamiltonian taking into account also the effect of a 
periodic driving. It contains two terms: one corresponding to an effective 
doublon hopping renormalized by the driving, and another one corresponding 
to an effective on-site chemical potential. This helped us understand 
unusual phenomena that constrain doublon motion. For example, 
Shockley-like edge states can be induced in any finite lattice by reducing 
the effective doublon hopping with the driving. These states may or may not 
compete against topological edge states depending on the dimensionality of 
the lattice. For 1D lattices, topological phases require the presence of 
chiral (sublattice) symmetry, which is spoiled by the on-site chemical 
potential. On the other hand, for 2D lattices threaded by a magnetic flux 
no symmetries are required, and topological edge states coexist with 
Shockley-like edge states. We demonstrate that edge states, either 
topological or not, can be used to produce the transfer of doublons between 
distant sites (on the edge) of any finite lattice. Furthermore, in 2D 
lattices with sites with different number of neighbors, doublon's dynamics 
can be confined to just one sublattice. We also analyze the feasibility of 
doublon experiments in noisy systems such as arrays of QDs, and estimate a 
doublon lifetime on the order of 10 ns for current devices.

For the analysis of quantum emitter dynamics, we have employed different 
techniques valid in the Markovian and non-Markovian regimes. When the 
emitters are spectrally tuned to one of the band gaps, the non-trivial 
topology of the bath leads to the emergence of chiral photon bound states, 
which are localized on the left or right of the QE depending on the sign of 
the dimerization constant $\delta$. This gives rise to directional 
interactions between the emitters. Specifically, when the emitter 
frequency is tuned to the middle of the inner band gap, the 
interaction between emitters coupled to different sublattices can be toggled
on and off by changing the sign of $\delta$. When the emitters are 
spectrally tuned to one of the bath's bands, different super/subradiant 
states appear depending on the sign of $\delta$. This leads to different 
behavior when a single photon scatters off one QE coupled both to the $A$ 
and $B$ cavities in a single unit cell, or two QEs coupled to different 
sublattices. Last, we analyze the effective spin Hamiltonians that
can be generated in the many-emitter case, and compute its phase diagram 
with exact diagonalization techniques. We find that for short-range 
interactions the emitters realize a valence bond solid phase, while for 
long-range interactions a double N\'eel order emerges.

One of the attractive points of our predictions is that they can be observed
in several platforms by combining tools that, in most of the cases, have 
been already implemented experimentally. Regarding the first part, doublons 
have been observed in several experiments using cold atoms trapped in 
optical lattices~\cite{winkler2006,preiss2015,tai2017}. Also, the SSH model 
has been realized in this kind of setups~\cite{atala2013}. As for the second
part, the photonic analogue of the SSH model has been implemented in 
several photonic platforms~\cite{malkova2009,jean2017,parto2018,zhao2018}, 
including some recent photonic crystal realizations~\cite{chen2018}. The 
latter are particularly interesting due to the recent advances in their 
integration with solid-state and natural atomic emitters (see 
Refs.~\cite{lodahl2015,chang2018} and references therein). Superconducting 
metamaterials mimicking standard waveguide QED are now being routinely built
and interfaced with one or many qubits in 
experiments~\cite{liu2017,mirhosseini2018}. The only missing piece is the 
periodic modulation of the couplings between cavities to obtain the SSH 
model, for which there are already proposals using circuit 
superlattices~\cite{goren2018}. Furthermore,
Quantum optical phenomena can be simulated in pure atomic scenarios by 
using state-dependent optical lattices. The idea is to have two different 
trapping potentials for two atomic metastable states, such that one state 
mostly localizes, playing the role of QEs, while the other state 
propagates as a matter-wave. This proposal~\cite{devega2008} has been 
recently used~\cite{krinner2018} to explore the physics of standard 
waveguide baths.   
Beyond these platforms, the bosonic analogue of the SSH model has also been 
discussed in the context of metamaterials~\cite{tan2014} or plasmonic and 
dielectric nanoparticles~\cite{kruk2017,pocock2018}, where the predicted 
phenomena could as well be observed.  

Topological matter is a very active research field in which 
important advances, both on theoretical and applied grounds, have been 
produced in recent years. The research here presented demonstrates the 
variety of phenomena that appear at the crossover between this and other 
fields of physics. This is a rather new and unexplored research direction,
which surely will provide exciting discoveries in the near future.
Prospective studies could investigate the dynamics of doublons in 1D 
lattices with higher topological invariants~\cite{perezgonzalez2019b}, or 
the use of topological edge states to transfer few-particle states other 
than doublons between distant regions of a lattice. It would also be 
interesting to analyze the dynamics of several doublons, or the dynamics 
of the Hubbard model in the intermediate interaction regime, where the 
interaction is of the order of the hopping.
Regarding the dynamics of quantum emitters, it would be interesting to 
study other topological baths. For example, the phenomena associated to 
the SSH bath would have analogues in higher dimensions considering 
photonic baths after higher-order topological 
insulators~\cite{benalcazar2017,xie2019}. As for the photonic SSH bath, 
we are currently working on an in-depth survey of the many-body phases 
that appear in each of the band gaps. Also considering other types of 
emitters (with a more complex level structure) which would allow for 
different effective spin interactions~\cite{douglas2015,gonzaleztudela2015}./\\                

\begin{otherlanguage}{spanish}
{\color{gray}
\noindent En esta tesis hemos estudiado problemas que generalizan la física 
de los aislantes topológicos. En la primera parte analizamos la dinámica de 
dublones en redes topológicas 1D y 2D. En la segunda parte, investigamos 
la dinámica de emisores cuánticos que interactúan con un baño común 
topológico tipo guía de ondas, concretamente, con un análogo fotónico del 
modelo SSH. %A continuación resumimos los mayores logros de nuestra investigación:

Para entender la dinámica de los dublones, hemos derivado un Hamiltoniano 
efectivo de una partícula que además incluye el efecto de una modulación
periódica del sistema (\emph{driving} en inglés). Este Hamiltoniano efectivo
contiene dos términos: uno se corresponde con el salto de dublones en la 
red, renormalizado por el \emph{driving}, y el otro se corresponde con un 
potencial químico local efectivo. Esto nos ha permitido entender fenómenos 
inusuales que constriñen la dinámica de los dublones. Por ejemplo, estados 
de borde de tipo Shockley pueden inducirse en cualquier red finita 
reduciendo el salto del dublón mediante el \emph{driving}. Estos estados 
pueden competir o no con estados de borde topológicos, en función de la 
dimensión de la red. En redes 1D, las fases topológicas requieren la 
presencia de simetría quiral (simetría de subred), la cual se rompe debido 
al potencial químico local. Por otro lado, en redes 2D atravesadas por un 
flujo de campo magnético, no se requiere ninguna simetría para tener fases 
topológicas, y los estados de borde topológicos pueden coexistir con estados
de borde de tipo Shockley. Demostramos que los estados de borde, ya sean 
topológicos o no, pueden usarse para transferir dublones entre sitios 
distantes (en el borde) de cualquier red finita. Además, en redes 2D con 
sitios con distinto índice de coordinación, la dinámica de los dublones 
puede confinarse a una única subred. También analizamos la posibilidad de 
hacer experimentos con dublones en sistemas ruidosos como son las cadenas de
puntos cuánticos, y estimamos para el dublón una vida media del orden de 10 
ns en dispositivos actuales.

Para el análisis de la dinámica de emisores cuánticos, hemos empleado 
distintas técnicas válidas en el régimen Markoviano y no-Markoviano. Cuando 
la frecuencia de los emisores se encuentra en uno de los \emph{band gaps}, 
la topología no trivial del baño produce la aparición de estados ligados de 
fotones que son quirales, es decir, que están localizados a la izquierda o 
derecha del emisor en función del signo de la constante de dimerización 
$\delta$. Esto da lugar a interacciones direccionales entre los emisores. 
En concreto, cuando la frecuencia de los emisores está ajustada al centro 
del \emph{band gap} interno, la interacción entre emisores acoplados a 
subredes distintas puede activarse y desactivarse cambiando el signo de 
$\delta$. Cuando la frecuencia de los emisores se encuentra en una de las 
bandas del baño, distintos estados super/subradiantes aparecen en función 
del signo de $\delta$. Esto produce un comportamiento distinto en función de
la topología del baño cuando un foton se dispersa a través de dos emisores 
acoplados a redes distintas. Por último, analizamos el Hamiltoniano de spin 
que puede generarse en el caso de muchos emisores, y calculamos su diagrama 
de fases mediante técnicas de diagonalización exacta. Encontramos que para 
interacciones de corto alcance los emisores realizan un sólido de enlaces de
valencia, mientras que para interacciones de largo alcance aparece un orden 
de tipo Néel doble.
 
Uno de los puntos atractivos de nuestras predicciones es que pueden 
observarse en varias plataformas combinando herramientas que, en la 
mayoría de los casos, ya han sido implementadas experimentalmente. Respecto 
a la primera parte, los dublones han sido observados en varios experimentos 
utilizando átomos ultrafríos atrapados en redes 
ópticas~\cite{winkler2006,preiss2015,tai2017}. Además, el modelo SSH ya ha 
sido realizado en este tipo de experimentos~\cite{atala2013}. Respecto a 
la segunda parte, el análogo del modelo 
SSH ha sido implementado en varias plataformas 
fotónicas~\cite{malkova2009,jean2017,parto2018,zhao2018}, incluidas algunas 
realizaciones de cristales fotónicos~\cite{chen2018} que son particularmente
interesantes debido a la reciente integración de emisores naturales y 
de estado sólido en las mismas (ver Refs.~\cite{lodahl2015,chang2018} y 
referencias allí mencionadas). Metamateriales superconductores que imitan 
guías de onda cuánticas se construyen ahora de forma rutinaria y ya hay 
experimentos en los que se acoplan con uno o varios 
qubits~\cite{liu2017,mirhosseini2018}. La única pieza que falta es la 
modulación periódica de los acoplos entre cavidades para obtener el modelo 
SSH, para lo cual ya hay propuestas utilizando superredes de 
circuitos~\cite{goren2018}. Asimismo, fenómenos de la óptica cuántica pueden simularse
en sistemas púramente atómicos utilizando redes ópticas dependientes de los 
estados cuánticos de los átomos. La idea es tener dos potenciales distintos 
para dos estados atómicos metaestables, de forma que un estado está mayormente
localizado, jugando el papel de los emisores, mientras que el otro 
estado se propaga como una onda de materia. Esta propuesta~\cite{devega2008}
ha sido utilizada recientemente~\cite{krinner2018} para explorar la física 
de baños de guía de ondas estándar. Más allá de estas plataformas, el 
análogo bosónico del modelos SSH también se ha discutido en el contexto de 
los metamateriales~\cite{tan2014} o de sistemas 
plasmónicos~\cite{kruk2017,pocock2018}, donde los fenómenos predichos 
podrían observarse también. 

La materia topológica es un campo de investigación muy activo en el que se 
han producido importantes avances tanto a nivel teórico como práctico en los
últimos años. Las investigaciones aquí presentadas demuestran la variedad de
fenómenos que aparecen al combinar este campo con otros campos de la física.
Esta es una dirección de investigación aún nueva e inexplorada que 
seguramente dará lugar a grandes descubrimientos en un futuro cercano. 
Estudios futuros podrían investigar la dinámica de dublones en redes 1D con 
invariantes topológicos más altos~\cite{perezgonzalez2019b}, o el uso de 
estados de borde topológicos para la transferencia de estados de pocas 
partículas distintos de los dublones entre regiones distantes de una red. 
También sería interesante analizar la dinámica de varios dublones, o la 
dinámica del modelo de Hubbard en el régimen de interacción intermedio, en 
el que ésta es del mismo orden que el salto de las partículas. Respecto a 
la dinámica de emisores cuánticos, sería interesante estudiar otros baños 
topológicos. Por ejemplo, los fenómenos descritos para el baño tipo SSH 
tendrían análogos en dimensiones mayores considerando baños similares a 
aislantes topológicos de orden más alto~\cite{benalcazar2017,xie2019}. En 
cuanto al baño fotónico SSH, estamos en estos momentos realizando un 
análisis en profundidad de las distintas fases que aparecen en cada uno de 
los band gaps. Además, estamos considerando también otros tipos de emisores 
(con una estructura de niveles más compleja) que permitirían generar 
distintas interacciones de spin efectivas~\cite{douglas2015,gonzaleztudela2015}.
}
\end{otherlanguage}

\printbibliography[notkeyword=own,heading=bibintoc,title={Bibliography}]

% \newrefcontext[sorting=ynt]
\newpage
\printbibliography[keyword=own,heading=subbibintoc,title={List of publications}]
% \nocite{*}

\end{document}